%% file: IB_corrections.tex
\documentclass[a4paper,11pt]{article}
\pdfoutput=1 % if your are submitting a pdflatex (i.e. if you have
\usepackage{jheppub} % for details on the use of the package, please
\usepackage{url}

\usepackage[latin1]{inputenc}
\usepackage[american]{babel}
\usepackage[OT1]{fontenc}
\usepackage{graphicx}
\usepackage{subfig}

\usepackage{pgf}
\usepackage{tikz}
\usepackage{graphicx}
\usepackage{amsmath}
\usepackage{nicefrac}
\usepackage{units}
\usepackage{textcomp}
\usepackage{float}
\usepackage{mathcomp} 
\usepackage{mathrsfs}
\usepackage{amssymb}
\usepackage{dsfont}
\usepackage{xcolor}
\usepackage{esvect}

\usepackage{tikz}

\usetikzlibrary{decorations.pathmorphing}
\usetikzlibrary{decorations.markings}

\tikzset{
    photon/.style={thick,black,decorate,decoration={snake,post length=0pt,pre
    length=2pt}}, 
    quark/.style={thick,black,postaction={decorate},
        decoration={markings,mark=at position .55 with
{\arrow[draw=black]{>}}}},
    thickquark/.style={very thick,black,postaction={decorate},
        decoration={markings,mark=at position .55 with {\arrow[black]{>}}}},
    thickquarkarrow/.style={very thick,black,postaction={decorate},
        decoration={markings,mark=at position .55 with {\arrow[black]{<}}}},
    gluon/.style={thick,decorate, draw=black,
        decoration={coil,amplitude=4pt, segment length=5pt,post length=0pt,pre
    length=2pt}},
    W/.style={thick,black,decorate,decoration={snake,segment length=5pt,post
    length=0pt,pre length=2pt}},
    H/.style={very thick,dashed}, 
}

\newcommand{\Psibar}{\overline{\Psi}}
\newcommand{\psibar}{\overline{\psi}}

\newcommand{\Oalpha}{\mathcal{O}(\alpha)}
\newcommand{\Oalphasquare}{\mathcal{O}(\alpha^2)}
\newcommand{\dde}{\frac{\partial}{\partial e}}
\newcommand{\ddesq}{\frac{\partial^2}{\partial e^2}}

\title{\boldmath Isospin breaking corrections to meson masses and the 
hadronic vacuum polarization: a comparative study}

\author[a]{P. Boyle,}
\author[b]{V.~G\"ulpers,}
\author[b]{J.~Harrison,}
\author[b]{A.~J\"uttner,}
\author[c]{C.~Lehner,}
\author[a]{A.~Portelli,}
\author[b]{C.~T.~Sachrajda,}

\affiliation[a]{School of Physics and Astronomy, University of 
Edinburgh,\\Edinburgh EH9 3JZ, United Kingdom}
\affiliation[b]{School of Physics and Astronomy, University of 
Southampton,\\Southampton SO17 1BJ, United Kingdom}
\affiliation[c]{Physics Department, Brookhaven National Laboratory,\\ Upton, NY 
11973, USA}

\emailAdd{V.M.Guelpers@soton.ac.uk}

\abstract{We calculate the strong isospin breaking and QED corrections to meson 
masses 
and the hadronic vacuum polarization in an exploratory study on a 
$64\times24^3$ lattice with an inverse lattice spacing of $a^{-1}=1.78$~GeV and 
an 
isospin symmetric pion mass of $m_\pi=340$~MeV. We include QED in an 
electro-quenched setup using two different 
methods, a stochastic and a perturbative approach. 
We find that the electromagnetic correction to the leading hadronic 
contribution to the anomalous magnetic moment of the muon is smaller than $1\%$ 
for the up quark and $0.1\%$ for the strange quark, 
although it should be noted that this is obtained using unphysical light 
quark masses. In addition to the results themselves,  we compare the precision 
which can be reached for the same computational cost using each method. Such a 
comparison is also made for the meson electromagnetic mass-splittings.
}

\usepackage{todonotes}

\begin{document}

\maketitle
\flushbottom

\setcounter{tocdepth}{3}
%\tableofcontents

%\newpage
\input{1_introduction}
\input{2_method}

\input{3_mesonmasses}

\input{4_hvp}
\input{5_conclusions}

\section*{Acknowledgements}
\enlargethispage{\baselineskip}
The authors warmly thank F. Sanfilippo for usefull discussions. This work has 
received funding from the  STFC 
Grant ST/L000296/1, the EPSRC Centre for
Doctoral Training in Next Generation Computational Modelling grant
EP/L015382/1 and
from the European Research Council under the European Union's Seventh Framework
Programme (FP7/2007-2013) / ERC Grant agreement 279757.
V.G. acknowledges partial support from the Horizon 2020 INVISIBLESPlus 
(H2020-MSCA-RISE-2015 -690575).
P.A.B. and A.P. are supported in part by UK STFC grant ST/L000458/1.
C.L. is supported in part by US DOE Contract $\#$AC-02-98CH10886(BNL) and in 
part through a DOE Office of Science Early Career Award.
This work used the DiRAC Blue Gene Q Shared Petaflop system at the
University of Edinburgh, operated by the Edinburgh Parallel Computing
Centre on behalf of the STFC DiRAC HPC Facility (www.dirac.ac.uk).
This equipment was funded by BIS National E-infrastructure capital
grant ST/K000411/1, STFC capital grant ST/H008845/1, and STFC DiRAC
Operations grants ST/K005804/1 and ST/K005790/1. DiRAC is part of
the National E-Infrastructure. Computing support for this work came partially 
from the Lawrence Livermore National Laboratory (LLNL) Institutional Computing 
Grand Challenge program.

\newpage
\appendix
\input{appendix}

\input{appendix_Oa2}
\input{appendix_coulomb}

\bibliographystyle{JHEP}
\bibliography{IB_corrections.bib}

\end{document}

%% file: 1_introduction.tex
\section{Introduction}
In recent years, lattice QCD has made remarkable progress in calculating 
quantities relevant to Standard Model phenomenology. Many of these calculations 
have reached a precision of $\lesssim1\%$ \cite{Aoki:2016frl}, e.g.\ the ratio 
$f_K/f_\pi$ of kaon and pion decay constants or the $K_{l3}$ form factor 
$f_+(0)$. Such lattice computations are usually done in the isospin 
symmetric limit with the masses of the up and down quarks equal ($m_u=m_d$).
However, 
two sources of isospin breaking (IB) are present in 
nature. The masses of 
up- and down quarks are different $m_d\neq m_u$, a correction which is of the 
order $\mathcal{O}((m_d-m_u)/\Lambda_\textrm{QCD})$. In addition, quarks carry 
an 
electric charge, and thus also interact electromagnetically. The latter not 
only applies 
to up and down quarks, but also to all other quark flavours. QED corrections 
are of 
$\Oalpha$, where $\alpha$ is the electromagnetic fine 
structure constant.
Both of these isospin breaking effects are expected to be of the order of $1\%$ 
and thus, can no 
longer be neglected 
in applications of lattice results to phenomenology 
at this level of precision.
\par
In the last few years significant 
progress has been made in directly including isospin breaking and QED 
corrections in lattice 
calculations.
So far computations including QED on the lattice have been mainly 
focused on determining electromagnetic corrections to spectral 
quantities such as hadron masses (see 
e.g. \cite{Blum:2007cy,Blum:2010ym,Borsanyi:2013lga,Borsanyi:2014jba,
Horsley:2015eaa,Horsley:2015vla,Basak:2016jnn,Fodor:2016bgu,
deDivitiis:2013xla,Giusti:2017dmp}). 
Pioneering work on the calculation of the QED correction to matrix elements has 
recently been published in
\cite{Carrasco:2015xwa,Lubicz:2016mpj,Lubicz:2016xro}. 
Another successful application of QCD$+$QED is the calculation of hadronic 
light-by-light scattering~\cite{Blum:2014oka,Blum:2015gfa,Blum:2016lnc}.
\par
Two methods are commonly used to include QED in lattice QCD computations. A 
non-perturbative method using stochastically generated $U(1)$ gauge 
configurations for the photon fields was first proposed in 
\cite{Duncan:1996xy}. We will refer to this method as the \textit{stochastic 
method} throughout the paper (see 
\cite{Blum:2007cy,Blum:2010ym,Borsanyi:2013lga,Borsanyi:2014jba,
Horsley:2015eaa,Horsley:2015vla,Basak:2016jnn,Fodor:2016bgu} for lattice 
QCD$+$QED calculations using the stochastic method). On the other hand, the 
electromagnetic coupling 
$\alpha$ is small in the low-energy regime, and thus, QED can be treated 
perturbatively. In \cite{deDivitiis:2013xla} the authors proposed to expand 
the Euclidean path integral in orders of $\alpha$ and 
explicitly calculate the leading order QED corrections. We will refer to this 
method as the \textit{perturbative method} in the following. To our knowledge, 
a direct comparison of results and statistical errors of both methods using the 
same setup and QCD 
gauge configurations has not yet been made.
\par
In this paper we present an exploratory study with unphysical quark masses, 
in which we calculate the QED correction with both the stochastic and the 
perturbative methods. This allows us to directly compare results and 
statistical precision at the same computational cost obtained with both methods. 
In this study, we work in 
an electro-quenched setup, i.e.\ we consider the sea quarks as 
electrically neutral and mass degenerate.
\par
Details of our strategy to include 
isospin breaking and QED corrections in the calculation are described 
in section \ref{sec:Strategy}. The setup of the calculation is given in 
section \ref{sec:setup}. In section \ref{sec:mesonmasses} we present the 
results 
for the isospin and QED corrections to meson masses, as a starting point for 
comparing the perturbative and the stochastic method. In particular, we compare 
the statistical precision obtained from both methods and we discuss how to 
extract the QED correction to meson masses to consistently compare results from 
the stochastic and the perturbative method. 
%\par
In section \ref{sec:hvp} we discuss results for the QED correction to 
the hadronic vacuum polarization (HVP). The HVP is the leading order hadronic 
contribution to the anomalous magnetic moment $a_\mu$ of the muon. 
We are currently observing a $3\sigma$ \cite{Olive:2016xmw} 
deviation between the experimentally 
measured value for $a_\mu$ and the Standard Model estimate. This has 
triggered 
increased efforts to determine the HVP in a lattice calculation (see e.g.\ 
\cite{Boyle:2011hu,DellaMorte:2011aa,Burger:2013jya,
Chakraborty:2014mwa, Bali:2015msa,Chakraborty:2016mwy, 
Borsanyi:2016lpl,DellaMorte:2017dyu}) 
aiming at a 
precision of $1\%$ to be competitive with the current most precise estimate 
\cite{Davier:2010nc,Hagiwara:2011af} from $e^+e^-\rightarrow$ hadrons. 
At this level of accuracy, isospin breaking corrections need 
to be included in the computation.
To our 
knowledge, the present work constitutes the first lattice calculation of the 
isospin 
breaking corrections to the HVP. However, we wish to emphasize that this is an 
exploratory study at 
unphysical quark masses and we do not attempt to quantify finite volume effects 
for the QED corrections to the HVP in this study. 
As for the meson masses, 
results and 
statistical errors for the QED correction to the HVP calculated with the 
stochastic and the perturbative method are compared.
Our main results and conclusions are summarized in section 
\ref{sec:conclusions}. Some preliminary 
results of our work have already been presented in \cite{Boyle:2016lbc}.

%% file: 2_method.tex
\section{Isospin Breaking on the Lattice}
\label{sec:Strategy}
In the following we give details on our strategies to include isospin breaking 
effects. We start with a discussion of the QCD$+$QED 
path integral in section~\ref{subsec:qcdqedpathintegral}. The stochastic and 
perturbative methods to calculate the QED corrections to hadronic observables 
are described in sections  \ref{subsec:stochmethod} and \ref{sec:pertmethod}, 
respectively.  In section \ref{subsec:strongIB}, we describe how we treat strong 
isospin breaking corrections, i.e.\ $m_u\neq m_d$.
\subsection{Lattice QCD$+$QED Path Integral}\label{subsec:qcdqedpathintegral}
In lattice QCD the expectation value of an observable $O$ is calculated in 
terms of the discretized Euclidean path integral, which is given 
by
\begin{equation}
  \left<O\right>_{0} = \frac{1}{Z_0}\! \int 
\!\!\mathcal{D}[U]\,\mathcal{D}[\Psi,\Psibar]\,\,O[\Psi,\Psibar ,U]\,\,
e^ { -S_{F,0}[\Psi,\Psibar , U]}\,e^{ - S_G[U]}\,,
\label{eq:pathintegral_0}
\end{equation}
with quark fields $\Psi$ and $\Psibar$ and $SU(3)$ gluon fields $U$. The 
subscript ``$0$'' on $\left<O\right>_0$ and $S_{F,0}$  denote that these 
quantities are without QED.
\par
However, quarks carry an electric charge and thus also interact 
electromagnetically. To account for QED effects we consider the Euclidean 
QED+QCD path integral
\begin{equation}
  \left<O\right> = \frac{1}{Z}\! \int 
\!\!\mathcal{D}[U]\,\mathcal{D}[A]\,\mathcal{D}[\Psi,\Psibar]\,\,O[\Psi,\Psibar 
,A, U]\,\,
e^ { -S_F[\Psi,\Psibar ,A, U]}\,\,e^{ - S_\gamma[A]}\,e^{ - S_G[U]}\,,
\label{eq:pathintegral}
\end{equation}
with photon fields $A$. 
In the following, expectation values without a 
subscript $\left<\cdot\right>$ denote the combined QED+QCD expectation value.
The observable $O$ 
can now, in general, also depend on the photon fields $A$ besides the quark 
fields $\Psi$, $\Psibar$ and the gauge fields $U$. The fermionic action 
$S_F[\Psi,\Psibar ,A, U]$ in 
\eqref{eq:pathintegral} also contains couplings of quarks to photons and can be 
obtained from the action $S_{F,0}[\Psi,\Psibar , U]$ by multiplying the $SU(3)$ 
gauge fields by appropriate $U(1)$ phases
\begin{equation}
 U_\mu(x) \rightarrow e^{-iq_feA_\mu(x)} U_\mu(x)\,,
\end{equation}
with the elementary charge $e$ and the charge $q_f$ of a given quark flavour, 
i.e.\ $\{q_u,q_d,q_s\}=\{2/3,-1/3,-1/3\}$.
We define the non-compact photon action as
\begin{equation}
S_{\gamma}\left[A\right]=\frac{1}{4}\sum_{x}\sum\limits_{\mu,\nu}
\left(\partial_{\mu}A_{\nu}\left(x\right)-\partial_{\nu}A_{\mu}
\left(x\right)\right)^{2}\,,
\label{eq:lattice_EM_action}
\end{equation}
with the forward derivative 
\begin{equation}
 \partial_\mu f(x) = f\left(x+\hat{\mu}\right)-f\left(x\right)\,.
 \label{eq:forward_deriv}
\end{equation}
Here and in the following we express all quantities in units of the lattice 
spacing $a$. The
Feynman gauge can be imposed in the photon action 
\eqref{eq:lattice_EM_action} by adding a gauge fixing term 
\begin{equation}
 S_{\gamma,\mathrm{Feyn.}}\left[A\right]  =  
S_{\gamma}\left[A_{\mu}\right]+\frac{1}{2}\sum_{x}\left(\sum_{\mu}\partial_{
\mu }
A_{\mu}\left(x\right)\right)^{2}\,.
\end{equation}
Using integration by parts, the Feynman gauge action can be written as
\begin{equation}
  S_{\gamma,\mathrm{Feyn.}}\left[A\right] = 
-\frac{1}{2}\sum\limits_x\sum\limits_\mu A_\mu(x) \partial^2 A_\mu(x)\,,
\label{eq:feynman_pos}
\end{equation}
with $\partial^2\equiv\sum_\mu \partial_\mu^*\partial_\mu\,$, where 
$\partial_\mu$ is the forward derivative \eqref{eq:forward_deriv} and 
$\partial_\mu^*$ the backward derivative defined by 
\begin{equation}
 \partial^*_\mu f(x) = f\left(x\right)-f\left(x-\hat{\mu}\right)\,.
 \label{eq:backward_deriv}
\end{equation}
One important point when including QED in the lattice calculation is the 
treatment of the zero-mode of the photon field.
This is associated with a shift symmetry 
of the photon action \eqref{eq:lattice_EM_action} 
\begin{equation}
 A_{\mu}\left(x\right)\rightarrow A_{\mu}\left(x\right)+c_{\mu}\,,
\end{equation}
which cannot be constrained by a gauge fixing condition. In our work, we choose 
to remove the spatial zero modes of the photon propagator on every time slice
\begin{equation}
 \sum_{\vec{x}}A_{\mu}\left(x_0,\vec{x}\right)=0\hspace{1cm}\textrm{for 
all}\hspace{0.2cm}\mu,x_{0}\,,
\end{equation}
or, in momentum space $\tilde{A}_{\mu}(k_0,\vec{k}=0)=0$. The formulation of 
QED resulting from this particular 
treatment of the zero-mode is called QED$_L$ and was first proposed 
in~\cite{Hayakawa:2008an}.
A discussion about different prescriptions of QED in a finite box with periodic 
boundary conditions can be found in 
\cite{Patella:2017fgk,Duncan:1996xy,Borsanyi:2014jba,
Gockeler:1989wj,Endres:2015gda,Lucini:2015hfa}.
\par
Throughout this paper we work in the electro-quenched approximation, i.e.\ 
when evaluating the path integral \eqref{eq:pathintegral} we neglect QED effects 
in the fermion determinant $\det(D[A,U])\equiv\det(D_0[U])$, where $D[A,U]$ 
and $D_0[U]$ are the Dirac operators with QED and without QED, respectively.
In the electro-quenched approximation effects from the electromagnetic vacuum 
polarization are neglected and, thus, sea quarks are electrically neutral. 
Effects from electro-quenching are $SU(3)$ and $1/N_c$ suppressed for $\Oalpha$ 
contributions and expected to be of the order of $\sim10\%$ 
\cite{Borsanyi:2013lga} of the QED correction.
\par
The stochastic \cite{Duncan:1996xy} and perturbative \cite{deDivitiis:2013xla} 
approach that we use to include electro-quenched QED in the calculation of the 
path integral \eqref{eq:pathintegral} are explained in detail in the following.

\subsection{Stochastic Method}
\label{subsec:stochmethod}
The stochastic method to include QED in lattice calculations has first been 
introduced in \cite{Duncan:1996xy}. Since then, this method has been used in 
several 
lattice QCD $+$ QED calculations (see 
e.g. \cite{Blum:2007cy,Blum:2010ym,Borsanyi:2013lga,Borsanyi:2014jba,
Horsley:2015eaa,Horsley:2015vla,Basak:2016jnn,Fodor:2016bgu}).
\par
In this study, we work in the electro-quenched approximation and, thus, 
the $U(1)$ photon gauge fields are generated independently of the 
$SU(3)$ gauge fields. This allows us to include QED using existing $SU(3)$ 
configurations.
In the electro-quenched approximation sea 
quarks are electrically neutral. Including electromagnetic effects for the sea 
quarks is computationally much more expensive, since it requires either the 
generation 
of new QED$+$QCD gauge configurations, or the calculation of reweighing factors 
and an accompanying increase in statistical variance. Lattice calculations with 
dynamical QED 
using the stochastic method have been done in 
\cite{Borsanyi:2014jba,Horsley:2015eaa,Horsley:2015vla}.
\par
In practice one stochastically draws appropriate $U(1)$ 
gauge 
configurations for the photon fields according to the Gaussian weight 
$\exp(-S_\gamma[A])$. The new link variables are then given 
as the $SU(3)$ gluon gauge links multiplied by the $U(1)$ phases
\begin{equation}
U_\mu(x)\rightarrow e^{-ieq_fA_\mu(x)} U_\mu(x)\,.
\label{eq:u3links}
\end{equation}
We define the lattice $U(1)$ photon fields at the mid-links of the lattice, 
i.e.\ we define $A_\mu(x)\equiv A_\mu(x+\hat{\mu}/2)$.
\par
We choose to initially generate the photon fields in the Feynman gauge due to 
the simple structure of the action in momentum space.
In momentum space the Feynman gauge action \eqref{eq:feynman_pos} is given by
\begin{equation}
S_{\gamma,\mathrm{Feyn.}}\left[A\right]=\frac{1}{2N}\sum\limits_{k,\vec{k}
\neq0}\hat{k}^{2}\sum_{\mu}\left|\tilde{A}_{\mu}\left(k\right)\right|^{2}\,,
\label{eq:feynman_mom}
\end{equation}
where $\tilde{A}_{\mu}(k)$ is the photon field in momentum space, $N$ is the 
total number of lattice points and the lattice momentum $ \hat{k}_{\mu}$ is 
given by
\begin{equation}
 \hat{k}_{\mu}=2\sin\left(\frac{k_{\mu}}{2}\right)\,.
\end{equation}
The sum over $k,\vec{k}\neq0$ in equation \eqref{eq:feynman_mom} indicates the 
removal of all spatial zero modes.
\par
Equation 
\eqref{eq:feynman_mom} implies, that all components $\tilde{A}_\mu(k)$ of 
the photon field can be drawn independently of each other from a Gaussian 
distribution with variance 
$2N/\hat{k}^{2}$.
\par
To check for gauge invariance in our calculation, we use photon fields in the
Feynman and the Coulomb gauge. A Feynman gauge photon field can be transformed 
into the
Coulomb gauge by using an 
appropriate projector \cite{Borsanyi:2014jba}
\begin{equation}
 \left(P_{C}\right)_{\mu\nu}  =  
\delta_{\mu\nu}-\left|\vec{\hat{k}}\right|^{-2}\hat{k}_{\mu}\left(0,\vec{\hat{k}
}\right)_{\nu} 
\hspace{1cm}\textrm{with}\hspace{0.3cm}\tilde{A}^\textrm{Coul}_\mu(k) =  
\left(P_{C}\right)_{\mu\nu} \tilde{A}^\textrm{Feyn}_\nu(k)\,.
\end{equation}
%\par
After generating the photon field in momentum space, it is converted 
to position space using a Fast Fourier Transform 
(FFT)\footnote{http://www.fftw.org/}. 
Once the photon field configurations are transformed into position space, they 
are multiplied with the $SU(3)$ gauge links according to equation 
\eqref{eq:u3links}. The calculation of hadronic observables then proceeds as 
in the case without QED, but using the combined QED$+$QCD gauge configurations. 
With the stochastic method QED corrections are calculated to all orders in 
$\alpha$ at once albeit in the electro-quenched approximation.
\par
Although the leading order QED corrections are of $\mathcal{O}(e^2)$, the 
statistical noise contains contributions at $\mathcal{O}(e)$, which would 
vanish in the limit of infinitely many QED configurations because of charge 
conjugation invariance. However, this 
$\mathcal{O}(e)$ noise can be exactly removed on every gauge configuration by 
averaging over calculations using $+e$ and $-e$ \cite{Blum:2007cy}. Since we 
are interested in QED 
corrections to hadronic quantities, we calculate correlation functions once 
without QED ($e=0$) and once with QED, while averaging over $+e$ and $-e$. 
Thus, the stochastic method requires $3$ inversions per quark flavour and 
source 
position ($e=0$, $+e$ and $-e$).
\subsection{Perturbative Method}
\label{sec:pertmethod}
In addition to the stochastic method to include QED in our lattice calculation, 
we use a perturbative method, adopting
the approach developed in \cite{deDivitiis:2013xla}. 
We will summarize this method in section \ref{subsubsec:PertIntro} and give 
details on our strategy to calculate the required correlation functions in 
section \ref{subsubsec:PartCalc}.
The perturbative method has been used in 
\cite{Carrasco:2015xwa,Lubicz:2016mpj} to determine the QED 
corrections to matrix elements.
%\par
%
%
%
\subsubsection{Introduction}
\label{subsubsec:PertIntro}
Since the electromagnetic coupling $\alpha$ is small in the low-energy regime, 
QED can be treated perturbatively. This is done by expanding the path integral 
\eqref{eq:pathintegral} as a series in the electromagnetic 
coupling
\begin{equation}
 \left<O\right> = \left<O\right>_{0} + 
\frac{1}{2}\,e^2\left.\frac{\partial^2}{\partial 
e^2}\left<O\right>\right|_{e=0} + \Oalphasquare\,\,.
\label{eq:eexpansion}
\end{equation}
At leading order, $\Oalpha$, one finds contributions with either two insertions 
of the conserved vector current $V^c_\mu$ or one insertion of the tadpole 
operator $T_\mu$ \cite{deDivitiis:2013xla}
\begin{equation}
 \left<O\right> = \left<O\right>_{0} - 
\frac{(eq_f)^2}{2} \left<O
T_\mu(x)\right>_{0}\Delta_{\mu\mu}(0) 
- \frac{e^2q_fq_{f'}}{2}\left<O 
V^c_\mu(x)V^c_\nu(y)\right>_{0}\Delta_{\mu\nu}(x-y)+ \Oalphasquare\,.
\label{eq:pathintexp}
\end{equation}
Note, that equation \eqref{eq:pathintexp} is only valid, when the operator $O$ 
does not depend on the electromagnetic coupling $e$. If the operator itself 
depends on $e$, this has to be taken into account, when expanding the path 
integral\footnote{This is not relevant for the QED correction to meson 
masses. However, for the HVP we use a setup with a conserved vector current at 
the 
sink, which in the lattice discretized theory contains a link variable, and 
thus, including QED, depends 
on $e$. A more detailed discussion can be 
found in section \ref{subsec:QEDHVP}.}. 
\par
The conserved vector current and the tadpole operator for the Domain Wall 
fermion action used in this work are given in \eqref{eq:consvectorcurrent} and 
\eqref{eq:tadpolop}, respectively. The photon propagator $\Delta_{\mu\nu}(x-y)$ 
in equation \eqref{eq:pathintexp} is given as 
\begin{equation}
 \Delta_{\mu\nu}(x-y) = \left<A_\mu(x) A_\nu(y)\right>_\gamma = \frac{\int 
\!\mathcal{D}[A]\,\,A_\mu(x) A_\nu(y)\,\,e^{-S_\gamma[A]}}{\int 
\!\mathcal{D}[A]\,\,e^{-S_\gamma[A]}}\,,
\end{equation}
with $\mu,
\nu=1,\ldots,4$. In the Feynman gauge the photon propagator is given by
\begin{equation}
  \Delta_{\mu\nu}(x-y) = 
\delta_{\mu\nu}\,\frac{1}{N}\sum\limits_{k,\vec{k}\neq0}\,\,
\frac{e^{ik\cdot(x-y)}}{\hat{k}^2}\,,
\label{eq:photonprop}
\end{equation}
where we subtract all spatial zero modes, i.e.\ we use the QED$_L$ formulation 
\cite{Hayakawa:2008an}
as in the stochastic approach above. In addition, to numerically check for 
gauge invariance of the observables studied in this work, we use the Coulomb 
gauge. The photon propagator in the Coulomb gauge is given in the appendix in 
equation \eqref{eq:coulomb_prop}.
\par
\begin{figure}[h]
 \centering
 \includegraphics[width=0.98\textwidth]{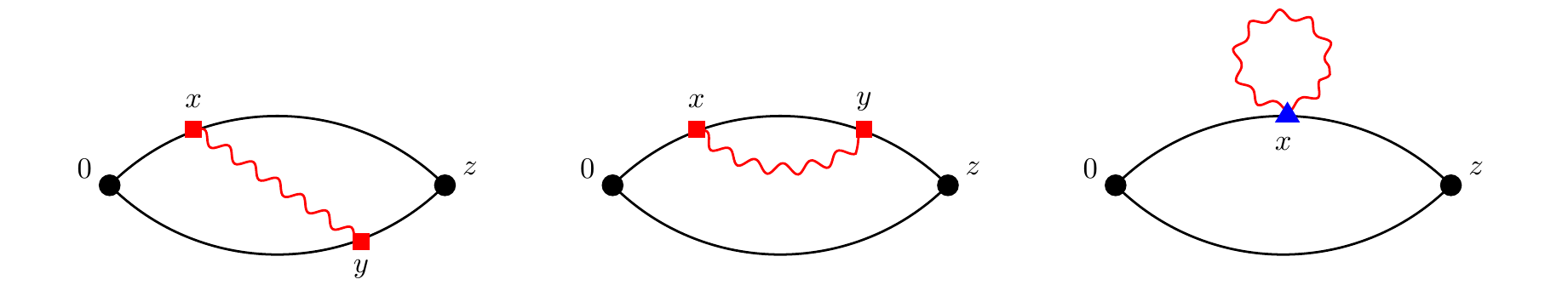}
 \caption{The three quark-connected diagrams that determine the leading order 
QED correction to mesonic two-point functions. The diagrams are from left to 
right: photon exchange diagram, quark self-energy diagram and tadpole diagram. 
Red 
squared vertices denote insertions of the conserved vector current, the blue 
triangle vertex an insertion of the tadpole operator.}
\label{fig:diagrams}
\end{figure}

For mesonic two-point functions one obtains from equation 
\eqref{eq:pathintexp} at leading order in $\alpha$ 
three different types of quark-connected Wick contractions: a photon exchange 
diagram, a quark self-energy diagram and a tadpole diagram. These diagrams are 
shown in figure \ref{fig:diagrams}. 
\par
We do not include any quark-disconnected diagrams in our study. In particular, 
we neglect diagrams that correspond to photons coupling to sea quarks. These 
diagrams would originate from an expansion of the fermion determinant, however, 
we work in the electro-quenched approximation where QED effects for the 
fermion determinant are neglected. In the 
perturbative method working in unquenched QED is possible by 
additionally calculating
the appropriate quark-disconnected diagrams. For the 
stochastic method unquenched QED requires the 
generation of combined QCD+QED gauge configurations at substantial extra cost. 
\par
\subsubsection{Numerical Calculation}
\label{subsubsec:PartCalc}
To illustrate how we calculate the diagrams shown in figure \ref{fig:diagrams}, 
we now consider as an example the photon exchange diagram for a charged kaon.
The 
corresponding correlation function is given by
\begin{equation}
 C^\textrm{exch}(z_0)  = 
\sum\limits_{\mu,\nu}\sum\limits_{\vec{z}}\!\sum\limits_{x,y}\!\textrm{Tr}\!
\left[S^s(z,x)\,\Gamma^c_\nu\,S^s(x,0)\,\gamma_5\,S^{u}(0,y)\,
\Gamma^c_\mu\, 
S^{u}(y,z)\,\gamma_5\right]\Delta_{\mu\nu}(x-y)\,,
\label{eq:K_exch}
\end{equation}
where $\Gamma^c_\mu$ denotes a conserved vector current insertion $V^c_\mu(x) 
\equiv\Psibar(x)\Gamma^c_\mu\Psi(x)$ and $S^{f}(0,x)$ is the propagator from 
$0$ to $x$ for a quark of flavour $f$.
\par
We calculate the correlation functions such as \eqref{eq:K_exch}
using sequential propagators. For this, the photon propagator has to be 
factorized into a factor that depends only on the position $x$ of one of the 
photon vertices and another factor that depends only on the position $y$ of 
the other photon vertex, such that sequential sources with insertions of 
the conserved vector current and a respective factor of $\Delta_{\mu\nu}(x-y)$ 
at $x$ or at $y$ can be constructed. This factorization can be achieved by 
inserting sets of stochastic sources in the photon propagator. In this work, we 
will do this in two different ways, which lead to different numerical costs and 
different statistical errors. One possibility is to use the same stochastic 
source for all Lorentz indices $\mu,\nu$ of the photon propagator and to 
calculate sequential propagators for every combination of ${\mu,\nu}$ 
separately. We will call this method \textit{single-$\mu$ insertion} in the 
following. On the other hand, one can use four different stochastic 
sources~\cite{deDivitiis:2013xla}
-- one for 
every Lorentz index $\mu={1,2,3,4}$ -- and to include the sum over $\mu$ or 
$\nu$ in equation \eqref{eq:K_exch} already in the sequential source. We will 
call this method \textit{summed-$\mu$ insertion} in the following. Both methods 
will be illustrated below.\par
We note, that it is also possible to use the stochastic 
photon fields generated for the stochastic method as an 
insertion at $x$ and $y$ \cite{Giusti:2017dmp} by using 
\begin{equation}
  \Delta_{\mu\nu}(x-y) = \left<A_\mu(x) A_\nu(y)\right>_\gamma\,.
\end{equation}
However, this is simply the exact $\Oalpha$-truncation of the stochastic 
method. Higher order $\Oalphasquare$ effects that differentiate the the two 
methods are, as we argue later, small and barely significant at this level of 
precision. Thus, we consider the setup of \cite{Giusti:2017dmp} effectively 
identical to the stochastic approach and we did not perform a dedicated 
calculation to reproduce it. 
\subsubsection*{Single-$\mu$ Insertion}
\begin{figure}[h]
\centering
\includegraphics[width=0.95\textwidth]{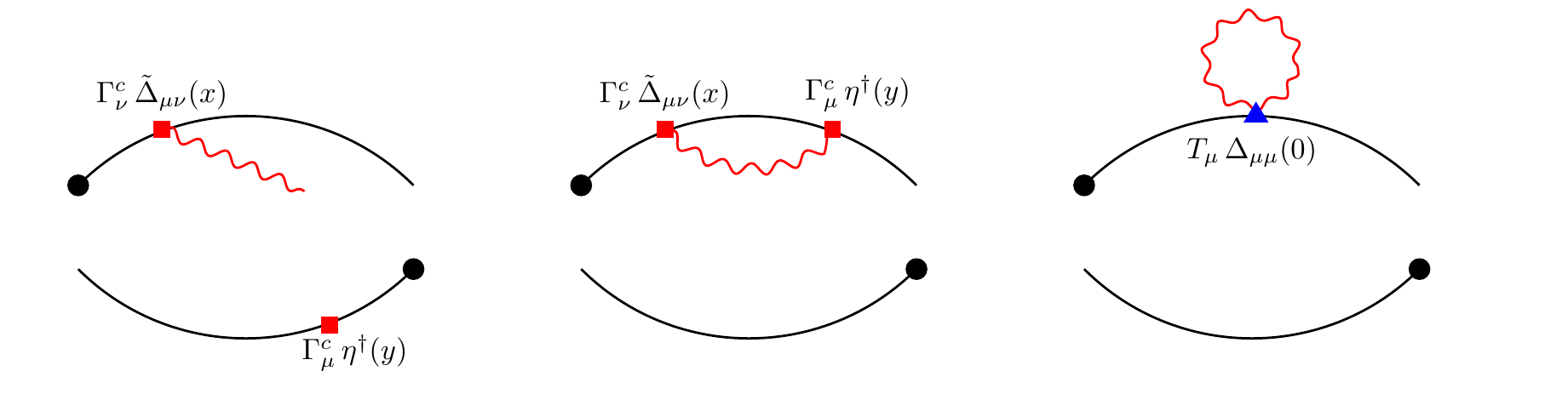}
\caption{Building blocks for the construction of the $\Oalpha$ QED 
correction diagrams for mesonic two-point function. From left to right: photon 
exchange diagram, quark self-energy diagram, tadpole diagram.}
\label{fig:buildig_blocks}
\end{figure}

For the numerical calculation of the correlation functions such as 
\eqref{eq:K_exch}, we rewrite the photon propagator as
\begin{equation}
 \Delta_{\mu\nu}(x-y) = 
\left<\sum\limits_u\Delta_{\mu\nu}
(x-u)\eta(u)\eta^\dagger(y)\right>_\eta\equiv\left<\tilde\Delta_{
\mu\nu}(x)\eta^\dagger(y)\right>_\eta\,,
\label{eq:phprop_stochsources}
\end{equation}
with stochastic sources $\eta$ that fulfil the condition
\begin{equation}
 \left<\eta(u)\eta^\dagger(y)\right>_\eta = \delta_{u,y}\,.
\end{equation}
Here, we choose complex $\mathbb{Z}_2$ noise sources $\eta(x)$, that have 
randomly 
picked entries from $\left\{\frac{1}{2}\left(\pm1\pm i\right)\right\}$ for 
every lattice site. The insertion of the set of stochastic sources in 
\eqref{eq:phprop_stochsources} allows to factorize the photon propagator with 
a factor $\tilde\Delta_{\mu\nu}(x)$
that only depends on the position $x$ of one of the photon vertices and another
factor $\eta^\dagger(y)$ that only depends on the position $y$ of the other
photon vertex. We calculate $\tilde\Delta_{\mu\nu}(x)$ using a Fast Fourier 
Transform.
The correlation function \eqref{eq:K_exch} for the photon exchange for a 
charged kaon can now be written as
\begin{align}
 C^\textrm{exch}(z_0) & = 
\left<\sum\limits_{\mu,\nu}\sum\limits_{\vec{z}}\!\sum\limits_{x,y}\!\textrm{Tr}
\!\left [ S^s(z ,
x)\,\Gamma^c_\nu\,\tilde\Delta_{\mu\nu}(x)\,S^s(x,0)\,\gamma_5\,S^{u}(0,y)\,
\Gamma^c_\mu\,\eta^\dagger(y)\, 
S^{u}(y,z)\,\gamma_5\right]\right>_\eta.
\end{align}
We construct this correlation function $C(z_0)$ using sequential propagators 
with insertions of the conserved vector current and either 
$\tilde\Delta_{\mu\nu}(x)$ or $\eta^\dagger(y)$
\begin{equation}
 C^\textrm{exch}(z_0) = 
\left<\sum\limits_{\mu,\nu}\sum\limits_{\vec{z}}\textrm{Tr}\left[\Sigma_{\mu\nu}
(z , 0)\,\gamma_5\,\Xi_\mu(0,
z)\,\gamma_5\right]\right>_\eta\,,
\end{equation}
with the sequential propagators
\begin{align}
\Sigma_{\mu\nu}(z,0)&= \sum\limits_{x} 
S^s(z,x)\,\Gamma^c_\nu\,\tilde\Delta_{\mu\nu}(x)\,S^s(x,0)\,,\\
\Xi_\mu(0,z)&=\sum\limits_{y} \,S^{u}(0,y)\,\Gamma^c_\mu\,  \eta^\dagger(y)
S^{u}(y,z)\,.
\end{align}
To build correlation functions of the type quark self-energy, we use appropriate
double sequential propagators, e.g.\ the quark self-energy diagram for a charged 
kaon 
with the photon attached to the $s$ quark is calculated as
\begin{align}
 C^\textrm{self}(z_0) & = 
\left<\sum\limits_{\mu,\nu}\sum\limits_{\vec{z}}\!\sum\limits_{x,y}\!\textrm{Tr}
\!\left[S^s(z,
y)\,\Gamma^c_\nu\,\tilde\Delta_{\mu\nu}(y)\,S^s(y,x)\,\Gamma^c_\mu\,
\eta^\dagger(x)\,S^{s}(x,0)\,\gamma_5\, 
S^{u}(0,z)\,\gamma_5\right]
\right>_\eta\\
& = 
\left<\sum\limits_{\mu,\nu}\sum\limits_{\vec{z}}\!\textrm{Tr}\!\left[\Lambda_{
\mu\nu}(z , 0)
\, \gamma_5\,S^{u}(0,z)\,\gamma_5\right]
\right>_\eta\,,
\end{align}
with the sequential propagator
\begin{equation}
\Lambda_{\mu\nu}(z,0) = \sum\limits_{x,y} S^s(z,
y)\,\Gamma^c_\nu\,\tilde\Delta_{\mu\nu}(y)\,S^s(y,x)\,\Gamma^c_\mu\,
\eta^\dagger(x)\,S^{s}(x,0)\,.
\end{equation}
The tadpole diagrams can be constructed from a 
sequential propagator with an insertion of the tadpole operator multiplied with 
the tadpole value $\Delta_{\mu\mu}(0)$ of the photon propagator, e.g.\ the 
tadpole diagram for a charged kaon with the photon attached to the $s$ quark is 
calculated as
\begin{align}
 C^\textrm{tad}(z_0) & = 
\sum\limits_{\mu}\sum\limits_{\vec{z}}\!\sum\limits_{x}\textrm{Tr}\!\left[
S^s(z,x)\,T_\mu\,\Delta_{\mu\mu}(0)\,S^s(x,0)\,\gamma_5\, 
S^{u}(0,z)\,\gamma_5\right]\\
&=\sum\limits_{\mu}\sum\limits_{\vec{z}}\textrm{Tr}\!\left[\Upsilon_\mu(z,
0)\,
\gamma_5\,S^{u}(0,z)\,\gamma_5\right]\,,
\end{align}
with the sequential propagator
\begin{equation}
\Upsilon_\mu(z,0) = 
\sum\limits_{x} S^s(z,x)\,T_\mu\,\Delta_{\mu\mu}(0)\,S^s(x,0)\,.
\end{equation}
The tadpole value $\Delta_{\mu\mu}(0)$ of the photon propagator can be 
calculated exactly for a given lattice size. All the 
required building blocks to construct the quark-connected diagrams for the 
$\Oalpha$ QED 
correction diagrams for mesonic two-point functions are shown in 
figure~\ref{fig:buildig_blocks}. \par 

The evaluation of the correlation functions shown in figure \ref{fig:diagrams} 
using sequential propagators as described above
requires a total of $17$ inversions per quark flavour and source position 
if 
the photon propagator is in the Feynman gauge, where only diagonal terms 
$\mu=\nu$ 
contribute (cf. equation \eqref{eq:photonprop}). 
These $17$ inversions are split as follows: $1$ inversion for the point-to-all 
propagator, $4$ sequential inversions with an insertion of
$\Gamma^c_\nu\,\tilde\Delta_{\mu\nu}(y)$ (for the Feynman gauge only $\mu=\nu$ 
is required), $4$ sequential inversions with insertion of 
$\Gamma^c_\mu\,\eta^\dagger(x)$, $4$ additional inversions to obtain the double 
sequential propagators and $4$ sequential inversions for the tadpole using 
$T_\mu\Delta_{\mu\mu}(0)$ as insertion.
If one uses a different gauge 
(e.g.\ Coulomb gauge) where also off-diagonal terms $\mu\neq\nu$ contribute, 
more inversions are required. Thus, in terms of numerical cost, the 
Feynman 
gauge is favourable for the perturbative approach with this setup to calculate 
sequential propagators. However, we 
also 
calculate the $\Oalpha$ QED correction using the Coulomb gauge on a subset 
of the statistics to check for gauge invariance. 
\par
We note, that the insertion of the photon propagator can be done using 
stochastic sources at both vertices by
\begin{equation}
 \Delta_{\mu\nu}(x-y) = \left<\sum\limits_{u,v}\Delta_{\mu\nu}
(v-u)\eta(u)\eta^\dagger(y)\zeta(v)\zeta^\dagger(x)\right>_{
\eta,\zeta}\equiv\left<\tilde{\tilde{\Delta}}_{
\mu\nu}\eta^\dagger(y)\zeta^\dagger(x)\right>_{\eta,\zeta}\,,
\end{equation}
with two sets of stochastic sources $\eta$ and $\zeta$. The correlation 
functions that determine the QED  corrections to the mesonic two-point 
functions are then calculated using sequential sources with 
appropriate insertions of either $\eta^\dagger$ or $\zeta^\dagger$. 
The photon propagator $\Delta_{\mu\nu}$ is included as 
\begin{equation}
\tilde{\tilde{\Delta}}_{\mu\nu} = \sum\limits_{u,v}\Delta_{\mu\nu}
(v-u)\eta(u)\zeta(v)\,,
\end{equation}
which, for a given combination of stochastic sources 
$\eta^\dagger$, $\zeta^\dagger$ and $\mu$, $\nu$, is simply an overall 
numerical factor which multiplies the remainder of the correlation function 
after all the quark contractions have been calculated.
This allows us to study e.g.\ different gauges or QED prescriptions without 
having 
to calculate new quark contractions, and thus, new quark inversions.
However, for the setup that we use in this exploratory study, this resulted in 
a significantly worse noise-to-signal ratio for the QED corrections. Inserting 
the photon propagator stochastically at both vertices we found the statistical 
error to be $\approx30$ times larger for the photon exchange diagram and  
$\approx60$ times larger for the quark self-energy diagram compared to using 
only one set of stochastic sources.
Thus, for the study presented here, we decided to use only one set of 
stochastic sources at one of the photon vertices.
\subsubsection*{Summed-$\mu$ Insertion}
\begin{figure}[h]
 \centering
 \includegraphics[width=0.8\textwidth]{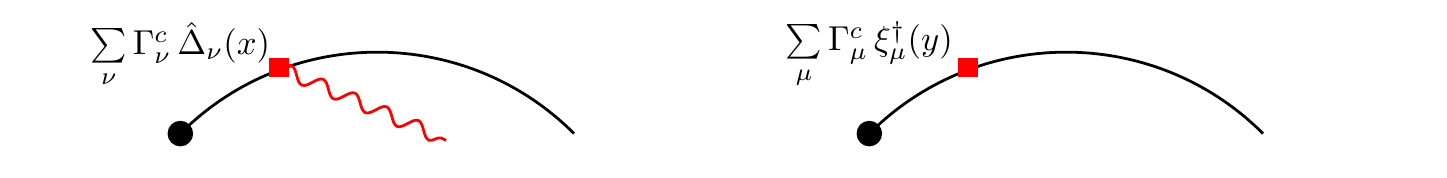}
 \caption{Sequential propagators for the photon exchange diagram using the 
summed-$\mu$-insertion.}
\label{fig:seqpropsummu}
\end{figure}

The number of inversions required for the construction of the diagrams shown in 
figure \ref{fig:diagrams} can be substantially reduced by using different 
stochastic sources for the $4$ Lorentz indices of the photon propagator 
\cite{deDivitiis:2013xla,Giusti:2017dmp}.
We start by rewriting the photon propagator as
\begin{equation}
 \Delta_{\mu\nu}(x-y) = \Big<\sum\limits_u\sum\limits_\sigma 
\Delta_{\sigma\nu}(x-u)\xi_\sigma(u)\xi^\dagger_\mu(y)\Big>_\xi = 
\left<\hat \Delta_\nu(x)\xi^\dagger_\mu(y)\right>_\xi\,,
\label{eq:hatDmunu}
\end{equation}
with stochastic sources
\begin{equation}
 \left<\xi_\sigma(u)\xi_\mu^\dagger(y)\right>_\xi = 
\delta_{uy}\delta_{\sigma\mu}\,.
\end{equation}
The photon exchange diagram \eqref{eq:K_exch} can now be written as
\begin{align}
 C^\textrm{exch}(z_0) & = 
\left<\sum\limits_{\mu,\nu}\sum\limits_{\vec{z}}\!\sum\limits_{x,y}\!\textrm{Tr}
\!\left [ S^s(z ,
x)\,\Gamma^c_\nu\,\hat\Delta_{\nu}(x)\,S^s(x,0)\,\gamma_5\,S^{u}(0,y)\,
\Gamma^c_\mu\,\xi_\mu^\dagger(y)\, 
S^{u}(y,z)\,\gamma_5\right]\right>_\xi\\
 &= \left<
\sum\limits_{\vec{z}}\textrm{Tr}\left[\hat\Sigma(z,0)\,\gamma_5\,\hat\Xi(0,
z)\,\gamma_5\right ]\right>_\xi\,,
\end{align}
with the sequential propagators
\begin{align}
 \hat\Sigma(z,0) & = \sum\limits_\nu \sum\limits_x S^s(z, 
x)\,\Gamma^c_\nu\,\hat\Delta_{\nu}(x)\,S^s(x,0)\,,\label{eq:seqpropsummu1}\\
\hat\Xi(0,z) &=\sum\limits_\mu \sum\limits_y 
S^{u}(0,y)\,\Gamma^c_\mu\,\xi_\mu^\dagger(y)\, 
S^{u}(y,z)\,.\label{eq:seqpropsummu2}
\end{align}
Each of these sequential propagators can be calculated with a single inversion 
using a sequential source with an insertion of either $\sum_\nu 
\Gamma^c_\nu\,\hat\Delta_{\nu}(x)$ or $\sum_\mu \Gamma^c_\mu\,\xi^\dagger_\mu$. 
The 
sequential propagators \eqref{eq:seqpropsummu1} and \eqref{eq:seqpropsummu2} 
are depicted in figure \ref{fig:seqpropsummu}.
\par
The calculation of the diagrams in figure \ref{fig:diagrams} requires a total 
of $5$ 
inversions per quark flavour and source position with the 
summed-$\mu$ insertion, $1$ for the point-to-all propagator, $1+1$ for the two 
sequential propagators \eqref{eq:seqpropsummu1} and \eqref{eq:seqpropsummu2}, 
$1$ additional inversion for the double 
sequential propagator for the quark self-energy diagram and $1$ inversion for 
the tadpole diagram using $\sum_\mu T_\mu \Delta_{\mu\mu}(0)$ as sequential 
insertion. Thus, the summed-$\mu$ insertion method is cheaper in computational 
cost compared to the single-$\mu$ insertion. However, the statistical error is 
expected to be larger, since the unwanted combinations, e.g.\ $\mu\neq\nu$ for 
the Feynman gauge, will contribute to the statistical noise.
\par
In this study, we use both, the single- and summed-$\mu$ insertion methods, and 
compare the statistical precision with the stochastic method to include QED in 
the lattice calculation.
\subsection{Strong Isospin Breaking}
\label{subsec:strongIB}
Even in the absence of QED, i.e.\ in pure QCD, isospin symmetry is broken by 
the different bare masses of up and down quarks.
In this work, we use two different strategies to account for effects from the 
strong isospin breaking, one by putting different values for the valence up- 
and down-quark masses, and one by expanding the Euclidean path integral in the 
quark mass 
\cite{deDivitiis:2011eh}. 
\par
Strong isospin breaking can be treated in a lattice 
calculation by simply using different values for up- and down-quark masses. 
In \cite{Fodor:2016bgu} the up- and down-quark mass difference has been 
determined in the $\overline{\textrm{MS}}$ scheme at $2$~GeV as
\begin{equation}
 m_u - m_d = -2.41(6)(4)(9)~\textrm{MeV}\,.
\label{eq:diffMSbar}
 \end{equation}
In section \ref{sec:setup} we specify the values we choose for the bare up- 
and down-quark mass for the setup used in this work to approximately reproduce 
the physical quark mass difference~\eqref{eq:diffMSbar}.
We include strong 
isospin breaking in a quenched setup, i.e.\ keeping the isospin symmetric 
sea-quark masses, to avoid having to generate new gauge configurations. 
\par
In addition, we use a strategy proposed in \cite{deDivitiis:2011eh} to account 
for 
strong isospin corrections. The idea is, to expand the path 
integral around the isospin symmetric light quark mass $\hat{m}$ 
\begin{equation}
 \left<O\right>_{m_f\neq \hat{m}} = \left<O\right>_{m_f=\hat{m}} + 
(m_{f} - \hat{m}) \left.\frac{\partial}{\partial 
m_f}\left<O\right>\right|_{m_f=\hat{m}} + \mathcal{O}\left((m_{f} - 
\hat{m})^2\right)\,,
\label{eq:mexpansion}
\end{equation}
where $m_f$ is either the mass of the down quark ($f=d$) or the up quark 
($f=u$). In this way, one explicitly calculates the leading isospin breaking 
correction $\mathcal{O}(m_f-\hat{m})$. Evaluating the derivative in equation 
\eqref{eq:mexpansion} one finds
\begin{equation}
  \left<O\right>_{m_f\neq \hat{m}} = \left<O\right>_{m_f=\hat{m}} - (m_{f} - 
\hat{m}) \left<O\,\mathcal{S}\right>_{m_f=\hat{m}}\,,
\label{eq:mexpansion2}
\end{equation}
with the scalar current
\begin{equation}
 \mathcal{S} = \sum\limits_x \overline{\psi}_f(x)\,\psi_f(x)\,,
 \label{eq:scalarcurrent}
\end{equation}
for a quark field $\psi_f$ of flavour $f$. A detailed derivation of equation 
\eqref{eq:mexpansion2} for the Domain Wall Fermion action used in this study 
can be found in appendix \ref{subsec:appendixstrongIB}.
\par
\begin{figure}[h]
 \centering
 \includegraphics[width=0.4\textwidth]{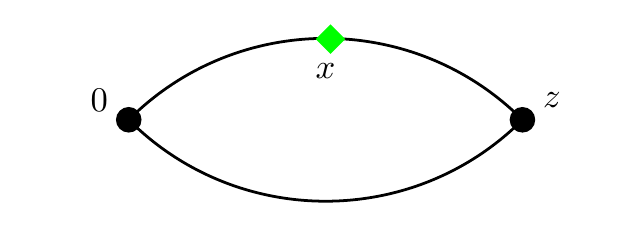}
 \caption{Quark-connected diagram for the strong isospin correction to a 
mesonic two-point function. The green diamond vertex denotes the insertion of 
the scalar current \eqref{eq:scalarcurrent}.}
 \label{fig:massinsertion}
\end{figure}

For a mesonic two-point function one finds at $\mathcal{O}(m_f-\hat{m})$ one 
type of quark-connected contribution, which is shown in figure 
\ref{fig:massinsertion}. Note, that we do not include strong isospin breaking 
effects for the sea quarks.
\par
To illustrate how we calculate the diagram shown in figure 
\ref{fig:massinsertion}, we consider as an example the strong isospin correction
for a kaon , which is, according to equation \eqref{eq:mexpansion2}, 
determined by a correlation function of the form
\begin{equation}
 C^\textrm{strongIB}_{K}(z_0) =  
\sum\limits_{\vec{z}}\!\sum\limits_{x}\!\textrm{Tr}\!\left[S^s(0,
z)\,\gamma_5\,S^{l}(z,x)\,S^{l}(x,0)\,\gamma_5\right]\,,
\label{eq:corrdiffKK}
\end{equation}
with a light quark propagator $S^l$ using the isospin symmetric quark mass 
$\hat{m}$.
We calculate \eqref{eq:corrdiffKK} using a sequential propagator
\begin{equation}
 C^\textrm{strongIB}_{K}(z_0)\! =  
\!\sum\limits_{\vec{z}}\!\textrm{Tr}\!\left[S^s(0,
z)\,\gamma_5\,\Omega(z,0)\,\gamma_5\right]\,\,\,\,\,\,\textrm{with}\,\,\,\, 
\Omega(z,0) =\! \sum\limits_{x} S^{l}(z,x)\,S^{l}(x,0).
\end{equation}
Both approaches for 
the inclusion of strong isospin breaking effects used in this study, are equal 
in terms of computational cost, i.e.\ number of inversions, since it requires 
either one inversion per gauge configuration and source position using a 
different quark mass for the down quark, or one inversion per gauge 
configuration and source position to calculate the sequential propagator 
$\Omega(z,0)$ using the isospin symmetric quark mass. However, using the 
expansion of the path integral in the mass is more flexible, since the 
deviation of the quark mass from the isospin symmetric mass $(m_f-\hat{m})$ is 
a free parameter, which is multiplied to the correlation function after all 
quark contractions have been computed. This allows for tuning the quark 
masses a posteriori, e.g.\ for fixing hadron masses to their physical value.
% \par
% 
% 
\section{Computational Setup}
\label{sec:setup}
For this study we use a $64\times24^3$ lattice with $N_f=2+1$ dynamical 
flavours of Shamir Domain Wall Fermions \cite{Kaplan:1992bt,Shamir:1993zy} with 
a 
Domain Wall height of $M_5=1.8$ and $L_s=16$, where $L_s$ is the length of the 
fifth dimension. For further details 
see \cite{Allton:2008pn,Aoki:2010dy}.
This gauge ensemble has been generated by the RBC/UKQCD collaboration using the 
Iwasaki gauge action \cite{Iwasaki:1984cj,Iwasaki:1985we}. 
The inverse lattice spacing of this 
ensemble has been determined without QED as $a^{-1}=1.78$~GeV 
\cite{Boyle:2017jwu}. The bare sea quark masses are $am_l=0.005$ and $am_s=0.04$ 
for light and strange quarks, respectively. With such quark masses the isospin 
symmetric pion mass on this QCD gauge ensemble is $m_\pi = 340$~MeV, thus, in 
this work we do not calculate at physical quark masses, even in the absence of 
QED. 
\par
In the present study, we use different values for the valence up- and 
down-quark masses. 
While keeping the valence up-quark at the same mass as 
the light quarks in the sea, we choose the valence down quark mass as
$am_d=0.005915$. Using the results from \cite{Blum:2014tka} to convert the bare 
mass difference $a(m_d-m_u)=0.000915$ to $\overline{\textrm{MS}}$, we find 
$m^R_d-m^R_u = 2.4$~MeV for $\overline{\textrm{MS}}$ at $2$~GeV, and thus 
we reproduce the physical mass difference given in \cite{Fodor:2016bgu} (cf. 
equation \eqref{eq:diffMSbar}).
\par
This choice ignores any QED effects in the renormalization of the quark
mass. While this is acceptable for the comparative study presented here,
more work is needed when aiming at physical predictions, see
e.g.~\cite{Borsanyi:2014jba,Horsley:2015vla}. For instance, we know that
there is
a small additive correction to the quark mass under renormalization in
our setup. The correction can be quantified in terms of the {\it
residual mass} which~\cite{Blum:2010ym} determined to be $m_{\rm
res}\approx 0.003$ on the above ensemble (see~\cite{Blum:2010ym} for
details). The residual mass is defined such that in the chiral limit
$m_f=-m_{\rm res}$ and it receives additional contributions in QCD+QED
which are of order $\mathcal{O}(\alpha m_{\rm res})$.
Moreover, the
multiplicative renormalization of the quark mass will receive QED
contributions at $\Oalpha$ which have not been taken into
account here.
\par
A consequence of these simplifications in our choice of parameters is
that the  neutral pion splitting in the chiral limit does not vanish for
finite $L_s$ \cite{Blum:2010ym} and indeed, in this work we find the neutral 
pion mass
shift due to QED to be sizeable. Since  we are only interested in a
comparative study of approaches to Lattice QCD+QED no attempt has been
made to correct for this effect.
Note that this effect is much more severe for lattice quark actions not
obeying chiral symmetry such as Wilson fermions \cite{Portelli:2010yn}.
\par
For the bare valence strange-quark mass we use $am_s=0.03224$ 
\cite{Blum:2014tka}, which, without QED, corresponds to the physical strange 
quark mass.
\par
Working with physical quark masses, requires to 
tune the quark masses to their physical values including QED. This could be 
done, for 
example, by tuning the up-, down- and strange-quark masses until the masses of 
charged pion and neutral and charged kaons agree with their experimentally 
measured values. In addition, this requires the determination of the lattice 
spacing 
including QED, which could be done by fixing another hadron mass to its 
physical value, e.g.\ the $\Omega$-baryon. However, since this is an 
exploratory study and mainly focused on the comparison of the stochastic and 
perturbative method for including QED, we have not retuned any of the quark 
masses in the presence of QED.
\par
We use $87$ QCD gauge configurations and $16$ source positions with 
$\mathbb{Z}_2$ wall sources \cite{Foster:1998vw,McNeile:2006bz,Boyle:2008rh} for 
the quark propagators. For the stochastic 
method we use one $U(1)$ QED configuration per QCD gauge configuration. For the 
perturbative method we use one $\mathbb{Z}_2$ noise for the insertion of the 
photon propagator per QCD gauge configuration and source position for the 
single-$\mu$ insertion method and one $\mathbb{Z}_2$ noise for every Lorentz 
index for the summed-$\mu$ insertion.

%% file: 3_mesonmasses.tex
\section{Isospin Breaking Corrections to Meson Masses}
\label{sec:mesonmasses}
As a starting point for comparing results from the stochastic and the 
perturbative method we calculate the isospin breaking corrections to meson 
masses.
Several other calculations of QED corrections to meson masses exist even  
at, or extrapolated to, the physical point, see e.g.\ 
\cite{Duncan:1996xy,Blum:2007cy,Blum:2010ym,Borsanyi:2013lga,
Borsanyi:2014jba, 
deDivitiis:2013xla,Horsley:2015eaa,Horsley:2015vla,Basak:2016jnn,Fodor:2016bgu,
Giusti:2017dmp} .
In this work, we use an exploratory setup with one gauge ensemble at 
non-physical quark masses. However, for the 
first time, we directly compare results from the stochastic and 
perturbative methods. We also explain that the QED correction to meson masses 
has to be determined in different ways for the stochastic and the perturbative 
data, to obtain results, which can properly be compared to each other.
\subsection{Extraction of the QED Correction to the Effective Mass}
\label{subsec:extractionofmasses}
The two-point correlation function for a pseudoscalar meson interpolation 
operator
$\left(\psibar_f\,\gamma_5\,\psi_{f'}\right)$ with quark flavours $f$ and $f'$ 
and vanishing spatial momentum $\vec{p}=0$, 
which is created at $0$ and annihilated at $x$, is given by 
\begin{equation}
  C(t) = 
\sum\limits_{\vec{x}}
\left<\psibar_{f'}(x)\,\gamma_5\,\psi_{f}(x)\,\,\,\psibar_{f}
(0)\,\gamma_5\,\psi_{f'}(0)\right>\,.
\end{equation}
Such a two-point correlation function has the following time-dependence for 
large Euclidean times, where excited-state contributions are suppressed 
\begin{equation}
 C(t) = A\left(e^{-m t} + e^{-m (T-t)}\right)\,, %+ \textrm{excited-state 
%contributions}
\label{eq:twopt}
 \end{equation}
for a lattice with time extend $T$ and periodic boundary conditions. The 
parameter $m$ that determines the leading exponential decay of \eqref{eq:twopt} 
is the mass of the ground state meson, whereas excited-state contributions are 
exponentially suppressed. 
In this study we are interested in the mass $m$ of 
the ground state. 
A common method to determine the mass of 
the ground state meson from a two-point correlation function $C(t)$ is to 
calculate an effective mass. In this work we use the definition of the 
effective mass, where one solves
\begin{equation}
\frac{C\left(t\right)}{C\left(t+1\right)}=\frac{\cosh\left(\left(t-T/2\right)m_{
\mathrm{eff}}\right)}{\cosh\left(\left(t+1-T/2\right)m_{\mathrm{eff}}\right)}\,,
\label{eq:cosheffmass}
\end{equation}
for $m_\mathrm{eff}$ at every $t$.\par
In the following, we discuss how to determine the QED correction to the 
effective mass. 
The effective 
mass including QED is given by the effective mass $m^0_\textrm{eff}$ without 
QED plus the QED correction $\delta m_\textrm{eff}$ %to the effective mass:
\begin{equation}
 m_\textrm{eff}(t) = m^0_\textrm{eff}(t) + \delta m_\textrm{eff}(t)\,.
\end{equation}
For the data from the stochastic approach the two-point function including QED 
contains corrections to all orders in $\alpha$ and has the form 
\eqref{eq:twopt} with $A=A_0+\delta A$ and $m=m_0+\delta m$.
Thus, the QED correction to the effective 
mass can be obtained by determining the effective mass according to 
equation \eqref{eq:cosheffmass} once for the two-point function with QED and 
once for the two-point function without QED and taking their difference 
\begin{equation}
 \delta m^\textrm{cosh}_\textrm{eff}(t) =  m_\textrm{eff}(t) - 
m^0_\textrm{eff}(t)\,.
\label{eq:coshmassmethod}
\end{equation}
In the following we refer to this method to extract the QED correction to the 
effective mass as the \textit{cosh-mass method}, which is the appropriate method 
to extract the QED correction using the stochastic data.
\par
On the other hand, the two-point function including QED can be expanded 
\cite{deDivitiis:2013xla} (neglecting the backwards propagating 
meson for simplicity)
\begin{equation}
 C(t) = C_0(t)+\delta C(t) 
 = (A_0+\delta 
A) e^{-m_0 t} \left(1 - \delta m\,t + \frac{1}{2}\delta m^2\,t^2 + 
\ldots\right)  \,.
\label{eq:deriv_ratio}
\end{equation}
In the perturbative approach, one explicitly only calculates the QED correction 
to the two-point function which are of $\Oalpha$. Keeping only terms which are 
of order $\alpha$ in equation~\eqref{eq:deriv_ratio} one finds 
\begin{equation}
 \delta C(t) = C_0(t) \left( \frac{\delta A}{A_0}- \delta m\, t \right)\,,
 \label{eq:Clinindm}
\end{equation}
for the QED correction $\delta C(t)$ from the perturbative data.
Equation \eqref{eq:Clinindm} implies, that the QED correction to the effective 
mass can be defined from the ratio of the QED correction $\delta C(t)$ to 
the two-point function and the two-point function $ C_0(t)$ without QED
\begin{equation}
 \delta m_\textrm{eff}^\textrm{ratio}(t) = \frac{\delta C(t)}{ C_{0}(t)} - 
\frac{\delta C(t+1)}{ C_{0}(t+1)}\,.
\label{eq:ratio_woperiodic}
\end{equation}
Equation \eqref{eq:ratio_woperiodic} can be extended to include the effects of 
the periodic boundary conditions
\begin{equation}
\begin{aligned}
\delta m^\textrm{ratio}_\textrm{eff}(t)& = \left[\frac{\delta C(t)}{ C_{0}(t)} 
- 
\frac{\delta C(t+1)}{ C_{0}(t+1)}\right] \\ 
&\times\frac{1}{ 
\left(\frac{T}{2}-t\right) 
\textrm{tanh}\left(m_0\left(\frac{T}{2}-t\right)\right) -  
\left(\frac{T}{2}-(t+1)\right) 
\textrm{tanh}\left(m_0\left(\frac{T}{2}-(t+1)\right)\right)}\,,
\end{aligned}
\label{eq:ratiomethod}
\end{equation}
using $m_0$ from a determination from the two-point function without QED as an 
input. In the following we refer to this method to extract the QED correction to 
the effective mass as the \textit{ratio method}, which is the appropriate 
method to extract the QED correction using the perturbative data.
\par
\begin{figure}[h]
  \centering
  \includegraphics[width=0.48\textwidth]{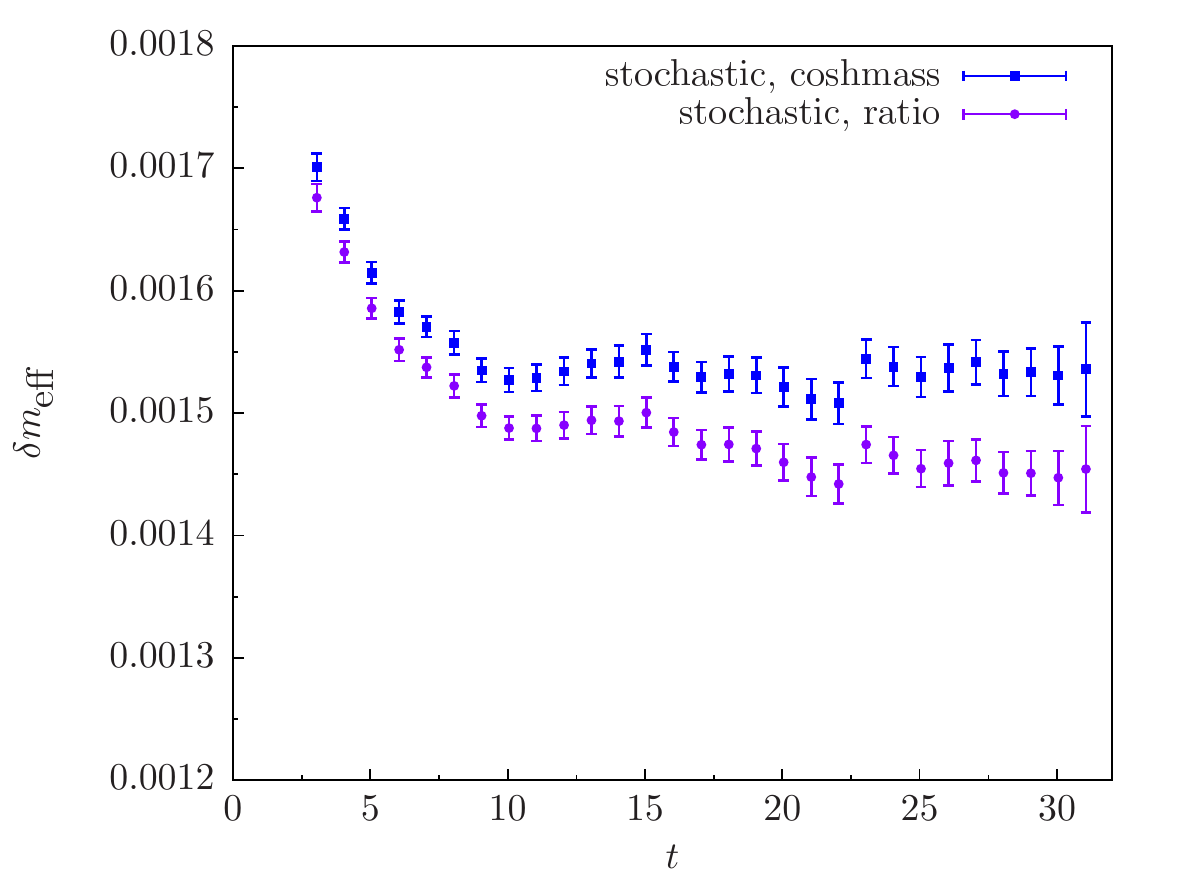}
  \includegraphics[width=0.48\textwidth]{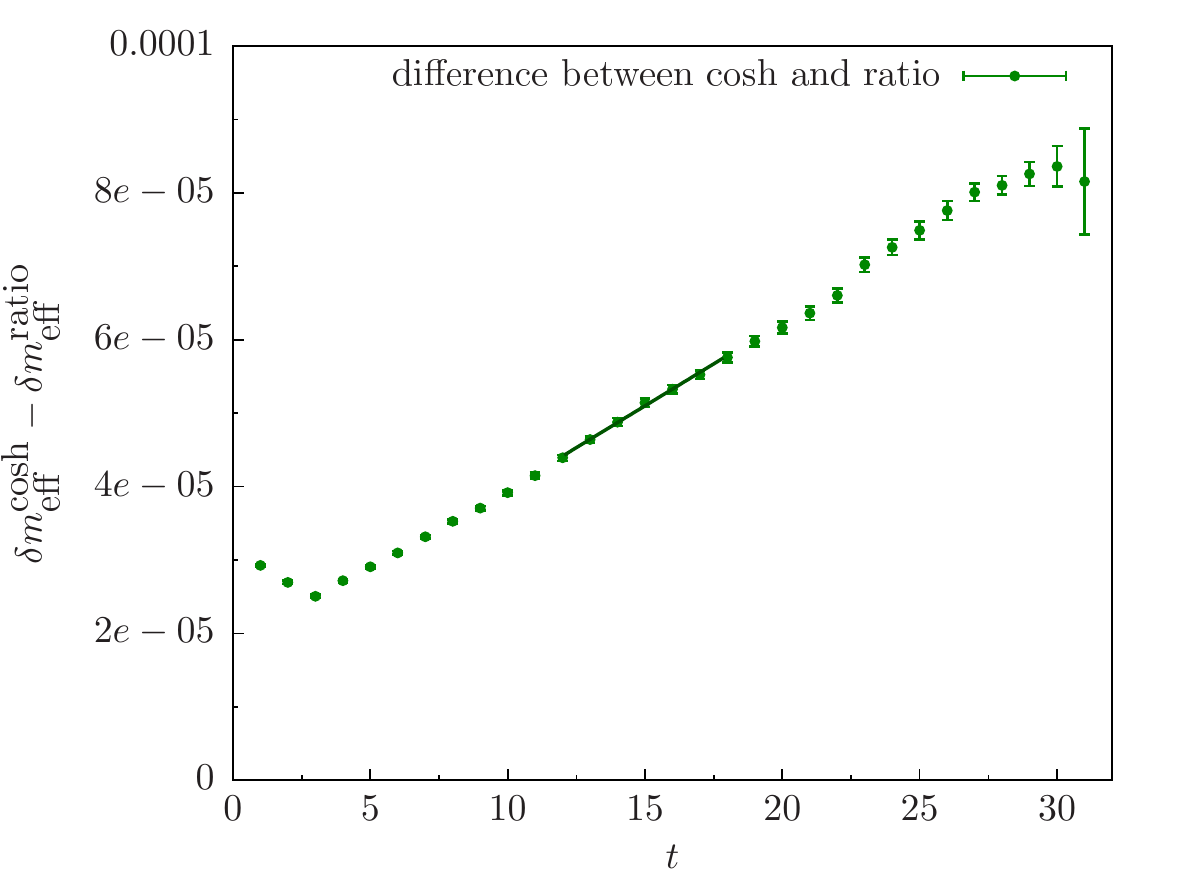}
  \caption{The QED correction to the effective mass of a charged kaon from the 
stochastic data. The plot on the left shows the results using the cosh-mass 
method (blue squares) and the ratio method (purple circles). The plot on the 
right shows the correlated difference between the results from cosh-mass and 
ratio method.}
 \label{fig:K+effmass}
 \end{figure}

When using the ratio method for the stochastic data, one has to take 
into account, that the QED correction to the two-point function includes QED 
corrections to all orders in $\alpha$. Keeping also higher order terms in the 
expansion \eqref{eq:deriv_ratio} one finds
\begin{equation}
 \frac{\delta C(t)}{C_0(t)} - \frac{\delta C(t+1)}{C_0(t+1)} = \delta m 
 \,- {{\underline{\delta m^2\,t - \frac{1}{2}\delta m^2 +   
\frac{\delta A}{A_0}\delta m}}}\,+\cdots\,,
\label{eq:ratiostoch}
\end{equation}
for the ratio method from the stochastic data. The underlined terms in 
\eqref{eq:ratiostoch} are included in the stochastic data, but not in the 
perturbative data. Thus, one expects the QED correction to the effective mass 
extracted using the cosh-mass method \eqref{eq:coshmassmethod} and the ratio 
method \eqref{eq:ratiomethod} from the stochastic data  to differ by 
\begin{equation}
\delta m_\textrm{eff}^\textrm{cosh} - \delta m_\textrm{eff}^\textrm{ratio} =  
\delta m^2\,t + \frac{1}{2}\delta m^2 - \frac{\delta A}{A_0}\delta 
m\,+\cdots\,\,.
\label{eq:diff_ratio_cosh}
\end{equation}
We indeed find this difference in our data as illustrated in figure 
\ref{fig:K+effmass}. The plot on the left-hand side shows the QED correction to 
the effective mass from the stochastic data extracted with the cosh-mass 
method (blue squares) and the ratio method (purple circles). We find a 
significant difference between the results from both extraction methods. The 
correlated difference is plotted on the right-hand side of figure 
\ref{fig:K+effmass}. We can numerically confirm, that this difference is given 
by equation \eqref{eq:diff_ratio_cosh} as expected.
\par
Thus, in the following we determine the QED correction to meson masses
using the cosh-mass method for the stochastic data and the ratio method for the 
perturbative data.

 \subsection{QED Correction to Meson Masses}
In the following we show results for the QED correction to meson masses. In 
subsection~\ref{subsubsec:mesonresults} the QED corrections to meson masses are 
determined and results from the perturbative and the stochastic method are 
compared. In subsection \ref{subsubsec:mesonstaterror} we compare the 
statistical errors on the results from both methods.
\subsubsection{Results}
\label{subsubsec:mesonresults}
\begin{figure}[h]
 \includegraphics[width=0.48\textwidth]{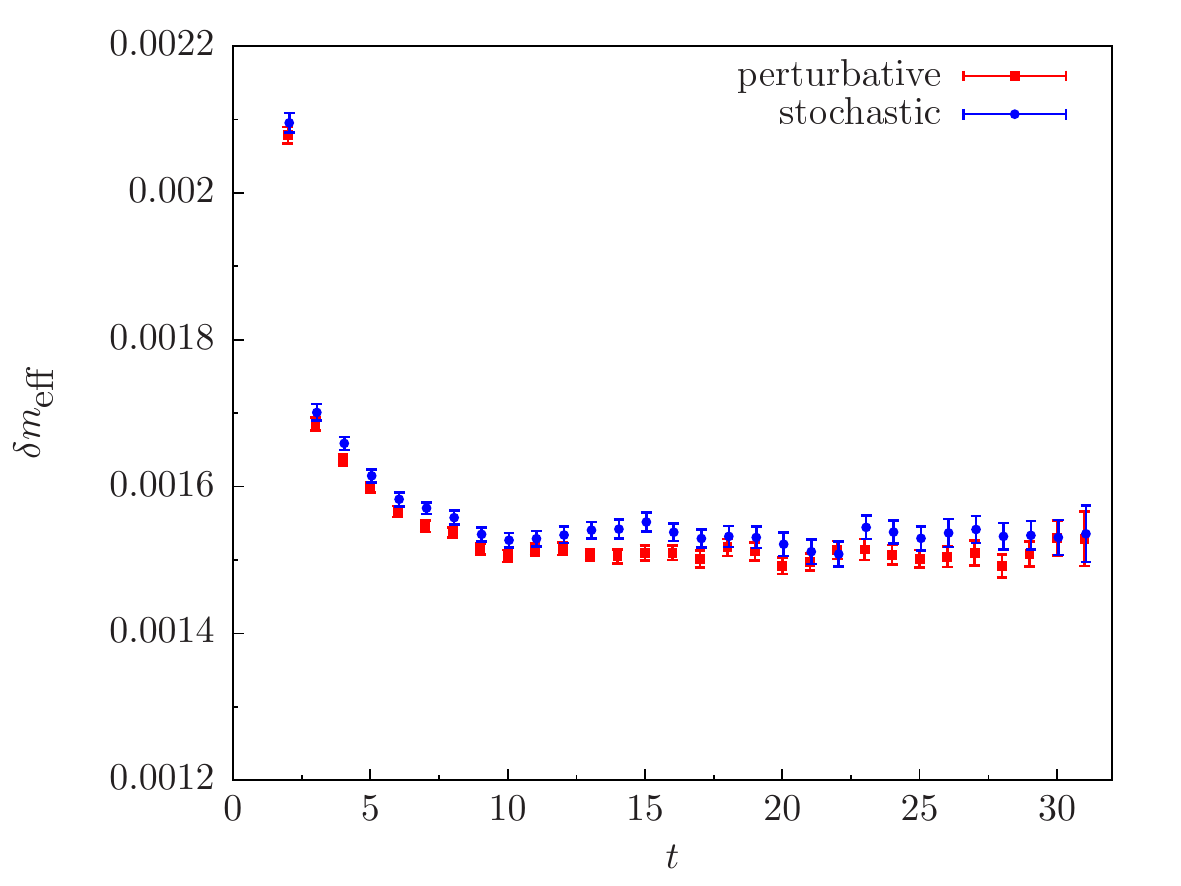}
 \includegraphics[width=0.48\textwidth]{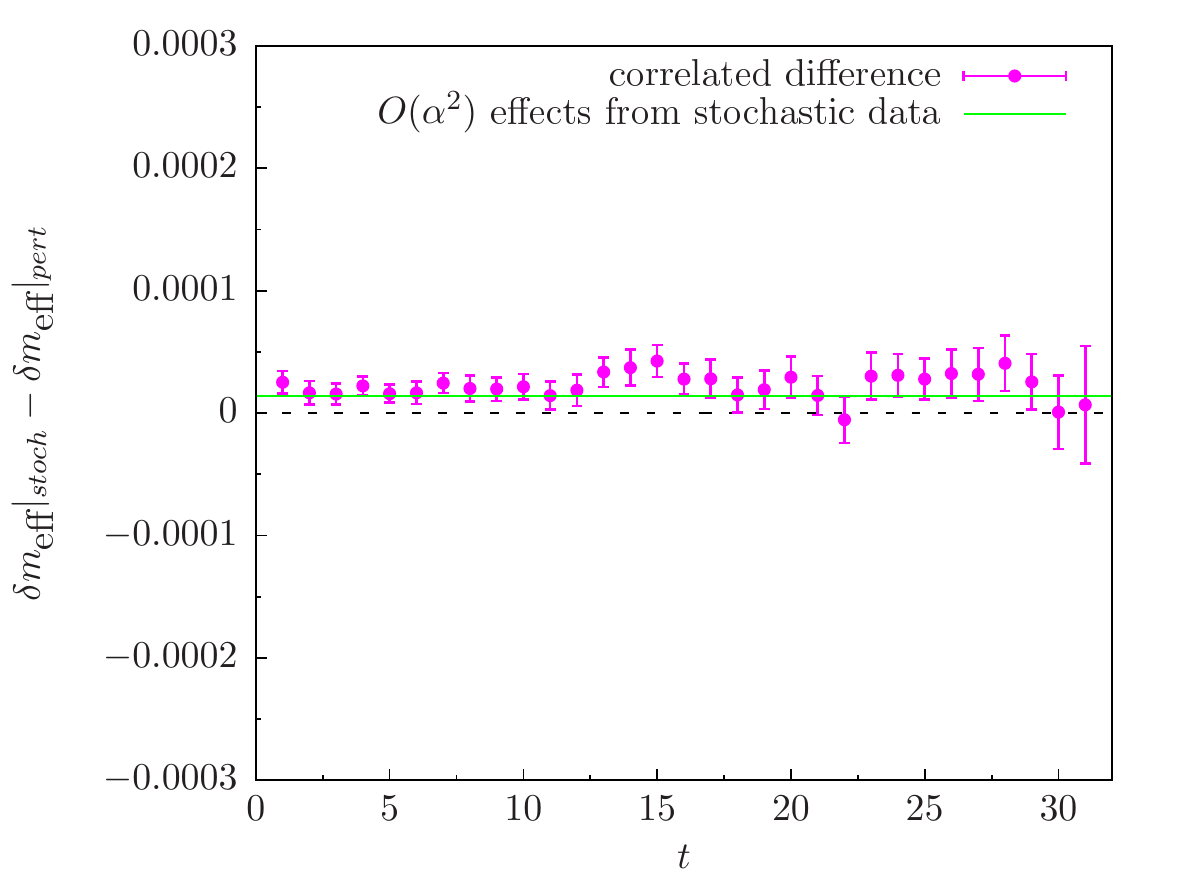}
 \caption{The QED correction to the effective mass of a charged kaon. The plot 
on the left shows a comparison between stochastic (blue circles) and 
perturbative (red squares) data, the plot on the right shows the correlated 
difference of both datasets. The solid green line shows the $\Oalphasquare$ 
effects which are included in the stochastic data, i.e.\ the expected 
discrepancy between stochastic and perturbative data.
}
\label{fig:KaonQEDeffmass}
 \end{figure}

The left-hand side of figure \ref{fig:KaonQEDeffmass} shows the QED correction 
to the effective mass of a charged kaon using the Feynman gauge for the photon 
fields. The red squares show results from the perturbative data using the 
ratio method \eqref{eq:ratiomethod} to extract the QED correction and the blue 
circles are results 
from the stochastic data using the cosh-mass method \eqref{eq:coshmassmethod}.
For the perturbative data the results shown have been calculated using the 
single-$\mu$ insertion, which, for the same amount of statistics, gives a 
smaller statistical error than the summed-$\mu$ insertion (see section 
\ref{subsubsec:mesonstaterror} for a detailed comparision of statistical 
errors).
\par
The plot on the right-hand 
side of figure \ref{fig:KaonQEDeffmass} shows the correlated difference between 
the stochastic and perturbative data. Both datasets are correlated since they 
have been calculated on the same QCD gauge configurations and the same source 
positions with the same $\mathbb{Z}_2$ wall sources for the quark propagators. 
Statistical errors are estimated using the bootstrap resampling method.\par
We find the correlated difference between both datasets to be non-zero at the 
level of $\approx 1.5\sigma$ and of the order of $1\%$ of the QED correction 
itself, which can be attributed to $\Oalphasquare$ effects.
To check this, we have 
repeated the calculation with the stochastic method using a second, larger 
value of the electromagnetic coupling $\alpha = \nicefrac{1}{4\pi}$. 
Using the results for the QED correction to the mass from two different values 
of the coupling $e$ and an ansatz $\delta m^{\textrm{stoch}}=\alpha m_1 + 
\alpha^2 m_2$, we can explicitly determine the $\Oalpha$- and the 
$\Oalphasquare$-correction which are included in the stochastic data. 
More details can be found in the appendix in section \ref{sec:Oalphasquare}.
The
$\Oalphasquare$-correction, which is included in the stochastic data, is 
shown by the solid green line on the right-hand side of 
figure \ref{fig:KaonQEDeffmass}. 
Thus, we find that the difference between the results from the perturbative and 
the stochastic method is described by the $\alpha^2$ contribution to the mass, 
which is only included in the stochastic data.
\par
To determine the QED correction to the mass of a meson, we fit a constant to 
the plateau region of the QED correction to the effective mass, such as the 
data in figure \ref{fig:KaonQEDeffmass}. The results are given in table 
\ref{tab:qedcorrmesonmasses} for the stochastic and the perturbative method. 
We give results for charged and neutral pions as well as charged and neutral 
kaons. Note, that we do not include the quark-disconnected diagram for the 
neutral pion. One also has to keep in mind, that our calculation is not using 
physical quark masses, and thus the results shown here are not at the physical 
point. The small but significant difference in the results from the 
stochastic and perturbative method for charged pion and kaon is due to higher 
order effects in $\alpha$, which are only included in the stochastic data. 
\par
QED in a finite box is subject to substantial finite volume effects. Although 
in this exploratory study we do not give results at the physical point 
and thus, 
correcting for finite volume effects is not strictly necessary, we include 
finite volume corrections for the meson masses to illustrate that they are 
significant. 
Finite volume effects for the QED correction to the meson masses are 
analytically known up to $\mathcal{O}(1/L^3)$ corrections and given 
by~\cite{Borsanyi:2014jba}
\begin{equation}
 m^2(L) \sim m^2 
\left\{1-q^2\alpha\left[\frac{\kappa}{mL}\left(1+\frac{2}{mL}\right)\right]
\right\}\,,
\label{eq:finVcorr}
\end{equation}
with $\kappa=2.837297$. $m(L)$ and $m$ are the meson masses including QED in 
finite and infinite volume, respectively. In table \ref{tab:qedcorrmesonmasses} 
we quote 
results in finite volume and results $\delta m^\textrm{inf V}$ in infinite 
volume, 
where finite volume effects have been accounted for using equation 
\eqref{eq:finVcorr}. 
\par
\begin{table}[h]
\centering
\begin{tabular}{|c||c|c||c|c|}
\hline
 & \multicolumn{2}{|c||}{stochastic} & \multicolumn{2}{|c|}{perturbative}\\
 & $\delta m$ $/$MeV & $\delta m^\textrm{inf V}$ $/$MeV & $\delta m$ $/$MeV & 
$\delta m^\textrm{inf V}$ $/$MeV\\
\hline
$\delta m_{\pi^+}^\gamma$ & $3.504\pm0.025$ & $4.597\pm0.025$ & $3.459\pm0.016$ 
& 
$4.552\pm0.016$\\
$\delta m_{\pi^0}^\gamma$ & $1.555\pm0.015$ & $1.555\pm0.015$ & $1.538\pm0.016$ 
& 
$1.538\pm0.016$\\
\hline
$\delta m_{K^+}^\gamma$ & $2.722\pm0.022$ & $3.699\pm0.022$ & $2.677\pm0.013$ & 
$3.653\pm0.013$ \\
$\delta m_{K^0}^\gamma$ & $0.547\pm0.005$ & $0.547\pm0.005$ & $0.548\pm0.005$ & 
$0.548\pm0.005$\\
\hline
\end{tabular}
\caption{Results for the QED correction to the meson masses from the stochastic 
and the perturbative method. For both methods the left column shows the result 
in finite volume and the right panel the result in infinite volume using 
\eqref{eq:finVcorr}. Note, that these results have not been obtained at the 
physical point. The large effect on the neutral pion mass is due to a
small
amount of residual chiral symmetry breaking in our Domain Wall Setup
(cf. discussion in section \ref{sec:setup} and in \cite{Blum:2010ym}).}
\label{tab:qedcorrmesonmasses}
\end{table}
\par
The QED correction to the meson masses has been previously calculated in an 
independent calculation \cite{Blum:2010ym} using a stochastic method on the 
same gauge ensemble, albeit different gauge configurations. A comparison of our 
results with the results from this independent calculation can serve as a cross 
check of our data. In table \ref{tab:cross-check} we show results for 
the squared mass splitting $\Delta m^{2}=(m_0+\delta m)^{2}-m_{0}^{2}$ for a 
pion. Both light quark masses in this comparison equal the light sea-quark 
mass, i.e.\ a bare mass of $m_l=0.005$. We find agreement 
between our results and the results from this previous calculation.
\begin{table}[h]
	\centering
	\begin{tabular}{|c|c|c|c|}
		\hline 
		$q_{1}$ & $q_{2}$ & $a^2\Delta m^{2}$ this work & 
$a^2 \Delta 
m^{2}$ from \cite{Blum:2010ym}\\
		\hline\hline
		2/3 & 2/3 & $\left(5.465\pm0.035\right)\times10^{-4}$ & 
$\left(5.406\pm0.064\right)\times10^{-4}$\\
		2/3 & -1/3 & $\left(7.677\pm0.052\right)\times10^{-4}$ & 
$\left(7.654\pm0.056\right)\times10^{-4}$\\
		-1/3 & -1/3 & $\left(1.341\pm0.009\right)\times10^{-4}$ & 
$\left(1.326\pm0.016\right)\times10^{-4}$\\
		\hline 
	\end{tabular}
	\caption{Comparison of pion squared mass splittings
	from the stochastic data and the results of a previous calculation in 
\cite{Blum:2010ym}. Results are given in lattice units. $q_1$ and $q_2$ denote 
the charges of the valence quarks in units of $e$. Both data sets use the 
Feynman gauge for the photon fields.}
	\label{tab:cross-check}
\end{table}
\par
\subsubsection{Comparison of Statistical Errors}
\label{subsubsec:mesonstaterror}
To compare the statistical errors on the QED correction to the effective mass 
between the stochastic and the perturbative data, one has to take into account, 
that these two datasets have not been obtained at the same numerical cost. For 
the 
stochastic data we need three inversions per quark flavour and source position 
($e=0$, $e$, $-e$) to obtain the QED correction. As described in section 
\ref{sec:pertmethod} in our setup the calculation of the QED correction with 
the perturbative method requires $17$ inversions per quark flavour and source 
position using the single-$\mu$ insertion, if the Feynman gauge is used for the 
photon propagator, and $5$ inversions using the summed-$\mu$ insertion. Thus, 
we find a $17/3$ or $5/3$ larger numerical cost for the perturbative method to 
obtain the same statistics than for the stochastic data.
\par
\begin{figure}[h]
 \centering
 \includegraphics[width=0.48\textwidth]{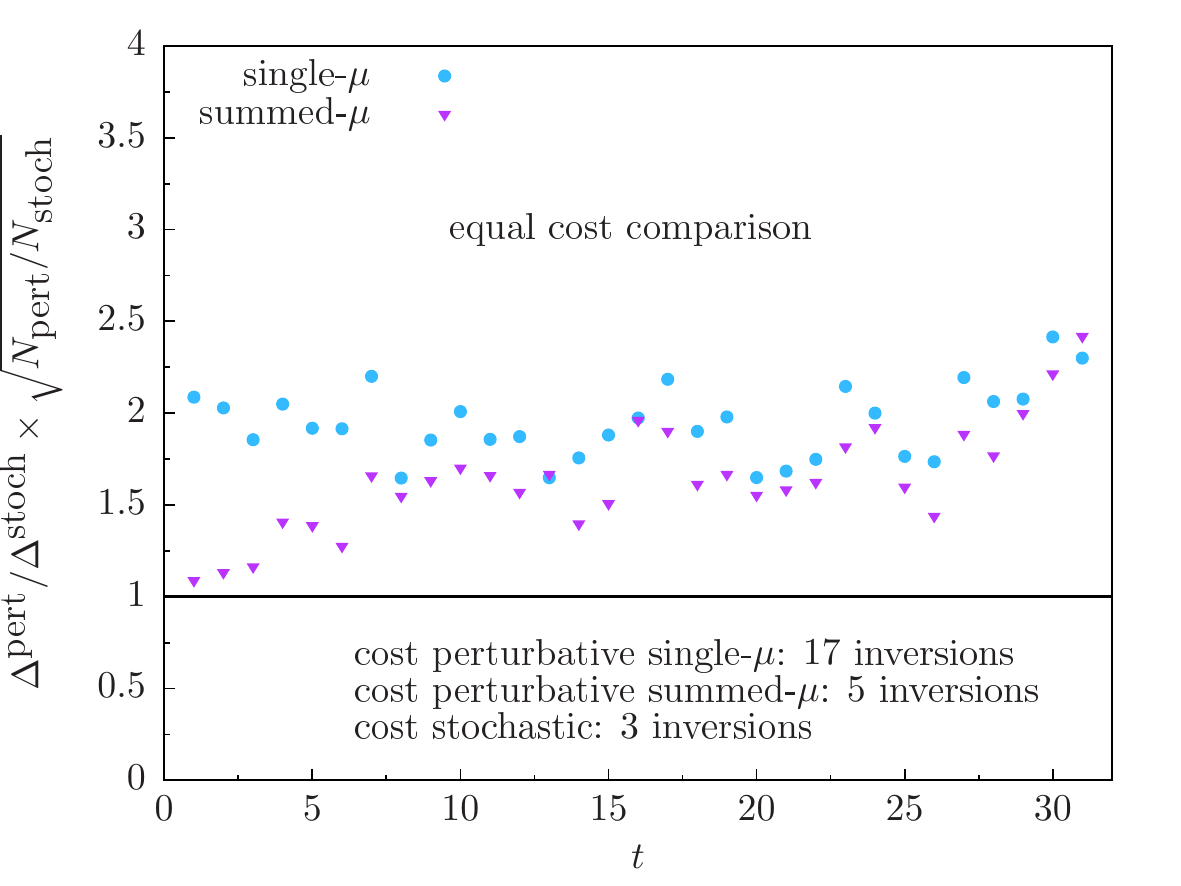}
\includegraphics[width=0.48\textwidth]{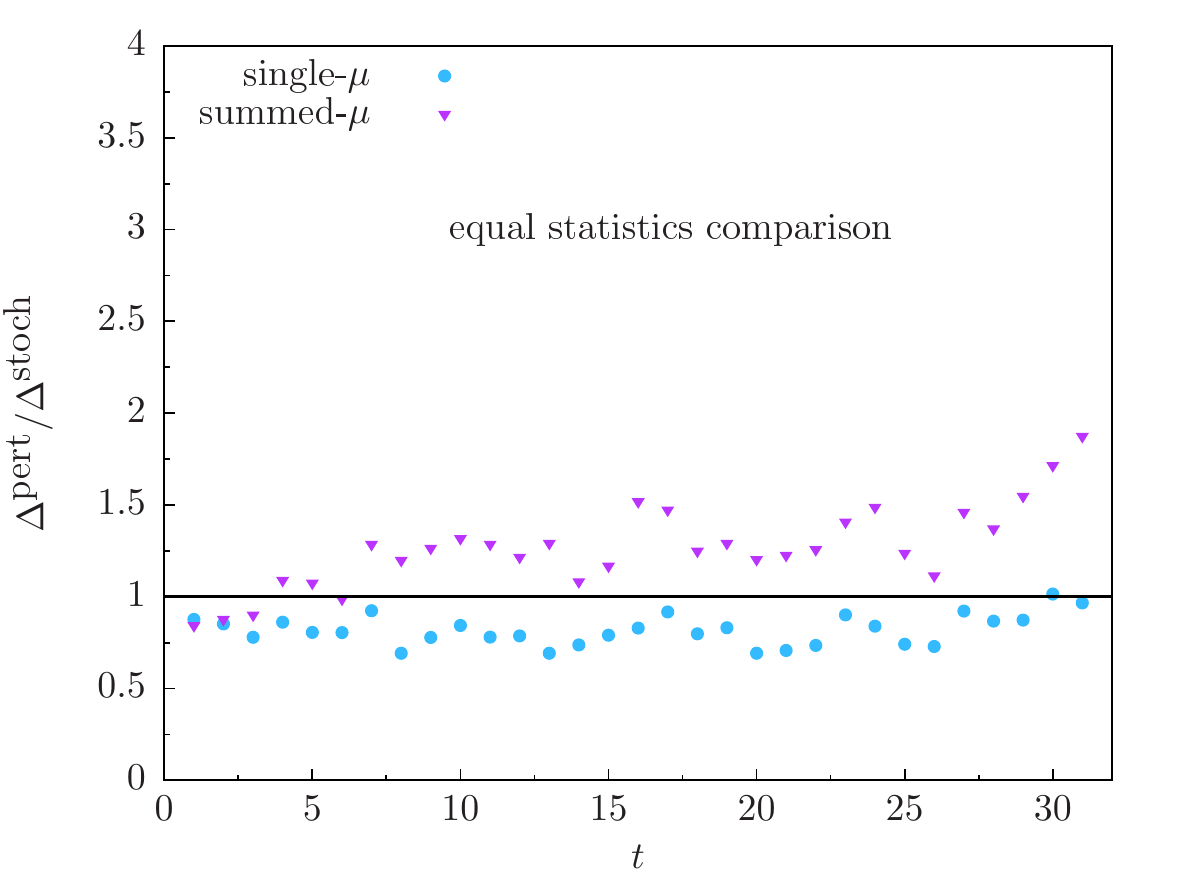}
\caption{Comparison of the statistical errors on the QED correction to the 
effective mass of a charged kaon between the stochastic and the perturbative 
data. Blue circles and purple triangles are using the single- or summed-$\mu$ 
insertion technique for the perturbative method, respectively. The plot on the 
left is an equal cost comparison, the plot on the right 
an equal statistics comparison.}
\label{fig:error_meson}
 \end{figure}
Figure \ref{fig:error_meson} shows a comparison between the statistical errors 
of the stochastic and the perturbative data. The plot on the left shows the 
error from the perturbative data divided by the error on the stochastic data, 
both scaled with their respective numerical cost (i.e.\ the number of 
inversions) to have an equal cost comparison. 
Blue circles and purple triangles are using the single- or summed-$\mu$ 
insertion technique for the perturbative method, respectively.
The black horizontal line shows the threshold above which the accuracy of the 
stochastic approach is superior to the one of the perturbative approach.
We find the perturbative method to give an error which 
is about a factor $1.5$ to $2$ larger than the error on the stochastic method 
for the same costs. 
Comparing the two different approaches for calculating the sequential 
propagators for the perturbative method, we find 
that at the same numerical cost the statistical error is smaller when using the 
summed-$\mu$ insertion.  
\par
The right-hand side of figure \ref{fig:error_meson} shows the ratio of the 
errors of perturbative and stochastic data with the same set of statistics. 
We 
find this ratio to be slightly smaller but close to one for the single-$\mu$ 
insertion and slightly larger then one for the summed-$\mu$ insertion, and 
thus finding 
similar statistical errors for the perturbative and stochastic data when using 
the same statistics for the QCD average. 
In summary, the ordering of statistical errors is
\begin{align}
& \Delta^\textnormal{stoch}<\Delta^{\textnormal{pert,summed-}\mu}
<\Delta^{\textnormal{pert,single-}\mu}& \hspace{1cm}&\textnormal{same cost}\\
&\Delta^{\textnormal{pert,single-}\mu}<\Delta^\textnormal{stoch}
<\Delta^{\textnormal{pert,summed-}\mu}&\hspace{1cm}&\textnormal{same 
statistics.}
\end{align}
For quenched QED, depending on whether the cost of QCD gauge configuration 
generation is to be included in a cost assesment, the optimal method to select 
will
either be the most precise for the same statistics if the cost of the 
measurement is sub-dominant, or if a sufficient ensemble of gauge configurations 
already exists it makes sense to 
select the most precise approach for fixed measurement cost.
\par
We note, that a cost comparison between stochastic and perturbative methods is 
less trivial in 
unquenched QED. While for the perturbative method, one needs to additionally 
calculate appropriate quark-disconnected diagrams, the stochastic method 
requires the generation of combined QCD$+$QED gauge configurations. 

\subsection{Strong Isospin Breaking Correction}
In the following we show results for the strong isospin breaking corrections 
to meson masses. 
%\par
As discussed above, we use two different approaches to account for strong 
isospin breaking, one by simply using different valence up- and down-quark 
masses when computing valence quark propagators, and one by expanding the path 
integral in the quark mass difference. When 
comparing results from these two approaches, one has to keep in mind, that, 
when choosing different values for up- and down-quark masses, we fixed the 
up-quark mass to the isospin symmetric mass $m_u=\hat{m}$ and changed the 
down-quark mass to be $m_d=\hat{m}+(m_d-m_u)$, where $(m_d-m_u)$ approximately 
corresponds to the physical light quark mass difference from 
\cite{Fodor:2016bgu}.
\par
In the following we focus on the strong isospin correction 
to the masses of charged kaon $K^+=\overline{s}\gamma_5 u$ and neutral kaon 
$K^0=\overline{s}\gamma_5 d$. 
In this context ``charged'' and ``neutral'' refers only to the quark content, 
not to electromagnetic charges. 
In particular, we consider the strong isospin contribution to the 
difference $\tilde{m}_{K^0}-\tilde{m}_{K^+}$ of the masses of charged and 
neutral kaon. 
Here, we define masses denoted by $\tilde{m}$ as masses that 
include strong isospin corrections, but no QED effects and 
\begin{equation}
 \tilde{m} \equiv m_{m_u=m_d} + \delta_{\textrm{s}} m\,,
\end{equation}
where $m_{m_u=m_d}$ is the isospin symmetric mass and $\delta_{\textrm{s}} m$ 
the strong 
isospin correction.
\par
We can obtain $\tilde{m}_{K^0}-\tilde{m}_{K^+}$ by simply calculating the 
effective mass according to equation~\eqref{eq:cosheffmass} once for a 
two-point 
function using a strange quark and a light quark with mass $m_u$ and once for a 
two-point function using a strange quark and a light quark with mass $m_d$ and 
taking 
their difference. \par
On the other hand, we obtain $\tilde{m}_{K^0}-\tilde{m}_{K^+}$ from 
the expansion of the path integral. According to equation 
\eqref{eq:mexpansion2} we find for the two-point correlation functions of 
charged and neutral kaon
\begin{equation}
 \tilde{C}_{K^0}(z_0) = \tilde{C}_{K^+}(z_0) - (m_d-m_u) 
C_{K}^\textrm{strongIB}(z_0) + 
\mathcal{O}\left((m_d-m_u)^2\right)\,,
\end{equation}
where $C_{K}^\textrm{strongIB}(z_0)$ is given by \eqref{eq:corrdiffKK}
\begin{equation}
 C^\textrm{strongIB}_{K}(z_0) =  
\sum\limits_{\vec{z}}\!\sum\limits_{x}\!\textrm{Tr}\!\left[S^s(0,
z)\,\gamma_5\,S^{u}(z,x)\,S^{u}(x,0)\,\gamma_5\right]\,.
\label{eq:corrdiffKK2}
\end{equation}
Since by expanding the path integral we only determine the strong isospin 
breaking correction which is linear in $\left(m_d-m_u\right)$, the difference $ 
\tilde{m}_{K^0}-\tilde{m}_{K^+}$ has to be extracted using
\begin{equation}
 \tilde{m}_{K^0}-\tilde{m}_{K^+} = - (m_d-m_u)\left(
\frac{C_{K}^\textrm{strongIB}(t)}{C_{K^+}(t)} - 
\frac{C_{K}^\textrm{strongIB}(t+1)}{C_{K^+}(t+1)}\right)\,,
\label{eq:ratiostrongIB}
\end{equation}
i.e.\ with the ratio method, as for the $\Oalpha$ corrections when using the 
perturbative method for QED (cf. section \ref{subsec:extractionofmasses}).
\par
\begin{figure}[h]
 \centering
 \includegraphics[width=0.48\textwidth]{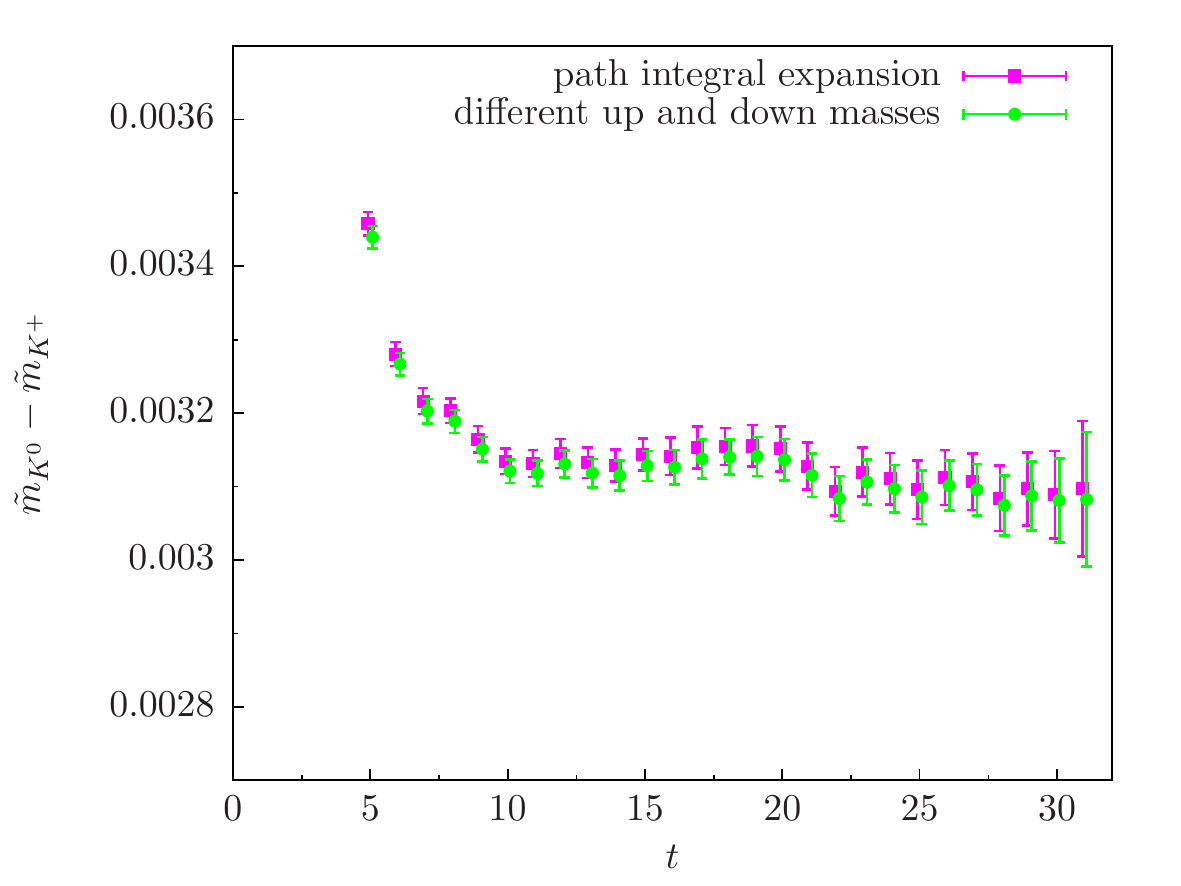}
 \caption{The strong isospin breaking contribution to the difference of the 
charged 
and neutral kaon masses. Green circle points are results 
using different bare 
quark masses for up and down quark, purple square points are results using the 
expansion of the path integral. Both data sets use the same $m_d-m_u$. 
}
\label{fig:KstrongIB}
\end{figure}

Figure \ref{fig:KstrongIB} shows the strong isospin correction to the 
difference of effective masses between 
charged and neutral kaon. The green circles show results using different 
up- and down-quark masses. Here we take the difference between the cosh 
effective mass of a charged and a neutral kaon. The purple square points show 
results using the path integral expansion and equation \eqref{eq:ratiostrongIB}.
\par
The strong isospin breaking contribution to the difference of the masses of 
neutral and charged kaons is determined by fitting a constant function to the 
plateau region of the data shown in figure \ref{fig:KstrongIB}. 
From these fits we obtain
\begin{align}
 \tilde{m}_{K^0}-\tilde{m}_{K^+} & = (5.551 \pm 
0.031)~\textrm{MeV}&\hspace{1cm}&\textrm{(different up and down masses)} \\
 \tilde{m}_{K^0}-\tilde{m}_{K^+} & = 
(5.575\pm0.033)~\textrm{MeV}&\hspace{1cm}&\textrm{(path integral expansion)}
\end{align}
for the data using different up- and down-quark masses and the path 
integral expansion, respectively. We find the statistical errors to be 
approximately the same for both methods to account for strong isospin breaking. 
Both methods have the same computational cost, since they require either one 
additional inversion with a second light quark mass or one additional inversion 
with a sequential insertion of the scalar current.
\par
The authors of \cite{deDivitiis:2013xla,deDivitiis:2011eh} showed, that the 
strong 
isospin breaking correction to the mass difference between a charged and a 
neutral pion vanishes at $\mathcal{O}(m_d-m_u)$, since the correlation 
functions $\tilde{C}_\pi^+ = \left<\pi^-\pi^+\right>$ and $\tilde{C}_\pi^0 = 
\left<\pi^0\pi^0\right>$ with $\pi^+ = u\gamma_5\overline{d}$ and 
$\pi^0=\frac{1}{\sqrt{2}}(u\gamma_5\overline{u}-d\gamma_5\overline{d})$ receive 
the same leading strong isospin correction 
\begin{equation}
 \tilde{C}_{\pi^+}(z_0) = \tilde{C}_{\pi^0}(z_0) = \tilde{C}_{\overline{u} 
u}(z_0) -
(m_d-m_u)\,C^\textrm{strongIB}_{\pi}(z_0)\,,
\end{equation}
with 
\begin{equation}
 C^\textrm{strongIB}_{\pi}(z_0) =  
\sum\limits_{\vec{z}}\!\sum\limits_{x}\!\textrm{Tr}\!\left[S^u(0,
z)\,\gamma_5\,S^{u}(z,x)\,S^{u}(x,0)\,\gamma_5\right]\,.
\label{eq:corrdiffpi2}
\end{equation}
However, the strong isospin breaking correction to the masses of neutral and 
charged pions differ at $\mathcal{O}\left((m_d-m_u)^2\right)$. When 
calculating pion correlation functions using different input bare masses 
for up and down quarks, we determine strong isospin correction at all orders in 
$(m_d-m_u)$. Using results from this approach we find a non-vanishing, albeit 
small mass difference  
\begin{equation}
 \tilde{m}_{\pi^+}-\tilde{m}_{\pi^0} = (0.1160 \pm 
0.0012)~\textrm{MeV}\,\hspace{1cm}\textrm{(different up and down masses)}.
\end{equation}

%% file: 4_hvp.tex
\section{Isospin Breaking Corrections to $a_\mu$}
\label{sec:hvp}
In the following, we determine the isospin breaking corrections to the 
anomalous magnetic moment of the muon $a_\mu$. In section \ref{subsec:amuintro} 
we give an introduction to $a_\mu$ and specify the setup we use to calculate 
the hadronic vacuum polarization. Results for the QED correction to the vector 
two-point function and the multiplicative renormalization $Z_V$ are given in 
sections \ref{subsec:QEDHVP} and \ref{subsec:dZV}, respectively. In section 
\ref{subsec:amusIB} we show results for the strong isospin breaking correction 
to $a_\mu$. 
Our results for the isospin breaking corrections to $a_\mu$ are 
summarized in section \ref{subsec:summaryHVP}.
\subsection{Introduction and Definitions}
\label{subsec:amuintro}
The anomalous magnetic moment of the muon $a_\mu$ has been experimentally 
measured with a precision of $\approx 0.5$~ppm at Brookhaven National 
Laboratory \cite{Bennett:2006fi} using polarized muons in a 
storage ring in a magnetic field. The Standard Model estimate 
\cite{Olive:2016xmw} of $a_\mu$ has been determined to the same level of 
accuracy. 
Thus, $a_\mu$ can serve as a high precision test of the Standard 
Model of particle physics. However, since many years a deviation of about 
$3\sigma$ persists between the experimental and theoretical estimates. This 
deviation might be a sign of new physics. Clearly, it is important to reduce 
the errors in both the experimental measurement and in the Standard Model 
calculation. From the 
experimental side, there are two upcoming experiments at Fermilab 
\cite{Venanzoni:2014ixa} and J-PARC \cite{Otani:2015jra}, both aiming to 
further reduce the experimental uncertainty. While the biggest contribution in 
the Standard Model estimate originates from the electromagnetic interaction, 
the largest contribution to the error comes from the strong interaction. The 
leading strong contribution to $a_\mu$ is given by the hadronic vacuum 
polarization (HVP). 
\par
Currently, the most precise theoretical estimate 
\cite{Davier:2010nc,Hagiwara:2011af} of the 
hadronic vacuum polarization uses the data from the cross section of 
$e^+e^-\rightarrow$ hadrons, and, thus, relies on experimental data. 
On the other hand, the HVP can be calculated from 
first principles using lattice QCD.
The hadronic 
vacuum polarization $\Pi(Q^2)$ is determined by the correlation function of two 
electromagnetic currents
\begin{equation}
\Pi_{\mu\nu}(Q^2) = \sum_x e^{iQ\cdot x}\left<j_\mu(x) j_\nu(0)\right> =  
(Q_\mu Q_\nu - Q^2\delta_{\mu\nu})\Pi(Q^2)\,,  
\end{equation}
with
\begin{equation}
 j_\mu = \frac{2}{3} \overline{u}\gamma_\mu u - \frac{1}{3} 
\overline{d}\gamma_\mu d - \frac{1}{3} 
\overline{s}\gamma_\mu s + \cdots\,.
\end{equation}
From the HVP form factor $\Pi(Q^2)$, the leading hadronic contribution to 
$a_\mu$ can be determined by \cite{Blum:2002ii}
\begin{equation}
a_\mu^{\textnormal{HVP}} = \left(\frac{\alpha}{\pi}\right)^2  
\int\limits_0^\infty \textnormal{d} Q^2\, 
K(Q^2)\,\left[\Pi(Q^2)-\Pi(0)\right]\,,%\hat{\Pi}(Q^2)
\label{eq:amu}
\end{equation}
with a kernel function $K(Q^2)$, which is known analytically.
\par
In recent years a lot of effort has been undertaken to determine the HVP 
contribution to $a_\mu$ using lattice calculations 
(see e.g\ 
\cite{Boyle:2011hu,DellaMorte:2011aa,Burger:2013jya,
Chakraborty:2014mwa, Bali:2015msa,Chakraborty:2016mwy, 
Borsanyi:2016lpl,DellaMorte:2017dyu}). 
However, to be competitive with the determination from $e^+e^-\rightarrow$ 
hadrons, a precision of $\lesssim1\%$ is required. At this level of 
precision, isospin breaking corrections can no longer be neglected. 
\par
In this work we achieve the first exploratory calculation of isospin breaking 
corrections to the HVP, using a setup identical to the one previously described 
for the meson mass splittings.
Note, that the QED corrections to 
$a_\mu$ are of 
the same order in $\alpha$ as the hadronic light-by-light scattering 
contribution (see 
\cite{Blum:2015gfa,Blum:2016lnc,Green:2015sra,Asmussen:2016lse} for lattice 
calculations of the hadronic light-by-light scattering).
\par
For the calculation of the HVP we choose a setup with a local vector current at 
the source and a conserved vector current at the sink
  \begin{equation}
 C_{\mu\nu}(x) = Z_V\,q_f^2 \left< V^c_{\mu}(x) V^\ell_{\nu}(0)\right>\,;
 \label{eq:Cmunu_corr}
\end{equation}
see \cite{Blum:2016xpd} for further details of the framework for our 
calculation of the 
hadronic vacuum polarization.
The conserved vector current $V^c_{\mu}$ for the Domain Wall Fermion 
formulation 
used in this work is given in equation \eqref{eq:consvectorcurrent}. The local 
vector current $V^\ell_{\nu}$ requires a multiplicative renormalization $Z_V$.
From the correlation function \eqref{eq:Cmunu_corr} we construct the HVP tensor 
as (see e.g.\ \cite{Blum:2016xpd})
\begin{equation}
 \Pi_{\mu\nu}(Q) = \sum_x e^{-iQ\cdot x} C_{\mu\nu}(x) - \sum_x C_{\mu\nu}(x)\,.
 \label{eq:hvptensor}
\end{equation}
In \eqref{eq:hvptensor} we have subtracted the zero-mode $\sum_x C_{\mu\nu}(x)$ 
of the vector-vector correlation function 
\cite{Bernecker:2011gh}, 
which vanishes in the infinite volume limit. In \cite{Blum:2016xpd} the 
authors showed that the zero-mode subtraction greatly reduces the statistical 
error on the HVP for low $Q^2$ when using $\mathbb{Z}_2$ Wall sources for the 
quark propagators. We also find such an improvement for the QED correction 
to the HVP, reducing the error for the smallest $Q^2$ by a factor of $\approx 
4$ for the up quark and $\approx 20$ for the strange quark.
\par
For the determination of the HVP form factor $\Pi(\hat{Q}^2)$ we use the spatial 
components of~\eqref{eq:hvptensor}
\begin{equation}
\Pi(\hat{Q}^2)  = \frac{1}{3} \sum_j \frac{\Pi_{jj}(Q)}{\hat{Q}^2}\,,
\end{equation}
with vanishing spatial momentum $\vec{Q}=0$.
\par
The QED correction to the hadronic vacuum polarization $\delta \Pi$ is 
determined by the QED correction to the correlation function $C_{\mu\nu}(x)$ 
\eqref{eq:Cmunu_corr}. Since we use the local vector current at the source, 
we have to apply the appropriate multiplicative renormalization $Z_V$. This 
multiplicative renormalization $Z_V$ itself receives a QED correction once 
electromagnetism is switched on. Thus, the QED correction to the 
local-conserved vector two-point function $C_{\mu\nu}(x)$ is 
given by
\begin{equation}
 \delta C_{\mu\nu}(x) = \delta Z_V\, q_f^2 \left< V^c_{\mu}(x) 
V^\ell_{\nu}(0)\right>_0 + Z^0_V q_f^2 \,\,\delta\left< V^c_{\mu}(x) 
V^\ell_{\nu}(0)\right>\,,
\label{eq:dCmunu}
\end{equation}
where $\left<V^c_{\mu}(x) 
V^\ell_{\nu}(0)\right>_0$ is the vector two-point function without QED, $Z_V^0$ 
the 
multiplicative renormalization without QED, $\delta Z_V$ the QED 
correction to $Z_V$ and $\delta\left< V^c_{\mu}(x) 
V^\ell_{\nu}(0)\right>$ the QED correction to the vector two-point function. 
It follows from equation \eqref{eq:dCmunu} that the QED correction to the HVP 
is given by
\begin{equation}
 \delta \Pi(\hat{Q}^2) = \delta^{Z_V}\Pi(\hat{Q}^2) + 
\delta^{V}\Pi(\hat{Q}^2)\,,
\end{equation}
where $\delta^{Z_V}\Pi(\hat{Q}^2)$ and $\delta^{V}\Pi(\hat{Q}^2)$ are the QED 
corrections 
from the correction to $Z_V$ and from the correction to the vector 
two-point function $\left< V^c_{\mu}(x)V^\ell_{\nu}(0)\right>$, respectively.
Similarly, we define the QED correction to $a_\mu$ as
\begin{equation}
 \delta a_\mu = \delta^{Z_V}\!a_\mu + 
\delta^{V}\!a_\mu\,.
\end{equation}
In general, $\delta a_\mu$ also receives a contribution from the 
QED correction to the lattice spacing. The lattice spacing enters in the kernel 
function $K(Q^2)$ in equation \eqref{eq:amu}, which depends on the muon mass. 
However, in this work we did not determine the lattice spacing in the presence 
of QED (cf. section \ref{sec:setup}).
\par
Our results for the QED correction $\delta^{V}\!a_\mu$ from the 
vector 
two-point function are presented in section \ref{subsec:QEDHVP} and results for 
the 
QED correction $\delta^{Z_V}\!a_\mu$ from the multiplicative 
renormalization are given in section \ref{subsec:dZV}.\par
\subsection{QED Correction to the Vector Two-Point Function}
\label{subsec:QEDHVP}
In this section we discuss the QED correction $\delta\left< 
V^c_{\mu}(x)V^\ell_{\nu}(0)\right>$ to the vector two-point function.
%\par
As described above, we use a conserved vector current $V^c_{\mu}$ at the sink 
when calculating the hadronic vacuum polarization. However, the conserved 
current depends on the link variables 
$U_\mu(x)$ (cf. equation \eqref{eq:consvectorcurrent}) and thus, in the 
presence of QED, 
on the photon fields and the 
electromagnetic charge $e$. In the following, we refer to the conserved vector 
current including the $U(1)$ photon fields as $V_\mu^{c,e}(x)$ to indicate the 
dependence on $e$. Thus, for the QED correction to the HVP we have to calculate 
the expectation value of an operator that itself depends on the electromagnetic 
coupling. This has to be taken into account when expanding the path integral 
for the perturbative method
\begin{equation}
 \left<V^{c,e}_{\mu}(x) V^\ell_{\nu}(0)\right> = \left<V^{c}_{\mu}(x) 
V^\ell_{\nu}(0)\right>_{0} + 
\frac{1}{2}\,e^2\left.\frac{\partial^2}{\partial 
e^2}\left<V^{c,e}_{\mu}(x) V^\ell_{\nu}(0)\right>\right|_{e=0} + 
\Oalphasquare\,\,. 
\label{eq:hvpeexpansion}
\end{equation}
This leads to two additional terms, that are not present in the QED 
correction to the meson masses. The corresponding diagrams are shown in figure 
\ref{fig:qeddiagrams_hvp}. A detailed derivation can be found in the appendix 
\ref{subsec:hvpexpansion}. The construction of these two terms does 
not require any additional inversions compared to the diagrams which we have
already considered for the meson masses.
\par
\begin{figure}[h!]
\centering
\includegraphics[width=0.7\textwidth]{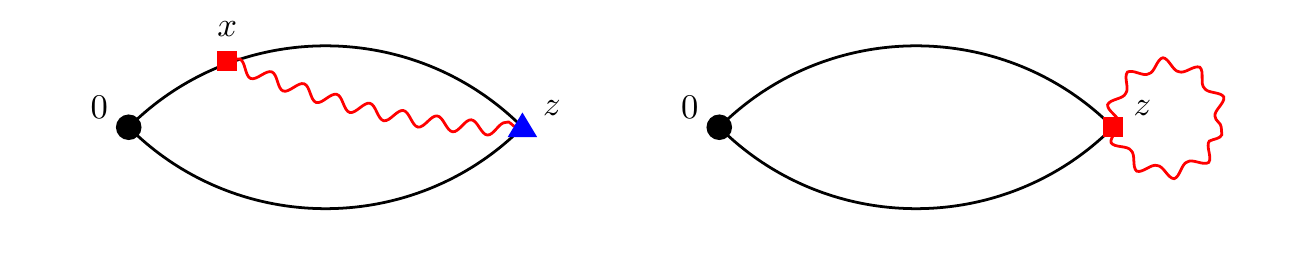}
 \caption{The two terms from the expansion of the conserved current at 
the sink. Red squared vertices and blue triangle vertices refer to insertions 
of the conserved vector current and the tadpole operator, respectively.}
 \label{fig:qeddiagrams_hvp} 
\end{figure}
\subsubsection{Results}
\label{subsubsec:hvpresults}
\begin{figure}[h]
 \centering
 \includegraphics[width=0.48\textwidth]{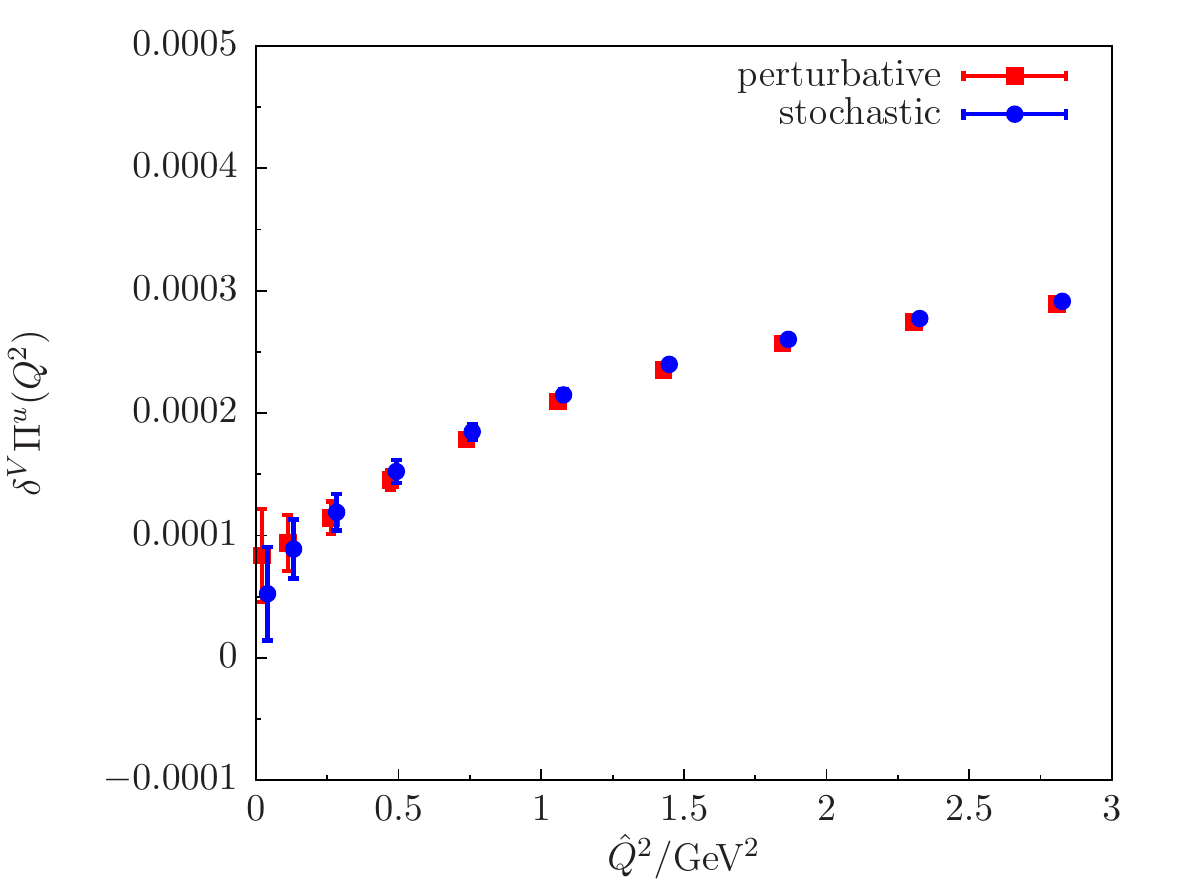}
\includegraphics[width=0.48\textwidth]{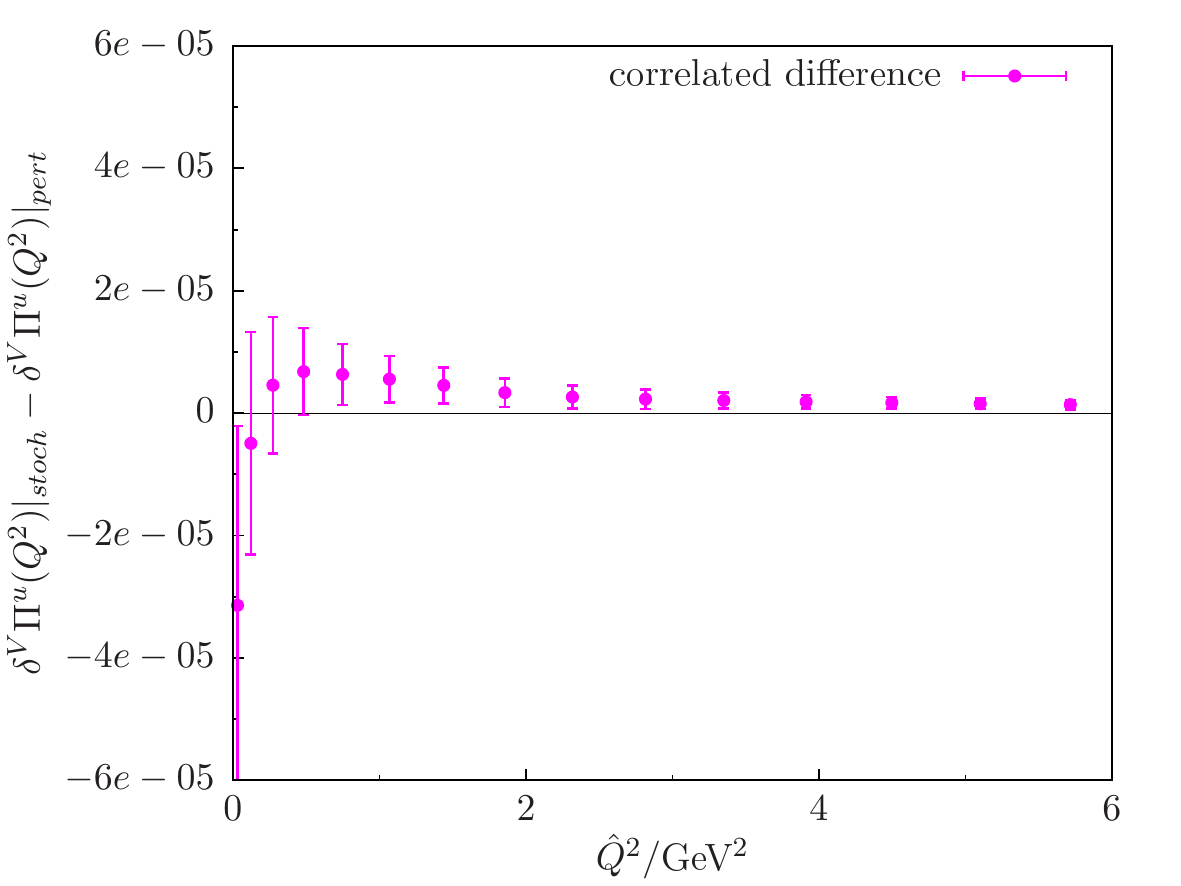}
\caption{The QED correction $\delta^V\Pi(\hat{Q}^2)$ to the HVP form factor for 
the up quark. The plot 
on the left shows results from the stochastic method (blue circles) and the 
perturbative method (red squares). The plot on the left shows the correlated 
difference between stochastic and perturbative data.}
\label{fig:qedhvp_u}
\end{figure}

\begin{figure}[h]
 \centering
 \includegraphics[width=0.48\textwidth]{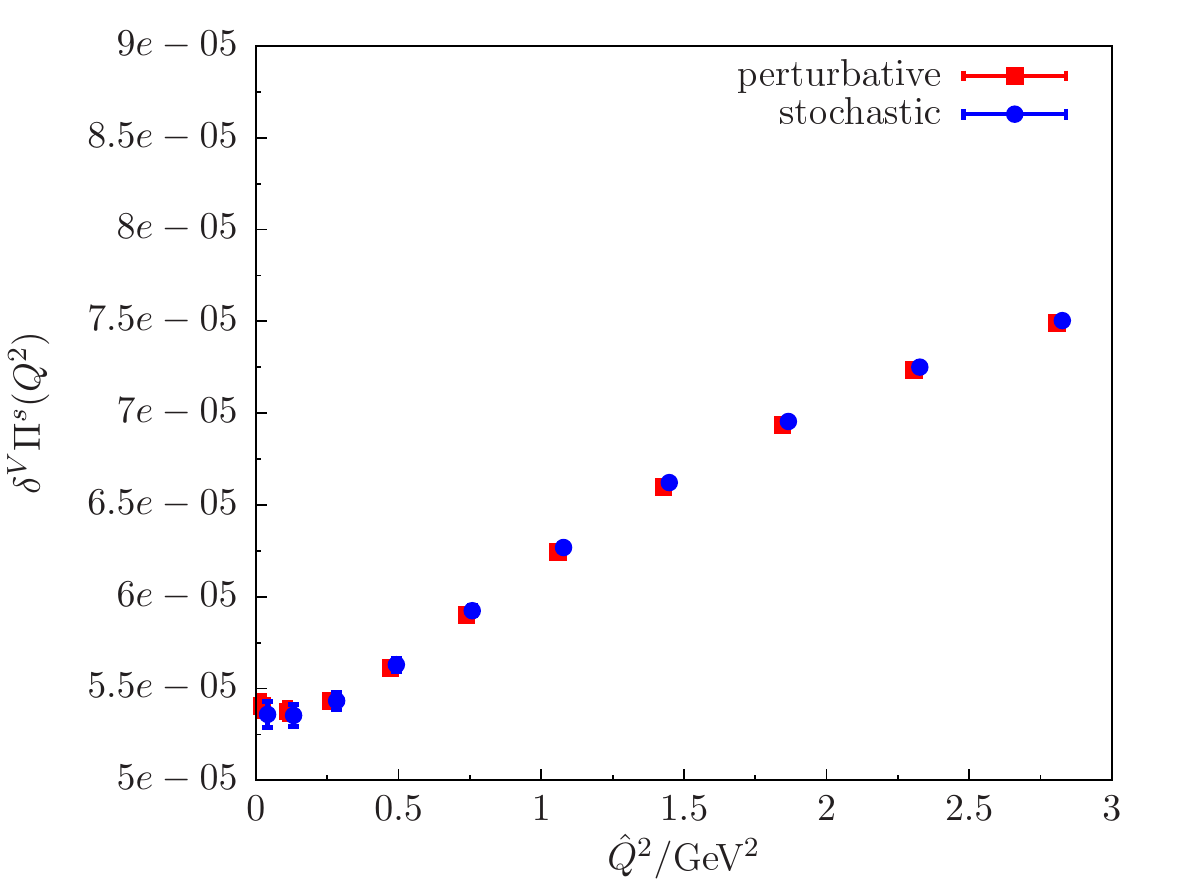}
\includegraphics[width=0.48\textwidth]{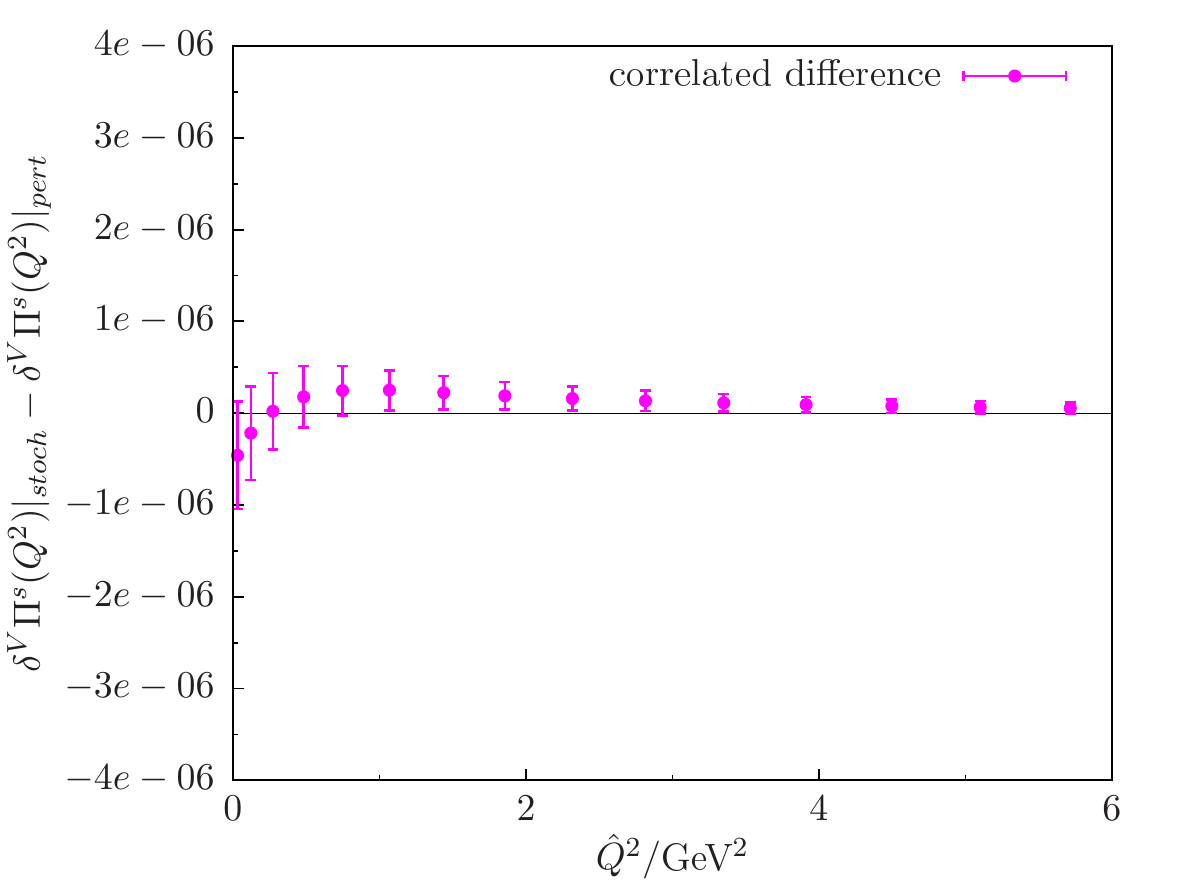}
\caption{The same as figure \ref{fig:qedhvp_u} for the strange quark}
\label{fig:qedhvp_s}
\end{figure}

Figures \ref{fig:qedhvp_u} and \ref{fig:qedhvp_s} show the QED correction 
$\delta^{V}\Pi(\hat{Q}^2)$ to 
the hadronic vacuum polarization form factor for up and strange quarks, 
respectively. The plots on the left-hand side of both figures show results from 
the perturbative and the stochastic method. 
For the perturbative data the results shown have been calculated using the 
single-$\mu$ insertion, which, for the same amount of statistics, gives a 
smaller statistical error than the summed-$\mu$ insertion (see section 
\ref{subsubsec:amucomerror} for a detailed comparision of statistical 
errors).
For the 
multiplicative renormalization $Z_V^0$ of the local vector current we use a 
value  
determined from the ratio of the local-conserved and the local-local vector 
two-point functions. Further details can be found in section 
\ref{subsec:dZV}, where we will also determine the QED correction to $Z_V$.
\par
\par
The plots on the right-hand side of figures \ref{fig:qedhvp_u} and 
\ref{fig:qedhvp_s} show the correlated difference between the data from 
perturbative and stochastic methods. We find the data from both methods differs 
for a large range of $Q^2$ at the level of about $1-1.5~\sigma$. 
In order to understand this small difference we perform 
a computation with a second value of the electromagnetic coupling 
$\alpha=\nicefrac{1}{4\pi}$ for the stochastic method, 
so that we can distinguish between the
leading 
and higher-order QED correction in the data from the stochastic method. We find 
that the deviation 
seen between stochastic and perturbative data for large $\hat{Q}^2$ to be 
consistent with 
$\Oalphasquare$ corrections. More details on the $\Oalphasquare$ QED 
corrections from the stochastic data are given in appendix 
\ref{sec:Oalphasquare}.
\par
The calculation of $a_\mu$ from the 
HVP form factor requires the subtracted HVP $\hat{\Pi}(\hat{Q}^2) 
= \Pi(\hat{Q}^2) -\Pi(0)$. For the QED correction we therefore need to 
determine $\delta^V\Pi(0)$.\par
The subtracted hadronic vacuum polarization $\hat{\Pi}(\hat{Q}^2)$ can be 
directly determined from the vector two-point function by 
\cite{Bernecker:2011gh}
\begin{equation}
 \hat{\Pi}(\hat{Q}^2) = \Pi(\hat{Q}^2) -\Pi(0) = 2\sum_x\, 
C_{jj}(x)\,\left[\frac{x_0^2}{2}-\frac{1-\cos(Qx_0)}{Q^2}\right]\,.
\label{eq:hvprenorm}
\end{equation}
Similarly, we calculate the QED correction to the subtracted HVP from the QED 
correction to the vector two-point function
\begin{equation}
 \delta^V \hat{\Pi}(\hat{Q}^2) = 2\sum_x\, 
\delta^V C_{jj}(x)\,\left[\frac{x_0^2}{2}-\frac{1-\cos(Qx_0)}{Q^2}\right]\,.
\label{eq:qedhvprenorm}
\end{equation}
The results for $\delta^V\hat{\Pi}$ can then be used to calculate the QED 
correction to $a_\mu$ according to equation \eqref{eq:amu}. We 
use a sine cardinal interpolation \cite{Blum:2016xpd} to obtain the HVP also 
at non-lattice momenta
\begin{equation}
 Q_0 = \frac{2\pi}{T} n_0\,,
\end{equation}
where $n_0$ can lie anywhere in $[-T/2,T/2)$ and not only on integer values. 
To obtain $a_\mu$ we integrate using the trapezoidal rule up to momenta 
$\hat{Q}^2\approx3$~GeV$^2$. 
The integrand in the integral to 
obtain $a_\mu$ (cf. equation \eqref{eq:amu}) is peaked at small momenta 
around the muon mass. Contributions from momenta $>3$~GeV$^2$ are very small, 
and we neglect these in this study.
The results for the QED corrections to $a_\mu$ from the QED 
correction to the vector two-point function are given in table 
\ref{tab:qedcorrhvp} alongside results without QED.
\par
\begin{table}[h]
\centering
\begin{tabular}{|c|c|c|c|c|}
\hline
 & $a^0_\mu\times10^{10}$ & $\delta^V\!a^\textrm{stoch}_\mu\times10^{10}$ & 
$\delta^V\!a^\textrm{pert}_\mu\times10^{10}$ & 
$\delta^V\!a_\mu^\textrm{stoch}-\delta^V\!a_\mu^\textrm{pert}$\\
\hline
$u$ & $310\pm18$ & $2.6\pm1.2$ & $0.7\pm1.2$ & $1.95\pm0.94$\\
$s$ & $48.49\pm0.23$ & $-0.0030\pm0.0014$ & $-0.0057\pm0.0014$ & 
$0.0027\pm0.0011$\\
\hline
\end{tabular}
\caption{The HVP contribution to $a_\mu$ without QED and the QED corrections 
$\delta^V\!a_\mu$ 
from stochastic and perturbative data at an isospin symmetric pion mass of 
$340$~MeV. Results have been obtained using 
equation \eqref{eq:hvprenorm}. The last column 
$\delta^V\!a^\textrm{stoch}-\delta^V\!a^\textrm{pert}$ shows the correlated 
difference between the results from both data sets. }
\label{tab:qedcorrhvp}
 \end{table}
We find the QED correction $\delta^V\!a_\mu$ for the up quark to 
be of 
the order of $\lesssim1\%$ of the value without QED. 
Results for the 
down quark can be obtained by multiplying the values for the up quark with the 
appropriate charge factor $1/4$ for $a_\mu^0$ and $1/16$ for $\delta^Va_\mu$. 
In contrast to the QED correction for the light quarks, we find the QED 
correction for the strange quark contribution to be negative. 
Although we find agreement between the HVP form factor from the 
perturbative and stochastic data (cf. figures \ref{fig:qedhvp_u} 
and \ref{fig:qedhvp_s}), we find the results for the QED correction 
$\delta^Va_\mu$ given in table \ref{tab:qedcorrhvp} to differ between the 
stochastic and perturbative approach by $2-3\sigma$. This is due to the 
$1-2\sigma$ deviation between both datasets for small $\hat{Q}^2$ in the HVP 
form factor. When calculating the subtracted HVP using equation 
\eqref{eq:hvprenorm} this difference gets enhanced over the whole $Q^2$ region 
for $\hat{\Pi}(\hat{Q}^2)$. 
\par
\begin{figure}[h]
\centering
\includegraphics[width=0.48\textwidth]{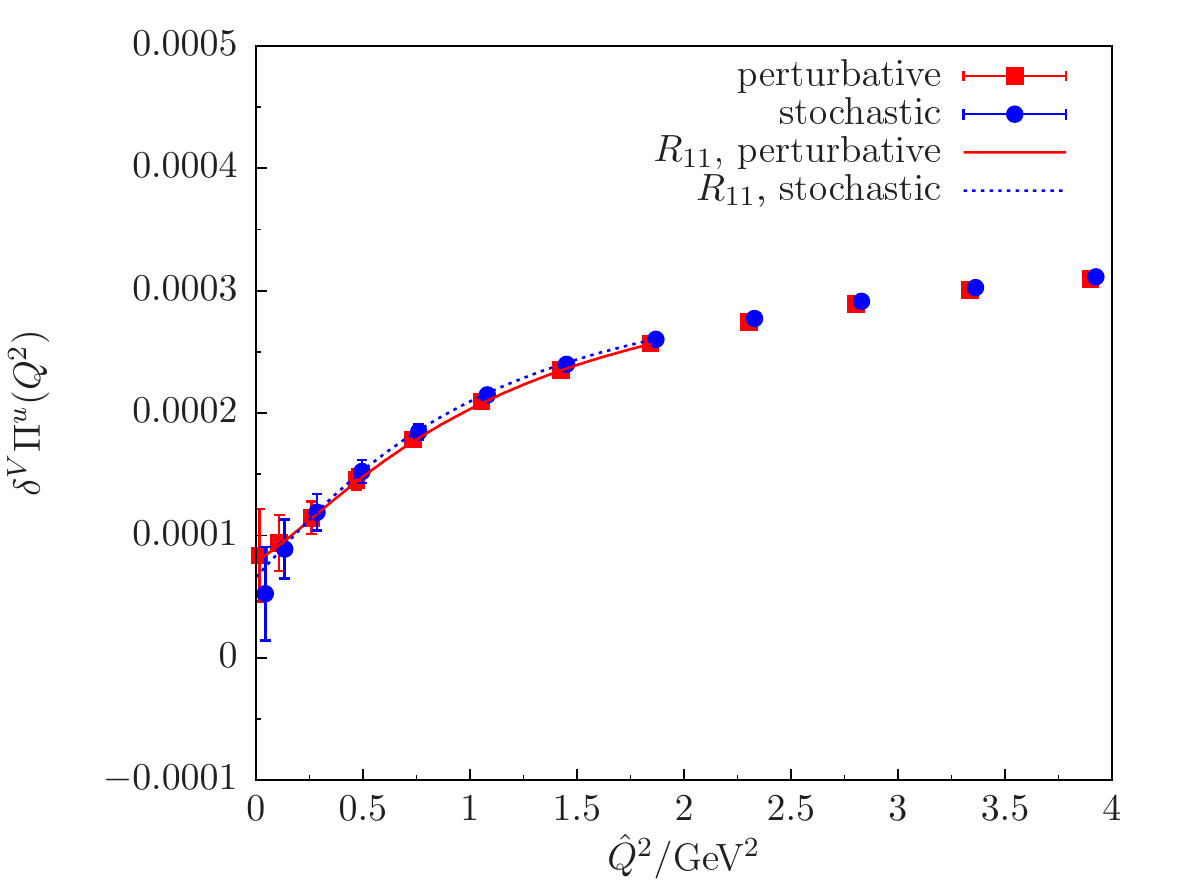}
\includegraphics[width=0.48\textwidth]{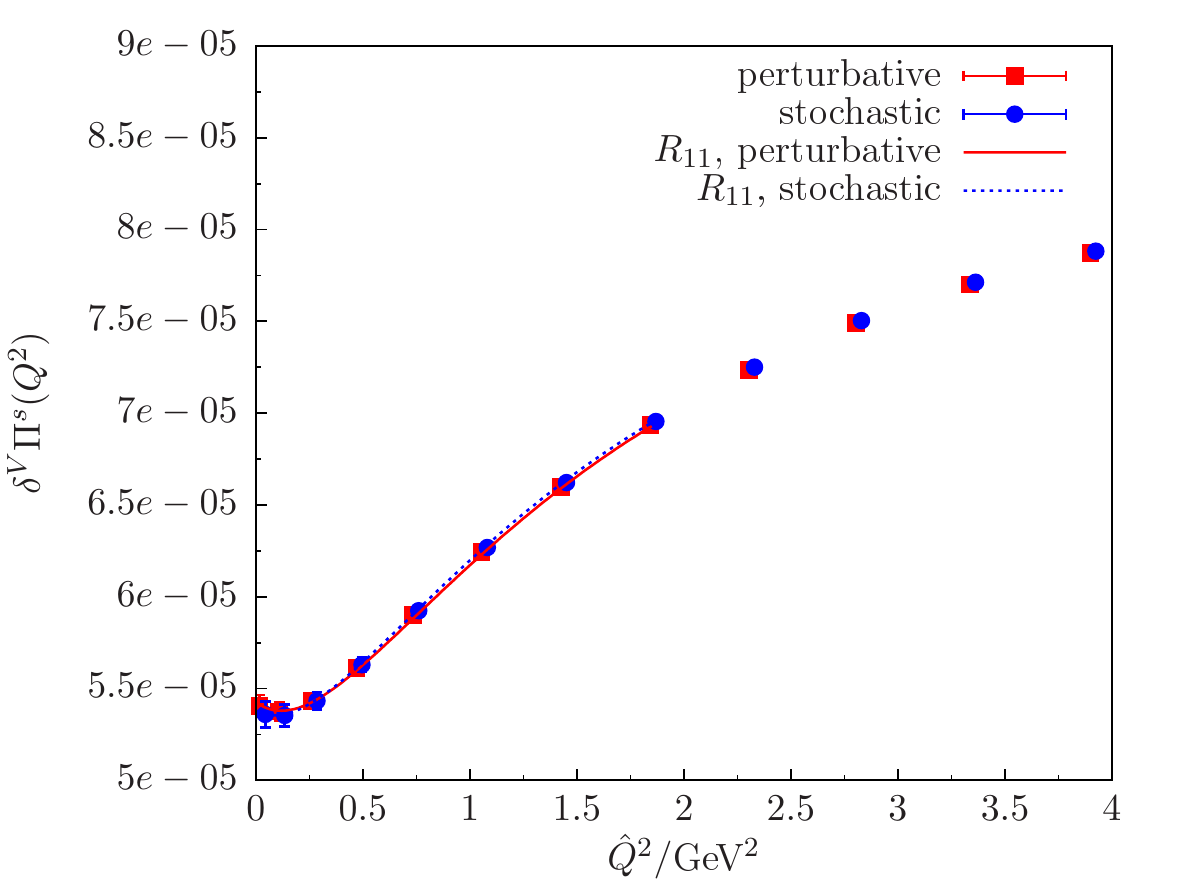}
\caption{The QED correction $\delta^V \Pi(\hat{Q}^2)$ to the HVP for the up 
quark (left) and the strange 
quark (right). The blue dashed line and the red solid line shows results of a 
Pad\'e fit of the form \eqref{eq:Pade11} to stochastic and perturbative data, 
respectively.}
\label{fig:R11}
\end{figure}

Another method to determine $\Pi(0)$ is to fit the 
HVP form factor to extrapolate to $Q^2=0$. Suitable fit functions are given by
Pad\'e approximants~\cite{Aubin:2012me}
\begin{equation}
 R_{mn}(\hat Q^2) = \Pi_0 + 
\hat{Q}^2\left(\sum\limits_{i=0}^{n-1}\,\frac{a_i}{b_i+\hat{Q}^2}+\delta_{
mn}\,c\right)\, \hspace{1cm}\textrm{with}\,\,\,\,n=m, m+1.
\end{equation}
In this work we use Pad\'e $R_{11}$, which has one pole
  \begin{equation}
 R_{11}(\hat{Q}^2) = \Pi_0 + 
\hat{Q}^2\left(\frac{a}{b+\hat{Q}^2}+c\right)\,.
\label{eq:pade11_0}
\end{equation}
To obtain a fit function for the QED correction from \eqref{eq:pade11_0}, we 
allow each parameter to receive a QED correction
\begin{align}
 R_{11}(\hat{Q}^2) &= R^0_{11}(\hat{Q}^2) + \delta^V\! R_{11}(\hat{Q}^2) = 
\Pi^0_0 + 
\delta^V \Pi_0+ 
\hat{Q}^2\left(\frac{a^0+\delta^V\!a}{b^0+\delta^V b+\hat{Q}^2} + 
c^0+\delta^V\!c\right) \\
& =\Pi^0_0 + \hat{Q}^2\left(\frac{a^0}{b^0+\hat{Q}^2}+ c^0\right) + 
\underbrace{\delta^V \Pi_0 + 
\hat{Q}^2\left(\frac{1}{b^0+\hat{Q}^2}\left[\delta^V\! 
a-\frac{\delta^V b\cdot a^0}{b^0+\hat{Q}^2}
\right ] + \delta^V\! c\right)}_{=\delta^V\! R_{11}}\,.\label{eq:Pade11}
\end{align}
 At $\Oalpha$ we find
 \begin{equation}
  \delta^V R_{11}(\hat{Q}^2) = \delta^V \Pi_0 + 
\hat{Q}^2\left(\frac{1}{b^0+\hat{Q}^2}\left[\delta^V\! 
a-\frac{\delta^V b\cdot a^0}{b^0+\hat{Q}^2}
\right ] + \delta^V\! c\right)\,,\label{eq:qedpade11}
 \end{equation}
as an ansatz for fitting the QED correction to the HVP. Since $\delta^V 
R_{11}(Q^2)$ also depends on the parameters $a^0$ and $b^0$ from the Pad\'e 
without QED, we perform a combined fit of the HVP without QED and the QED 
correction $\delta^V \Pi(\hat{Q}^2)$. Results of these fits are shown in figure 
\ref{fig:R11} for the QED 
correction to the HVP for the up and the strange quark. The dashed blue curve 
shows the fit result for the stochastic data (blue circles), the solid red 
curve shows the fit result for the perturbative data (red squares). 
\par
For the calculation of $a_\mu$ according to equation \eqref{eq:amu} we use the 
fit result of the Pad\'e in the fit range, which is indicated by the range in 
which the Pad\'e function is plotted in figure~\ref{fig:R11}. For higher $Q^2$ 
we use the data and trapezoidal rule for the integration. As before, we 
integrate up to $\approx 3$~GeV$^2$.
\par
The results for $a_\mu$ without QED as well as the QED 
corrections from perturbative and stochastic data using the Pad\'e $R_{11}$ are 
given in table \ref{tab:qedcorrhvp_pade}. We find the results for the QED 
correction to $a_\mu$ for the up quark to be smaller than the values in 
table \ref{tab:qedcorrhvp} determined using equation \eqref{eq:hvprenorm}. 
For the strange quark we again find a negative QED correction to $a_\mu$, 
which is in agreement with the results given in table \ref{tab:qedcorrhvp}. 
We find the results from the perturbative and the stochastic data 
to differ by $1-2\sigma$. This difference is not 
as pronounced as when using equation \eqref{eq:hvprenorm} (cf. results in 
table \ref{tab:qedcorrhvp}), since the deviation between both data sets at 
small 
$\hat{Q}^2$ is reduced by the Pad\'e fit as one can see on figure 
\ref{fig:R11}.
\par
 \begin{table}[h]
\centering
\begin{tabular}{|c|c|c|c|c|}
\hline
 & $a^0_\mu\times10^{10}$ & $\delta^V a^\textrm{stoch}_\mu\times10^{10}$ & 
$\delta^V 
a^\textrm{pert}_\mu\times10^{10}$ & 
$\delta^V\!a_\mu^\textrm{stoch}
-\delta^V\!a_\mu^\textrm{pert}$\\
\hline\hline
$u$ & $318\pm11$ & $0.65\pm0.31$ & $0.37\pm0.33$ & 
$0.27\pm0.26$\\
$s$ & $47.98\pm0.25$ & $-0.0030\pm0.0012$ & $-0.0049\pm0.0011$ & 
$0.0019\pm0.0010$\\
\hline
\end{tabular}
\caption{The HVP contribution to $a_\mu$ without QED and the QED corrections 
from stochastic and perturbative data at an isospin symmetric pion mass of 
$340$~MeV. Results have been obtained using 
Pad\'e $R_{11}$.}
\label{tab:qedcorrhvp_pade}
 \end{table}
\par 
The difference of the results for $\delta^V\!a_\mu$ using the two different 
methods to determine $\hat{\Pi}(\hat{Q}^2)$ discussed 
here is shown in table \ref{tab:diffBMPade} (i.e.\ the difference between 
results from tables \ref{tab:qedcorrhvp} and \ref{tab:qedcorrhvp_pade}). 
This difference mainly arises from the large statistical errors on the 
QED correction for small $\hat{Q}^2$. A better resolution of the QED correction 
in this region would allow for a more reliable determination of $\Pi(0)$ and 
thus $\hat{\Pi}(\hat{Q}^2)$. 
\par
\begin{table}[h]
\centering
\begin{tabular}{|c|c|c|}
\hline
& stoch $\times 10^{10}$ & pert $\times 10^{10}$\\
\hline\hline
$u$ & $2.0\pm1.1$ & $0.3\pm1.0$\\
$s$ & $0.00003\pm0.00063$ & $-0.00080\pm0.00058$ \\
\hline
\end{tabular}
 \caption{Correlated difference of results obtained from using equation 
\eqref{eq:hvprenorm} or Pad\'e $R_{11}$ for determining $\hat{\Pi}(\hat{Q}^2)$}
\label{tab:diffBMPade}
\end{table}
\par
\begin{figure}[h]
  \centering
  \includegraphics[width=0.4\textwidth]{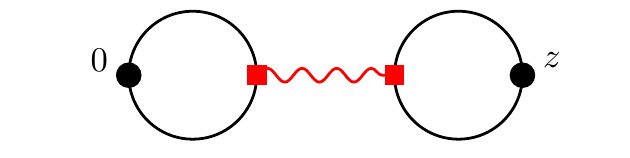}
  \caption{Quark-disconnected diagram for the QED correction to $a_\mu$.}
  \label{fig:disc}
\end{figure}

Currently we do not include any quark-disconnected diagrams in the calculation 
of QED corrections. However, 
one type of quark-disconnected diagrams is also present in the electro-quenched 
approximation. A sketch of this diagram is shown in figure   
\ref{fig:disc}. Note, that in this context one is only interested in the case, 
where the two quark lines are additionally connected by gluons. If the two quark 
lines are only connected by the photon and not by gluons this diagram is 
conventionally counted as a higher order HVP contribution (see e.g.\ 
\cite{Jegerlehner:2009ry}), not as a QED correction to the leading order HVP.
We include this diagram neither in the stochastic nor in the perturbative data.
\par
The counterpart of this diagram without QED, i.e.\ without a photon coupling 
the two quark loops, is $SU(3)$ suppressed (see 
\cite{Francis:2014hoa,Chakraborty:2015ugp,Blum:2015you} for lattice QCD 
calculations of the quark-disconnected contribution to the HVP and 
\cite{DellaMorte:2010aq,Bijnens:2016ndo} for estimates of the disconnected HVP 
in chiral perturbation theory). However, when
including a photon to obtain the QED correction shown in figure 
\ref{fig:disc} the corresponding correlation function is no longer $SU(3)$ 
suppressed. Thus, this quark-disconnected diagram might give a large QED 
correction to the HVP. We plan to include this contribution in future 
calculations. 
Note, that the diagram shown in figure \ref{fig:disc} 
determines the mixing of  $\rho$- and $\omega$-mesons. 
\subsubsection{Comparison of Statistical Errors}
\label{subsubsec:amucomerror}
\begin{figure}[h!]
\centering
 \includegraphics[width=0.48\textwidth]{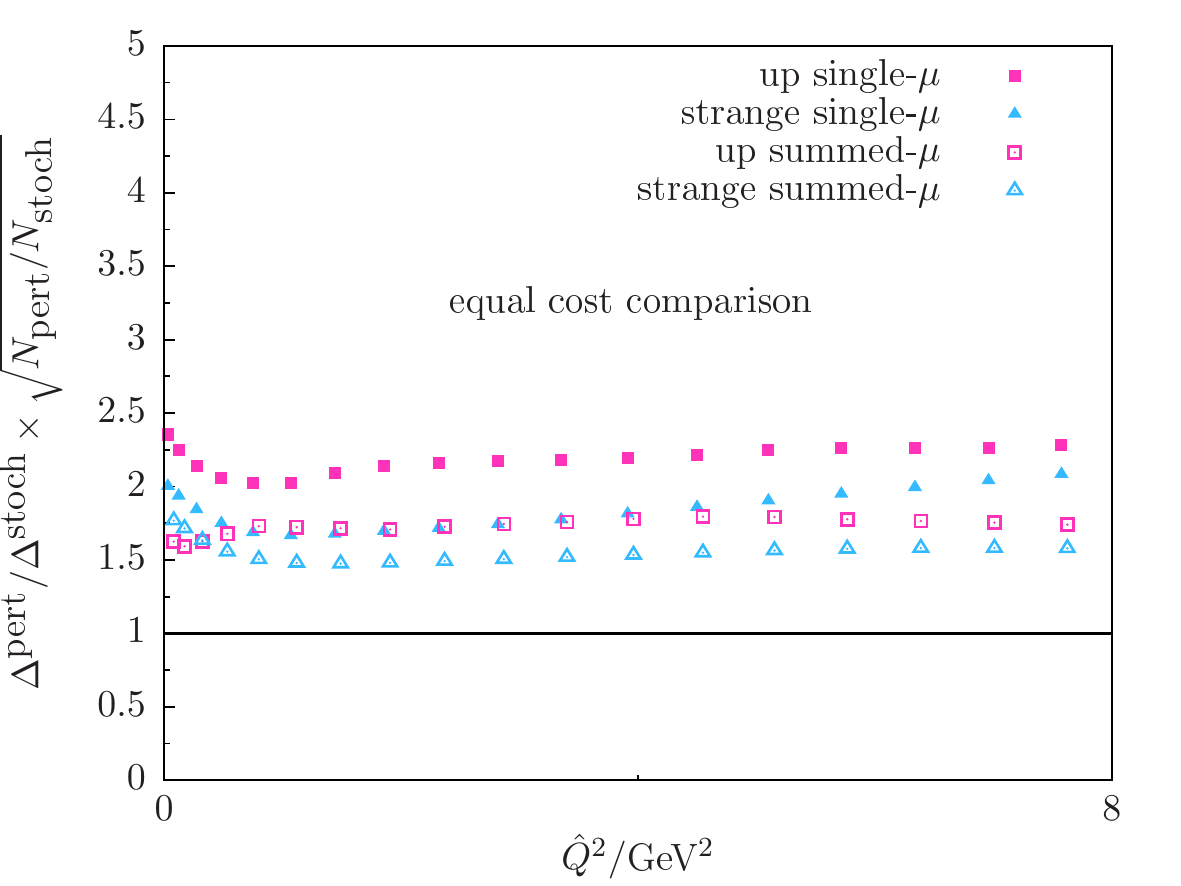}
\includegraphics[width=0.48\textwidth]{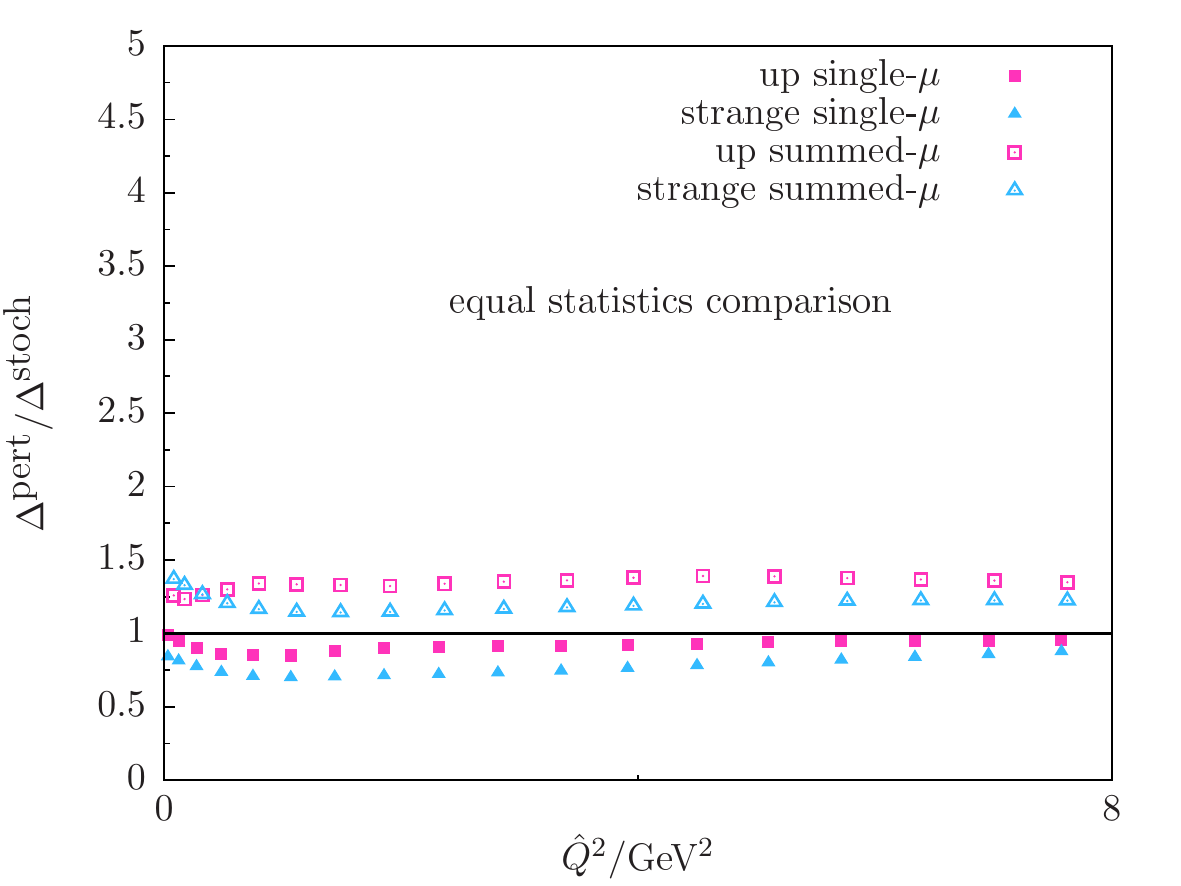}
\caption{Ratio of statistical errors of perturbative and stochastic data. The 
plot on the left shows an equal cost comparison, the plot on the right shows 
an equal statistics comparison. Closed and open symbols refer to 
the single- or summed-$\mu$ insertion technique for the perturbative method, 
respectively. Purple squares show results for the up quark, blue triangles 
results for the strange quark.}
\label{fig:staterrorhvp}
\end{figure}

To compare the statistical errors between the perturbative and the stochastic 
data, we consider their ratio.
Figure 
\ref{fig:staterrorhvp} shows the error on the perturbative data divided by 
the error on the stochastic data. For the plot on the left-hand side, the 
errors are scaled by the total number of inversions used in each case to obtain 
an equal cost comparison. 
Closed and open symbols denote results from the the single- or summed-$\mu$ 
insertion technique for the perturbative method, respectively.
The horizontal black line shows ``1'' where both methods 
would give the same precision with the same numerical cost. However, we 
find the statistical error from the perturbative method to be larger then the 
error from the stochastic method. Comparing the two different approaches for 
calculating the sequential 
propagators for the perturbative method, we find 
that at the same numerical cost the statistical error is smaller when using the 
summed-$\mu$ insertion (open symbols). \par
The plot on the right-hand 
side of figure \ref{fig:staterrorhvp} shows the ratio of errors in an equal 
statistics comparison. We find this ratio to be smaller but close to one for the 
single-$\mu$ 
insertion and slightly larger then one for the summed-$\mu$ insertion. 
\par 
We find the same ordering of statistical errors as for the QED correction 
to meson masses
\begin{align}
& \Delta^\textnormal{stoch}<\Delta^{\textnormal{pert,summed-}\mu}
<\Delta^{\textnormal{pert,single-}\mu}& \hspace{1cm}&\textnormal{same cost}\\
&\Delta^{\textnormal{pert,single-}\mu}<\Delta^\textnormal{stoch}
<\Delta^{\textnormal{pert,summed-}\mu}&\hspace{1cm}&\textnormal{same 
statistics.}
\end{align}
One has to keep in mind, that our study is done using unphysical quark masses 
and this might be a mass dependent finding. Indeed we observe a trend in an 
increasing ratio of errors from the perturbative over the stochastic method as 
the quark mass is decreased, suggesting that this ratio might even be larger 
for physical quark 
masses.

\subsection{QED Correction to $Z_V$}
\label{subsec:dZV}
The calculation of the HVP using a local current at the source (cf. equation 
\eqref{eq:Cmunu_corr}) requires the determination of the appropriate 
multiplicative renormalization $Z_V$. When including QED in the lattice 
calculation also $Z_V$ obtains an electromagnetic correction
\begin{equation}
 Z_V = Z_V^0 + \delta Z_V\,.
\end{equation}
This results in a further correction
$\delta^{Z_V}\Pi(\hat{Q}^2)$  to the HVP at $\Oalpha$. In this work, we 
determine the multiplicative renormalization from local-conserved and 
local-local vector two-point functions. We define
\begin{equation}
 C^{lc}_0(t) = 
\frac{1}{3}\sum\limits_{\mu=1}^3\sum\limits_{\vec{x}}
\left<V_\mu^c(x)V_\mu^l(0)\right>_0
\hspace{0.5cm}\textrm{and}\hspace{0.5cm}
C^{ll}_0(t) = 
\frac{1}{3}\sum\limits_{\mu=1}^3\sum\limits_{\vec{x}}
\left<V_\mu^l(x)V_\mu^l(0)\right>_0\,,
\end{equation}
as the local-conserved and local-local vector two-point functions without QED 
and 
\begin{equation}
 C^{lc}(t) = 
\frac{1}{3}\sum\limits_{\mu=1}^3\sum\limits_{\vec{x}}
\left<V_\mu^{c,e}(x)V_\mu^l(0)\right>
\hspace{0.5cm}\textrm{and}\hspace{0.5cm}
C^{ll}(t) = 
\frac{1}{3}\sum\limits_{\mu=1}^3\sum\limits_{\vec{x}}
\left<V_\mu^l(x)V_\mu^l(0)\right>\,,
\end{equation}
as the local-conserved and local-local vector two-point functions with QED. \par
\begin{figure}[h]
\centering
 \includegraphics[width=0.48\textwidth]{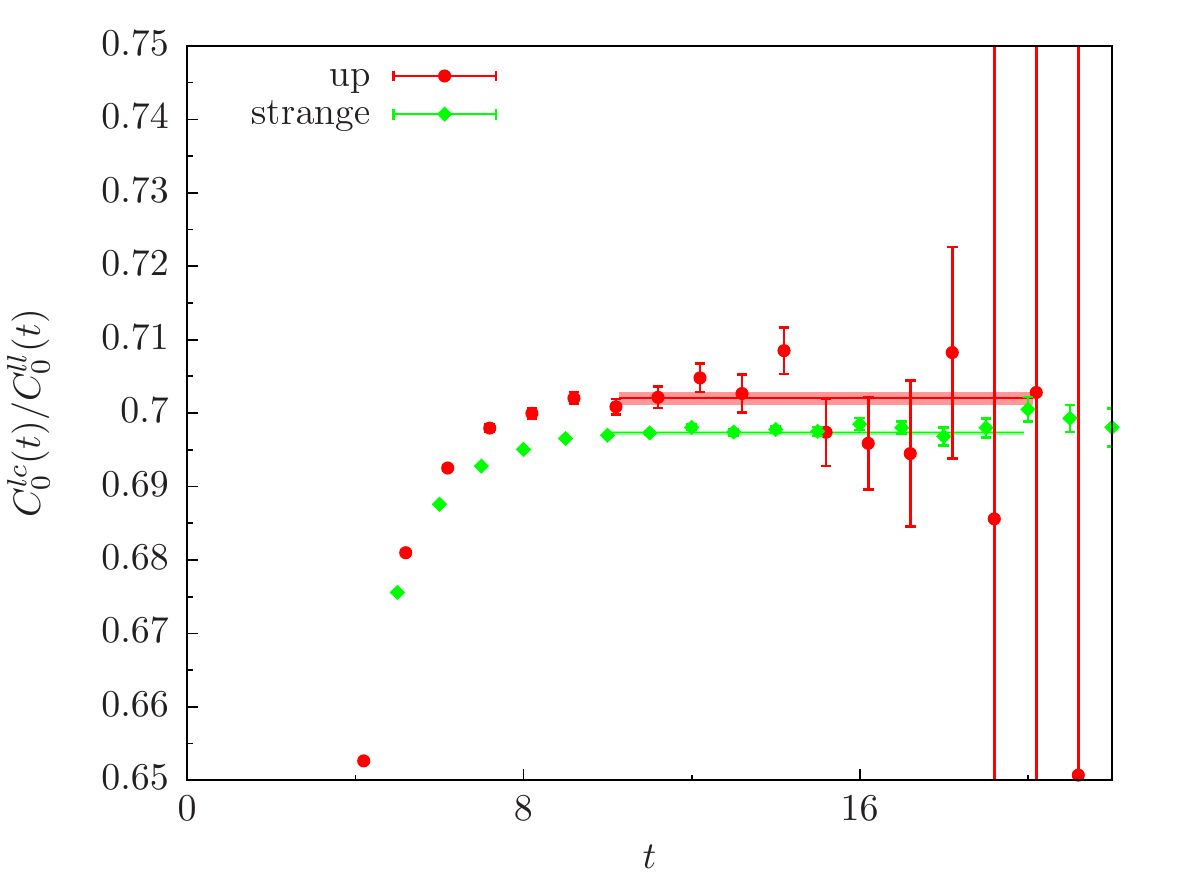}
\caption{The ratio of the local-conserved and local-local vector two-point 
function without QED for up (red circles) and strange 
(green diamonds).}
\label{fig:Zv0}
\end{figure}

The renormalization of the vector current without QED can be determined from 
the large time behaviour of the ratio of the local-conserved and local-local 
vector two-point functions
\begin{equation}
 Z_V^0 = \frac{ C^{lc}_0(t)}{C^{ll}_0(t)}\,.
\end{equation}
This ratio is shown in figure \ref{fig:Zv0} for the up and strange quark.
$Z_V^0$ is determined by fitting a constant to the plateau region in the data 
as indicated in figure \ref{fig:Zv0}. The results of these fits are given in 
table \ref{tab:resultsZV}.
\par
\begin{figure}[h]
 \centering
 \includegraphics[width=0.48\textwidth]{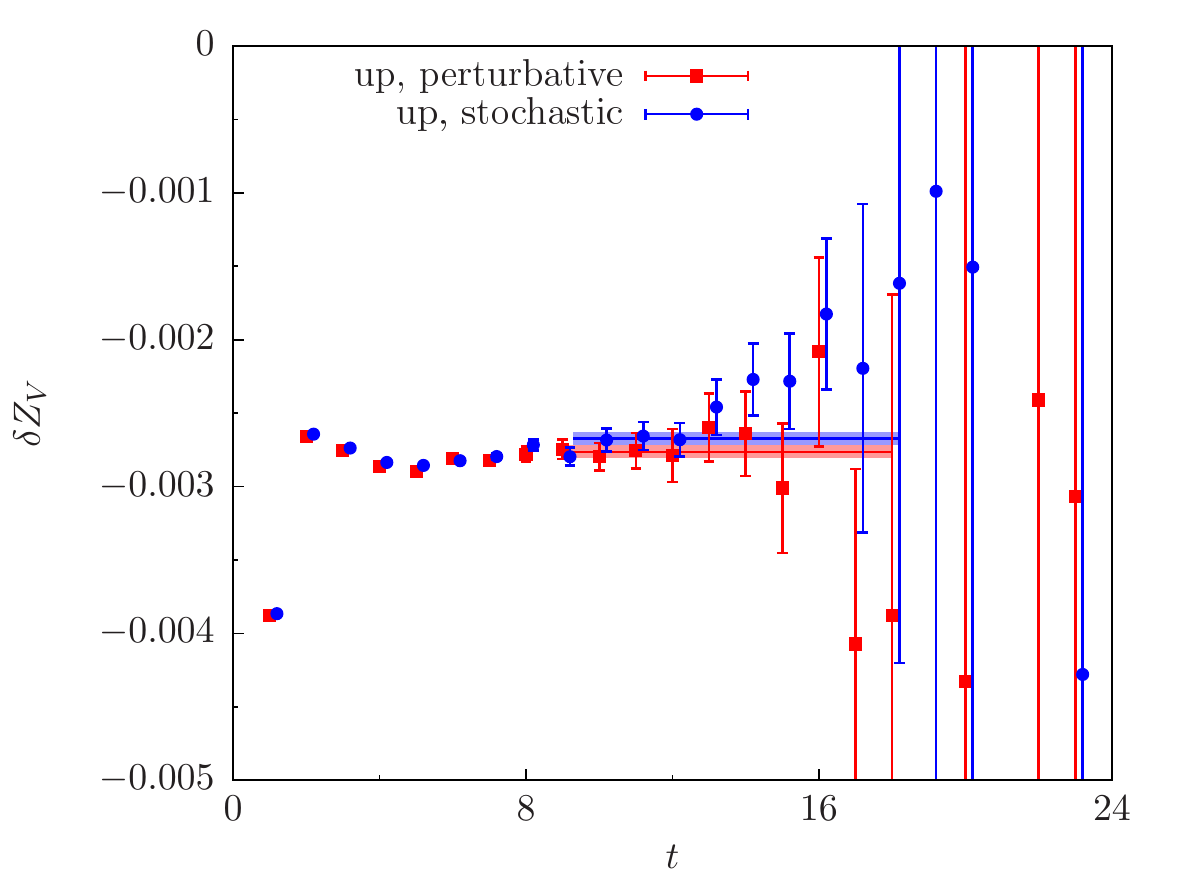}
\includegraphics[width=0.48\textwidth]{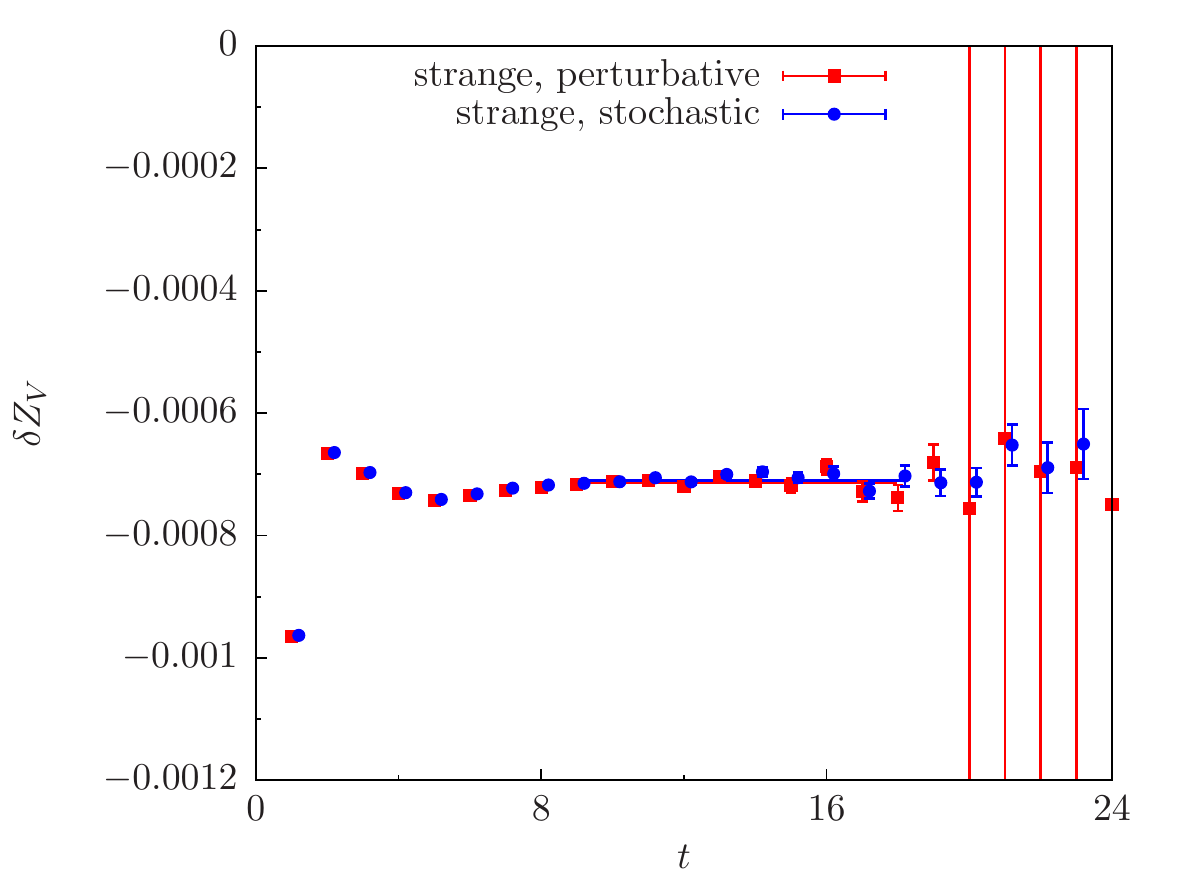}
\caption{The QED correction to $Z_V$. The plot on the left shows the 
results of equation \eqref{eq:dZv} for the up quark, the plot on the right for 
the strange quark. Red squares and blue circles denote data from the 
perturbative and the stochastic method, respectively.}
\label{fig:dZV}
 \end{figure}

The multiplicative renormalization of the vector current including QED is given 
by
\begin{equation}
 Z_V   = \frac{C^{lc}(t)}{C^{ll}(t)} = \frac{C^{lc}_0(t)+\delta 
C^{lc}(t)}{C^{ll}_0(t)+\delta 
C^{ll}(t)} = \frac{C^{lc}_0(t)}{C^{ll}_0(t)}+ 
\underbrace{\left(\frac{\delta 
C^{lc}(t)}{C^{ll}_0(t)}-\frac{C^{lc}_0(t)}{C^{ll}_0(t)}\frac
{\delta C^ { ll }(t) } { C^{ll}_0(t)}\right)}_{=\delta Z_V} + 
\Oalphasquare\label{eq:Zv0+1}\,,
\end{equation}
with the QED corrections to the local-conserved $\delta C^{lc}(t)$ and the 
local-local $\delta C^{ll}(t)$ vector two-point function. Equation 
\eqref{eq:Zv0+1} implies that the QED correction to $Z_V$ can be determined by 
\begin{equation}
 \delta Z_V = \frac{\delta 
C^{lc}(t)}{C^{ll}_0(t)}-\frac{C^{lc}_0(t)}{C^{ll}_0(t)}\frac
{\delta C^ { ll }(t) } { C^{ll}_0(t)}\,.
\label{eq:dZv}
\end{equation}
The results for $\delta Z_V$ using equation \eqref{eq:dZv} are shown in figure 
\ref{fig:dZV}. The plot on the left shows data for the up quark, the plot on 
the right data for the strange quark. A constant has been fitted to the plateau 
region of the data to obtain $\delta Z_V$. The results from these fits are 
given 
in table \ref{tab:resultsZV} alongside the results for $Z_V^0$. We find the QED 
correction to $Z_V$ to be negative and smaller than $0.5\%$ for the up quark 
and even smaller for the strange quark, where the QED correction is more 
suppressed due to the smaller charge factor.
\par
\par
\begin{table}[h]
\centering
\begin{tabular}{|c|c|c|c|}
 \hline
 & $Z_V^0$ & $\delta Z_V^\textrm{stoch}$ & $\delta Z_V^\textrm{pert}$ \\
 \hline\hline
 up & $0.70209\pm0.00083$  & $-0.002674\pm0.000043$& $-0.002756\pm0.000044$\\
 strange & $0.69737\pm0.00017$  & 
$-0.0007102\pm0.0000016$& $-0.0007139\pm0.0000016$\\
  \hline
\end{tabular}
\caption{Results for the multiplicative renormalization of the vector current 
without QED $Z_V^0$ and the QED correction $\delta Z_V$ from the perturbative 
and the stochastic data.} 
\label{tab:resultsZV}
\end{table}
In table \ref{tab:QEDcorrdueZV} we give results for the additional QED 
correction to $a_\mu$ due to the QED correction to $Z_V$. For the up
quark we find the correction to $a_\mu$ from $\delta Z_V$ to be of the same 
order but with a different sign than $\delta^V\!a_\mu$, the correction from the 
QED correction to the 
vector two-point function itself (cf. results in tables \ref{tab:qedcorrhvp} 
and 
\ref{tab:qedcorrhvp_pade}). For the strange quark both QED corrections have the 
same sign, but the QED correction to $a_\mu$ from $\delta Z_V$ is about an
order of magnitude bigger.
\par
\begin{table}[h]
\centering
\begin{tabular}{|c|c|c|}
\hline
 & $\delta^{Z_V}\! a^\textrm{stoch}_\mu \times10^{10}$ & $\delta^{Z_V}\! 
a^\textrm{pert}_\mu \times10^{10}$ 
\\
\hline\hline
up & $-1.212\pm0.052$ & $-1.249\pm0.047$\\
strange & $-0.04886\pm0.00028$& 
$-0.04911\pm0.00027$\\
\hline
\end{tabular}
\caption{The QED correction to $a_\mu$ due to the QED correction 
to the multiplicative renormalization $Z_V$ for the local-vector current.}
\label{tab:QEDcorrdueZV}
\end{table}
\subsection{Strong Isospin Breaking Correction}
\label{subsec:amusIB}
\begin{figure}[h]
 \centering
 \includegraphics[width=0.48\textwidth]{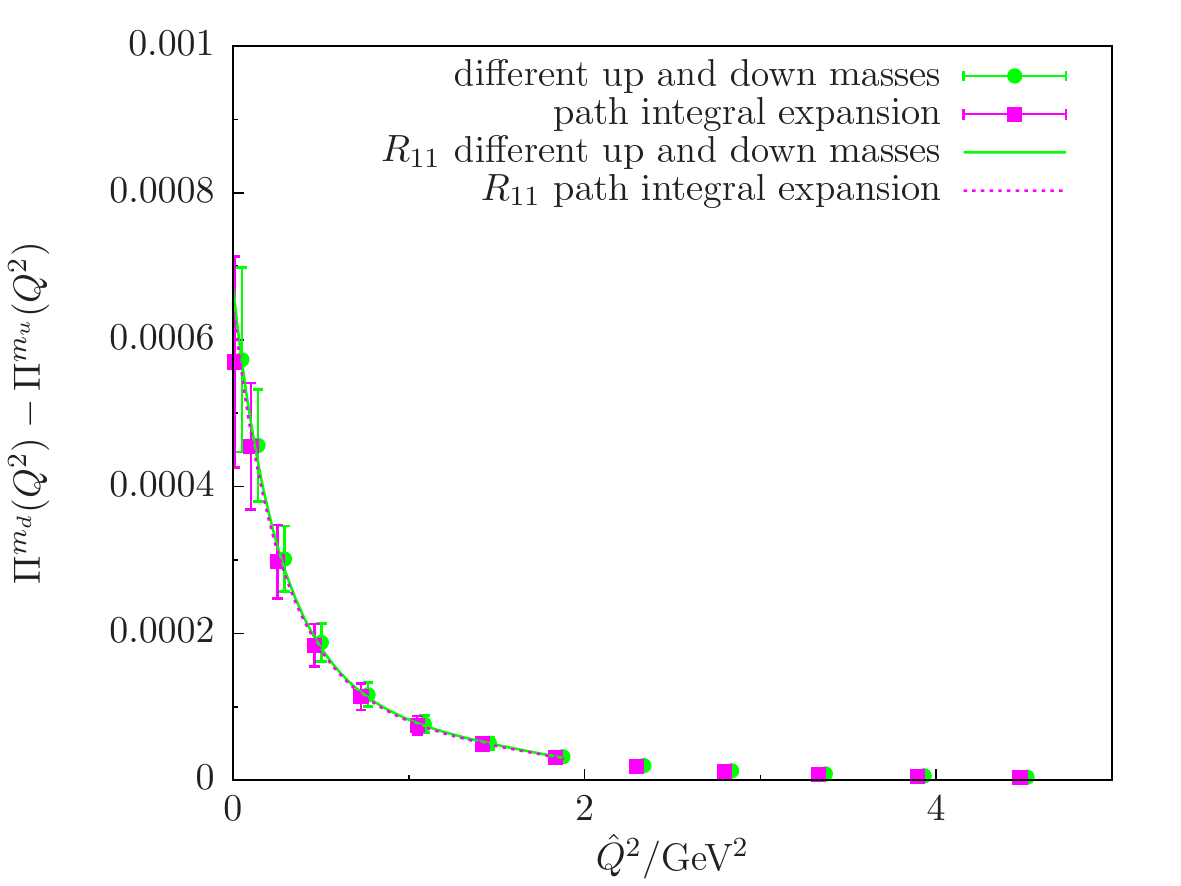}
 \caption{Strong isospin correction to the HVP form factor. Green circles show 
results obtained using different masses for up and down quark. Purple squares 
show results using the path integral expansion. The solid green line and dashed 
purple line show results from a Pad\'e fit.}
\label{fig:HVP_strongIB}
\end{figure}

To determine the strong isospin breaking corrections to $a_\mu$ we will in the 
following look at the HVP for the down quark $\Pi^d(\hat{Q}^2)/q_f^2$, which is 
determined by the correlation function (cf. equation \eqref{eq:Cmunu_corr})
  \begin{equation}
 C_{\mu\nu}(x)/q_f^2 = Z_V\, \left< V^c_{\mu}(x) V^\ell_{\nu}(0)\right>\,.
\end{equation}
We compute the strong isospin correction to $\Pi^d(\hat{Q}^2)/q_f^2$ by 
either using the difference of the HVP calculated with different masses for up 
and down quarks, $\Pi^{m_d}(\hat{Q}^2) - \Pi^{m_u}(\hat{Q}^2)$, or by 
using the expansion of the path integral. In the latter, the strong isospin 
breaking correction is given by
\begin{equation}
 \delta_\textrm{s} C_{\mu\nu}(x)/q_f^2 = -Z_V\, (m_d-m_u)\left< 
V^c_{\mu}(x) V^\ell_{\nu}(0)\,\mathcal{S}\right>_{m_u=m_d}\,,
\label{eq:hvp_strongIB_expansion}
\end{equation}
with the scalar current $\mathcal{S}$.
In figure \ref{fig:HVP_strongIB} the strong isospin correction to the HVP is 
plotted against $\hat{Q}^2$. Green circles show the difference of the HVP 
calculated using different masses for up and down quark. The purple squares 
show results obtained from the expansion of the path integral in the quark 
mass, i.e.\ equation \eqref{eq:hvp_strongIB_expansion}. We find the data sets 
from both methods to account for strong isospin to agree with each other.
\par
To determine the strong isospin breaking correction $\delta_\textrm{s}a_\mu$ to 
the anomalous magnetic moment of the muon from the data shown in figure 
\ref{fig:HVP_strongIB}, we use either equation \eqref{eq:hvprenorm} to 
determine $\hat{\Pi}(\hat{Q}^2)$ or a Pad\'e fit.
The Pad\'e fit can be done in a similar way as for the 
QED correction, i.e.\ assigning a strong isospin breaking correction to each of 
the parameters in the Pad\'e function. Thus, for $R_{11}$ we obtain
 \begin{equation}
  \delta_\textrm{s} R_{11}(\hat{Q}^2) = \delta_\textrm{s} \Pi_0 + 
\hat{Q}^2\left(\frac{1}{b^0+\hat{Q}^2}\left[\delta_\textrm{s} 
a-\frac{\delta_\textrm{s} b\cdot a^0}{b^0+\hat{Q}^2}
\right ] + \delta_\textrm{s} c\right)\,,\label{eq:qedpade11_strong}
 \end{equation}
as an ansatz to fit the data for the strong isospin correction to the HVP. The 
results of these fits are shown in figure \ref{fig:HVP_strongIB} by the solid 
green line for the data using different masses for up and down quark and the 
dashed purple line for the data using the path integral expansion.
\par
The results for the strong isospin breaking correction to $a_\mu$ are given in 
table \ref{tab:strongIB_amu}. Using the Pad\'e fits we are able to resolve the 
strong isospin breaking correction to $a_\mu$. When comparing the results in 
table \ref{tab:strongIB_amu} with the value in the isospin symmetric limit (cf. 
table \ref{tab:qedcorrhvp_pade}) $a^u_\mu/q_u^2=(716\pm25)\times 10^{-10}$, we 
find that the strong isospin correction 
is $\delta_\textrm{s}a_\mu/a^u_\mu\approx0.9\%$. 
\par
\begin{table}[h]
\centering
\begin{tabular}{|c|c|c|}
 \hline
  & $\delta_\textrm{s}a_\mu/q_f^2$ using \eqref{eq:hvprenorm} & 
$\delta_\textrm{s}a_\mu/q_f^2$ using Pad\'e $R_{11}$\\
\hline\hline
different u and d masses & $(-6.1\pm 8.8)\times10^{-10}$ & $(-6.7\pm 
1.6)\times10^{-10}$ \\
path integral expansion & $(-7.2\pm 7.4)\times10^{-10}$& $(-6.4\pm 
1.7)\times10^{-10}$\\
\hline
\end{tabular}
\caption{Strong isospin breaking correction to $a_\mu$. The middle column shows 
results using equation \eqref{eq:hvprenorm} to obtain $\hat{\Pi}(\hat{Q}^2)$, 
the right column results using Pad\'e $R_{11}$.}
\label{tab:strongIB_amu}
\end{table}
In addition, we have to determine the strong isospin breaking correction to the 
multiplicative renormalization $Z_V$, which can be obtained by comparing 
results for $Z_V$ using either $m_u$ or $m_d$ as the valance quark mass. We 
find this correction to be very small $\delta_\textrm{s} Z_V / Z_V = 
(Z^{m_u}_V - Z^{m_d}_V)/ Z_V \approx 0.02\%$.
\subsection{Summary IB Corrections to $a_\mu$}
\label{subsec:summaryHVP}
Our results for the QED corrections to the anomalous magnetic 
moment of the muon are summarized in table~\ref{tab:QEDcorr}. 
The first column $a_\mu^0$ shows 
the result in the isospin symmetric limit for up and strange quarks (Note, that 
in the 
isospin symmetric limit, the contribution from the down quark is simply $1/4$ 
of the up quark). 
Results for the QED 
correction $\delta^Va_\mu$ from the vector two-point function are given in 
columns two to five and the QED correction $\delta^{Z_V}a_\mu$ from the 
multiplicative renormalization in columns six and seven. 
For the QED correction $\delta^Va_\mu$ we have determined the subtracted vacuum 
polarization $\hat{\Pi}(\hat{Q}^2)$ with two different methods, either using 
equation \eqref{eq:hvprenorm} or using Pad\'e $R_{11}$ to obtain $\Pi(0)$ and 
we quote both results separately. We find results from both techniques for 
determining $\hat{\Pi}(\hat{Q}^2)$ to differ especially for the up quark. 
This difference mainly arises from
the large statistical errors on the QED correction for small $\hat{Q}^2$. A 
reduction of the statistical error in the low $\hat{Q}^2$ region is required to 
achieve a more reliable determination of $\Pi(0)$.

\begin{table}[h]
\centering
\footnotesize
 \begin{tabular}{|c||c||c|c|c|c||c|c|}
 \hline
 & $a_\mu^0\times10^{10}$ & 
\multicolumn{4}{|c||}{$\delta^Va_\mu\times10^{10}$}& 
\multicolumn{2}{|c|}{$\delta^{Z_V}a_\mu\times10^{10}$}\\
 & & stoch, \eqref{eq:hvprenorm} & stoch, $R_{11}$ & pert, \eqref{eq:hvprenorm} 
& pert, $R_{11}$ & stoch & pert\\
\hline\hline
$u$ & $318(11)$ & $2.6(1.2)$ & $0.65(31)$ & $0.7(1.2)$ & 
$0.37(33)$&$-1.212(52)$&$-1.249(47)$\\
$s$ & $47.98(25)$ & $-0.0030(14)$ & $-0.0030(12)$ & 
$-0.0057(14)$ & $-0.0049(11)$&$-0.04886(28)$ & $-0.04911(27)$\\
\hline
\end{tabular}
\caption{Summary of our results for the QED correction to $a_\mu$ with an 
isospin symmetric pion mass of $\approx 340$~MeV. Results are 
shown for the stochastic and perturbative method.}
\label{tab:QEDcorr}
\end{table}
\par
The total QED correction to $a_\mu$ is given by the sum of the two 
contributions $\delta^Va_\mu$ and 
$\delta^{Z_V}a_\mu$. We have not added these 
contributions in 
order to illustrate, 
that the statistical error is dominated by the QED correction $\delta^Va_\mu$ 
that originates from the QED correction to the vector two-point function, 
while the QED correction to the multiplicative renormalization $Z_V$ is 
determined very precisely. 
\par
We find the overall QED correction to $a_\mu$ to be 
smaller than $1\%$ for the 
up quark, where the QED contribution is enhanced by the charge factor 
compared 
to the down and strange quark. For the strange quark we find the QED 
correction 
to be about $0.1\%$ of the isospin symmetric result. 
\par
The above findings are for unphysical sea and valence light quark masses and 
QED corrections to $a_\mu$ might be larger at the physical point.
\par
Our results for the strong isospin breaking correction are summarized in table 
\ref{tab:strongIBsummary}. We have accounted for strong isospin breaking by 
either using different masses for the valence up and down quark or by using a 
path integral expansion in $(m_u-m_d)$. For both datasets we have determined 
the 
subtracted vacuum 
polarization $\hat{\Pi}(\hat{Q}^2)$ with two different methods, either using 
equation \eqref{eq:hvprenorm} or using Pad\'e $R_{11}$ to obtain $\Pi(0)$ and 
both results are quoted separately in table \ref{tab:strongIBsummary}. We find 
the strong isospin correction to be $0.9\%$ of the isospin symmetric result.
\par
\begin{table}[h]
\centering
\footnotesize
 \begin{tabular}{|c|c|c|c|c|}
 \hline
 $a_\mu/q_f^2\times10^{10}$ & 
\multicolumn{4}{|c|}{$\delta_s a_\mu/q_f^2\times10^{10}$}\\
 & diff masses, eq. \eqref{eq:hvprenorm} & diff masses, $R_{11}$ & expansion, 
eq. \eqref{eq:hvprenorm} & expansion, $R_{11}$ \\
\hline\hline
$716\pm25$ & $-6.1\pm8.8$ &$-6.7\pm1.6$ & $-7.2\pm7.4$& 
$-6.4\pm1.7$\\
\hline
\end{tabular}
\caption{Summary of the results for the strong isospin breaking correction.}
\label{tab:strongIBsummary}
\end{table}

%% file: 5_conclusions.tex
\section{Conclusions and Outlook}
\label{sec:conclusions}
In this work we have calculated the isospin breaking corrections to meson 
masses 
and the hadronic vacuum polarization in an exploratory study on a 
$64\times24^3$ lattice with an inverse lattice spacing of $a^{-1}=1.78$~GeV 
using unphysical quark masses.\par
We have included electromagnetic effects in the lattice calculation with
two different approaches, a stochastic and a perturbative approach. To our 
knowledge, this work is the first direct comparison of results obtained 
from these two methods. In both methods, we have treated QED in an 
electro-quenched setup, i.e.\ we have considered the sea quarks as electrically 
neutral.
\par
As a starting point for comparing the stochastic and perturbative methods we 
have calculated the QED 
correction to meson masses. We have shown, that these QED corrections have to 
be extracted differently from stochastic or perturbative data, taking into 
account, that the stochastic data contains QED corrections to the correlation 
functions from all orders in $\alpha$, while the perturbative data only 
include $\Oalpha$ corrections. We find the results from the perturbative and 
the stochastic method to be consistent with each other up to small deviations, 
which are of $\Oalphasquare$. Albeit the $\Oalphasquare$ corrections are small, 
we are able to resolve these with the statistics used in this study.
\par
In this work we have determined for the first time the QED corrections to the 
hadronic vacuum polarization and its contribution to the anomalous 
magnetic moment of the muon $a_\mu$. We have 
calculated the QED correction to the local-conserved vector two-point function 
and to the multiplicative renormalization $Z_V$ for the local vector current 
used in our setup to calculate the HVP. An overview over our results for the 
QED correction to the HVP is presented in table \ref{tab:QEDcorr}. In total, we 
find the QED correction 
to $a_\mu$ to be $<1\%$ for the up quark and $0.1\%$ for the strange 
quark. However, one has to keep in mind that this 
calculation has not been done using physical quark masses.
In addition, we have determined the strong isospin correction to the HVP, which 
we find to be $\approx 0.9\%$. 
An important conclusion from this is, that when aiming at a calculation of 
$a_\mu$ with a precision of $1\%$,
QED and strong isospin breaking corrections would need to be included.
\par
Our data allows us to directly compare the statistical precision obtained 
from the stochastic and the perturbative method. We find that for the QED 
correction to the meson masses as well as for the HVP the stochastic method 
results in a statistical error which is about a factor of $1.5-2$ smaller than 
the statistical error from the perturbative method for the same numerical cost. 
Thus, the 
stochastic method is favourable for the particular choice of 
simulation parameters and quantities considered in this 
work. However, this might differ for a study using unquenched QED, where a cost 
comparison between both methods is less trivial.
\par
In this work we have not made any 
attempt to include finite volume corrections for the QED correction to the HVP.
Finite volume corrections with photons in a finite 
box can be substantial (cf.\ e.g.\ the results in \cite{Borsanyi:2014jba} and 
in table \ref{tab:qedcorrmesonmasses}) and 
thus need to be taken into account. We are currently investigating the finite 
volume corrections to the QED correction to $a_\mu$ to include those in our 
calculations.  
\par 
Having successfully completed this exploratory study with unphysical 
quark masses, a calculation of the QED corrections to the HVP at physical quark 
masses is under active investigation. This present work has demonstrated the 
methods and feasibility to enable the future work.

%% file: appendix.tex
\section{Domain Wall Action and Currents}
The Domain Wall fermion action used in this work is given by 
\cite{Furman:1994ky,Blum:2000kn}
\begin{equation}
 S_{F,0}[\Psi,\Psibar , U] = - \sum\limits_{x,x'} 
\sum\limits_{s,s'=0}^{L_s-1} \Psibar(x,s) D_0(x,s;x',s') \Psi(x',s')
\label{eq:DWFaction}
\end{equation}
with
\begin{equation}
  D_0(x,s;x',s') = \delta_{s,s'}D_0^{\parallel}(x,x') + \delta_{x,x'} 
D_0^{\perp}(s,s')
\end{equation}
and
\begin{align}
 D_0^{\parallel}(x,x')  = &
\frac{1}{2}\sum\limits_{\mu=1}^4\left[(1-\gamma_\mu) 
U_\mu(x)\delta_{x+\mu,x'} + (1+\gamma_\mu) 
U^\dagger_\mu(x')\delta_{x-\mu,x'} + (M_5-4)\delta_{x,x'}\right]\\
 D_0^{\perp}(s,s')  = &
\frac{1}{2}\left[(1-\gamma_5)\delta_{s+1,s'}+(1+\gamma_5)\delta_{s-1,s'}
-2\delta_{s,s'}\right ]
\label{eq:Dparallel}\\
& - \frac{m_f}{2}\left[(1-\gamma_5)\delta_{s,L_s-1}\delta_{0,s'} 
+(1+\gamma_5)\delta_{s,0}\delta_{L_s-1,s'}\right]\,.
\label{eq:Dperp}
\end{align}
To include couplings of photon fields to the quarks in the fermionic action 
\eqref{eq:DWFaction} one has to replace the gauge links in \eqref{eq:Dparallel} 
and \eqref{eq:Dperp} as follows
\begin{equation}
\begin{aligned}
 U_\mu(x)&\rightarrow e^{-ieq_fA_\mu(x)} U_\mu(x)\\
 U^\dagger_\mu(x)&\rightarrow e^{ieq_fA_\mu(x)} U^\dagger_\mu(x)\,.
\end{aligned}
\label{eq:linkQED_app}
\end{equation}

The conserved vector current $V_\mu^c(x)$ and the tadpole operator $T_\mu(x)$ 
are given by
\begin{equation}
 V_\mu^c(x) =\!\! \sum\limits_{s=0}^{L_s-1} 
\frac{1}{2}\left[\Psibar(x+\mu,s)(1+\gamma_\mu) U^\dagger_\mu(x) 
\Psi(x,s) - \Psibar(x,s)(1-\gamma_\mu) U_\mu(x) 
\Psi(x+\mu,s)\right]
 \label{eq:consvectorcurrent}
\end{equation}
\begin{equation}
 T_\mu(x) = \!\!\sum\limits_{s=0}^{L_s-1} 
\frac{1}{2}\left[\Psibar(x+\mu,s)(1+\gamma_\mu) U^\dagger_\mu(x) 
\Psi(x,s) + \Psibar(x,s)(1-\gamma_\mu) U_\mu(x) 
\Psi(x+\mu,s)\right]\,.
 \label{eq:tadpolop}
\end{equation}

\section{Expansion of the Path Integral}
The expansion of the expectation value of an observable $O$ in the 
electromagnetic coupling $e^2$ is given as
\begin{equation}
 \left<O\right> = \left<O\right>_{0} + 
\frac{1}{2}\,e^2\left.\frac{\partial^2}{\partial 
e^2}\left<O\right>\right|_{e=0} + \Oalphasquare\,.
\label{eq:eexpansion_app}
\end{equation}
The leading order electromagnetic correction is thus determined by 
\begin{equation}
 \frac{\partial^2}{\partial e^2} \left<O\right> =  \frac{\partial^2}{\partial 
e^2}\left[\frac{1}{Z}\! \int 
\!\!\mathcal{D}[U]\,\mathcal{D}[A]\,\mathcal{D}[\Psi,\Psibar]\,\,O\,\,
e^ { -S_F[\Psi,\Psibar ,A, U]}\,\,e^{ - S_\gamma[A]}\,e^{ - S_G[U]}\right]\,.
\label{eq:deviationO}
\end{equation}
In the electro-quenched approximation we do not include QED in the fermion 
determinant $\det D[U,A] \equiv\det D_0[U]$, and consequently the partition 
function $Z$ does not depend on the electromagnetic coupling $e$.
\subsection{Meson Two-Point Functions}
For a meson two-point function the observable $O$ is of the form $O = 
(q_f\Gamma q_f')(q_f\Gamma' q_f')$ for two quark flavours $f$ and $f'$, and 
does 
not depend on the elementary charge $e$. Thus, equation \eqref{eq:deviationO} 
can be written as 
\begin{equation}
 \frac{\partial^2}{\partial e^2} \left<O\right> =\frac{1}{Z}\! \int 
\!\!\mathcal{D}[U]\,\mathcal{D}[A]\,\mathcal{D}[\Psi,\Psibar]\,\,O\,\,\left( 
\frac{\partial^2}{\partial 
e^2}e^ { -S_F[\Psi,\Psibar ,A, U]}\right)\,\,e^{ - S_\gamma[A]}\,e^{ - 
S_G[U]}
\label{eq:derivativepath}
\end{equation}
In the following we drop the dependence of $S_F$ on the fields $\Psi,\Psibar ,A, 
U$ for simplicity, i.e.\ $S_F\equiv S_F[\Psi,\Psibar ,A, U]$. The derivative in 
\eqref{eq:derivativepath} can be written as
\begin{align}
 \frac{\partial^2}{\partial e^2} e^ { -S_F} = e^{ 
-S_F} \left[\left(\dde S_F\right)\left(\dde S_F\right) - 
\ddesq S_F\right]\,.
\end{align}
For the Domain Wall action \eqref{eq:DWFaction} including QED (cf. 
\eqref{eq:linkQED_app}) one finds
\begin{equation}
\begin{aligned}
 \dde S_F  = -\sum\limits_{x,\mu} iq_f \sum\limits_{s=0}^{L_s-1} 
\frac{1}{2}\Big[&\Psibar(x+\mu,s)(1+\gamma_\mu) 
e^{ieq_fA_\mu(x)} U^\dagger_\mu(x) 
\Psi(x,s) \\
&- \Psibar(x,s)(1-\gamma_\mu) e^{-ieq_fA_\mu(x)} U_\mu(x) 
\Psi(x+\mu,s)\Big] A_\mu(x)
 \label{eq:ddeS}
\end{aligned}
\end{equation}
and
\begin{equation}
\begin{aligned}
 \ddesq S_F  = \sum\limits_{x,\mu} q^2_f \sum\limits_{s=0}^{L_s-1} 
\frac{1}{2}\Big[&\Psibar(x+\mu,s)(1+\gamma_\mu) 
e^{ieq_fA_\mu(x)} U^\dagger_\mu(x) 
\Psi(x,s) \\
&+ \Psibar(x,s)(1-\gamma_\mu) e^{-ieq_fA_\mu(x)} U_\mu(x) 
\Psi(x+\mu,s)\Big] A_\mu(x)A_\mu(x)\,.
 \label{eq:ddesqS}
\end{aligned}
\end{equation}
Inserting this into \eqref{eq:derivativepath} yields
\begin{align}
 \frac{\partial^2}{\partial e^2} \left<O\right> \Big|_{e=0}\!\! = - q_f q_f' 
\!\!\! \sum\limits_{x,\mu;y,\nu}\!\!\!\left<O V_\mu^c(x) V_\nu^c(y) 
A_\mu(x)A_\nu(y)\right> -  q_f^2\sum\limits_{x,\mu}\! 
\left<OT_\mu(x)A_\mu(x)A_\mu(x)\right>
\end{align}
with the conserved vector current \eqref{eq:consvectorcurrent} and the tadpole 
operator \eqref{eq:tadpolop}. Using
\begin{equation}
 \left<A_\mu(x)A_\nu(y)\right>_\gamma = \Delta_{\mu\nu}(x-y)
\end{equation}
one finds for the expansion \eqref{eq:eexpansion_app} of the path integral at 
$\Oalpha$
\begin{equation}
 \left<O\right> = \left<O\right>_{0} - 
\frac{(eq_f)^2}{2} \left<O 
T_\mu(x)\right>_{0}\,\,\Delta_{\mu\mu}(0) 
- \frac{e^2q_fq_{f'}}{2}\left<O 
V^c_\mu(x)V^c_\nu(y)\right>_{0}\,\,\Delta_{\mu\nu}(x-y)\,,
\label{eq:pathintexp_app}
\end{equation}
where $\left<\cdot\right>_0$ is the expectation value over fermionic and 
gluonic fields.

\subsection{HVP}
\label{subsec:hvpexpansion}
For the QED correction to the HVP in the perturbative method we have to expand 
the path integral for an operator of the form
\begin{equation}
 O = V_\mu^{c,e}(z)V_\nu^l(0)\,,
 \label{eq:qedvectorcorrelator}
\end{equation}
with a local vector current $V_\nu^l$ and a conserved vector current 
$V_\mu^{c,e}$, which, including QED, is given by
\begin{equation}
\begin{aligned}
 V_\mu^{c,e}(x) =  \!\!\sum\limits_{s=0}^{L_s-1} 
\frac{1}{2}\big[&\Psibar(x+\mu,s)(1+\gamma_\mu) e^{ieq_f 
A_\mu(x)}U^\dagger_\mu(x) 
\Psi(x,s)\\
&- \Psibar(x,s)(1-\gamma_\mu) e^{-ieq_f 
A_\mu(x)}U_\mu(x) 
\Psi(x+\mu,s)\big]\,.
\end{aligned}
\end{equation}
Taking into account the explicit dependence of the operator on the 
electromagnetic coupling, one has to calculate
\begin{equation}
\begin{aligned}
 \frac{\partial^2}{\partial e^2} \left<O\right> =\frac{1}{Z}\! \int 
\!\!\mathcal{D}[U]\,\mathcal{D}[A]\,\mathcal{D}[\Psi,\Psibar]\,&\left[\,O\left( 
\frac{\partial^2}{\partial 
e^2}e^ { -S_F[\Psi,\Psibar ,A, U]}\right)+ \left(\ddesq 
O\right) e^ { -S_F[\Psi,\Psibar ,A, U]}\right. \\
&\,\,\,\,\,\,+ 2 \left.\left(\dde 
O\right)\left(\dde e^ { -S_F[\Psi,\Psibar ,A, U]}\right) \right]\,\,e^{ - 
S_\gamma[A]}\,e^{ - S_G[U]}
\end{aligned}
\label{eq:derivativepath_hvp}
\end{equation}
The derivatives of the conserved vector current $V_\mu^{c,e}$ with respect to 
$e$ are given by
\begin{equation}
\begin{aligned}
 \dde V_\mu^{c,e}(z) = iq_f\!\!\sum\limits_{s=0}^{L_s-1} 
\frac{1}{2}\big[&\Psibar(z+\mu,s)(1+\gamma_\mu) e^{ieq_f 
A_\mu(z)}U^\dagger_\mu(z) 
\Psi(z,s)\\
&+ \Psibar(z,s)(1-\gamma_\mu) e^{-ieq_f 
A_\mu(z)}U_\mu(z) 
\Psi(z+\mu,s)\big] A_\mu(z)\\
\end{aligned}
\end{equation}
and 
\begin{equation}
\begin{aligned}
 \ddesq V_\mu^{c,e}(z) = -q^2_f\!\!\sum\limits_{s=0}^{L_s-1} 
\frac{1}{2}\big[&\Psibar(z+\mu,s)(1+\gamma_\mu) e^{ieq_f 
A_\mu(z)}U^\dagger_\mu(z) 
\Psi(z,s)\\
&- \Psibar(z,s)(1-\gamma_\mu) e^{-ieq_f 
A_\mu(z)}U_\mu(z) 
\Psi(z+\mu,s)\big] A_\mu(z)A_\mu(z)\,.
\end{aligned}
\end{equation}
Thus, in total, one finds for the expansion of the path integral for the 
operator \eqref{eq:qedvectorcorrelator}
\begin{equation}
\begin{aligned}
 \left<V_\mu^{c,e}(z)V_\nu^l(0)\right> = &\left<V_\mu^c(z)V_\nu^l(0)\right>_{0} 
- \frac{(eq_f)^2}{2} \left<V_\mu^c(z)V_\nu^l(0)\, 
T_\mu(x)\right>_{0}\,\,\Delta_{\mu\mu}(0) \\
&- \frac{(eq_f)^2}{2}\left<V_\mu^c(z)V_\nu^l(0)
V^c_\mu(x)V^c_\nu(y)\right>_{0}\,\,\Delta_{\mu\nu}(x-y)\\
&-(eq_f)^2\left<T_\mu(z)V_\nu^l(0)V_\sigma^c(x)\right>_0 
\,\,\Delta_{\mu\sigma}(z-x)\\
&-\frac{(eq_f)^2}{2}\left<V_\mu^c(z)V_\nu^l(0)\right>_{0}\,\,\Delta_{\mu\mu}(0) 
+ \Oalphasquare\,.
\label{eq:pathintexphvp_app}
\end{aligned}
\end{equation}
The second term on the right-hand side of the first line of 
\eqref{eq:pathintexphvp_app} is the tadpole diagram, the term on the second 
line gives rise to the quark self-energy and the photon exchange diagram. The 
terms 
in the third and fourth line are the two terms shown in figure 
\ref{fig:qeddiagrams_hvp} and originate from the expansion of the operator.

\subsection{Strong Isospin Breaking}
\label{subsec:appendixstrongIB}
For determining the strong isospin correction using the path integral 
expansion, we have to calculate
\begin{equation}
 \left<O\right> = \left<O\right>_{m_s=\hat{m}} + 
(m_f - \hat{m}) \left.\frac{\partial}{\partial 
m_f}\left<O\right>\right|_{m_f=\hat{m}} + \mathcal{O}((m_f - \hat{m})^2)\,.
\label{eq:mexpansionapp}
\end{equation}
The derivative of the expectation value with respect to $e$ is given by
\begin{equation}
 \frac{\partial}{\partial m_f} \left<O\right> = \frac{1}{Z}\! \int 
\!\!\mathcal{D}[U]\,\mathcal{D}[\Psi,\Psibar]\,\,O\left(- 
\frac{\partial}{\partial 
m_f} S_F[\Psi,\Psibar , U] \right)e^ { -S_F[\Psi,\Psibar , U]}\,\,e^{ - 
S_G[U]}\,.
\end{equation}
For the Domain Wall Fermions used in this work, we find
\begin{equation}
 \begin{aligned}
\frac{\partial}{\partial m_f} S_F &= \sum\limits_{x,x'} 
\delta_{x,x'}\sum\limits_{s,s'=0}^{L_s-1} 
\Psibar(x,s)\left[\frac{1}{2} 
(1-\gamma_5)\delta_{s,L_s-1}\delta_{0,s'}+\frac{1}{2} (1+\gamma_5)\delta_{ 0,s 
}\delta_{s',L_s-1}\right] \\
&=\sum\limits_{x}\left[\Psibar(x,L_s-1)\frac{1}{2} 
(1-\gamma_5)\Psi(x,0)+\Psibar(x,0)\frac{1}{2} (1+\gamma_5)\Psi(x,L_s-1)\right]\\
&=\sum\limits_x \overline{\psi}(x)\psi(x)
 \end{aligned}
\end{equation}
with four dimensional fields $\overline{\psi}$, $\psi$. Thus, we find for the 
path integral expansion \eqref{eq:mexpansionapp}
\begin{equation}
  \left<O\right> = \left<O\right>_{m_f=\hat{m}} - 
(m_f - \hat{m}) \left<O\mathcal{S}\right>_{m_f=\hat{m}} + \mathcal{O}((m_f 
- \hat{m})^2)
\end{equation}
with the scalar current $\mathcal{S}=\sum\limits_x \overline{\psi}(x)\psi(x)$.

%% file: appendix_Oa2.tex
\section{$\Oalphasquare$ Effects}
\label{sec:Oalphasquare}
\subsection{Meson Masses}
The results from the stochastic and the perturbative method are expected to 
differ by effects which are of $\Oalphasquare$. Indeed, when comparing the QED 
correction to the effective mass (cf. figure \ref{fig:KaonQEDeffmass}), we find 
a deviation between both datasets of about $1\%$ of the QED correction itself. 
This deviation is consistent with $\Oalphasquare$ corrections. To explicitly 
check this, we have repeated the calculation with the stochastic method for a 
second larger value of the electromagnetic coupling $\alpha=1/4\pi$.
\par
\begin{figure}[h]
 \includegraphics[width=0.48\textwidth]{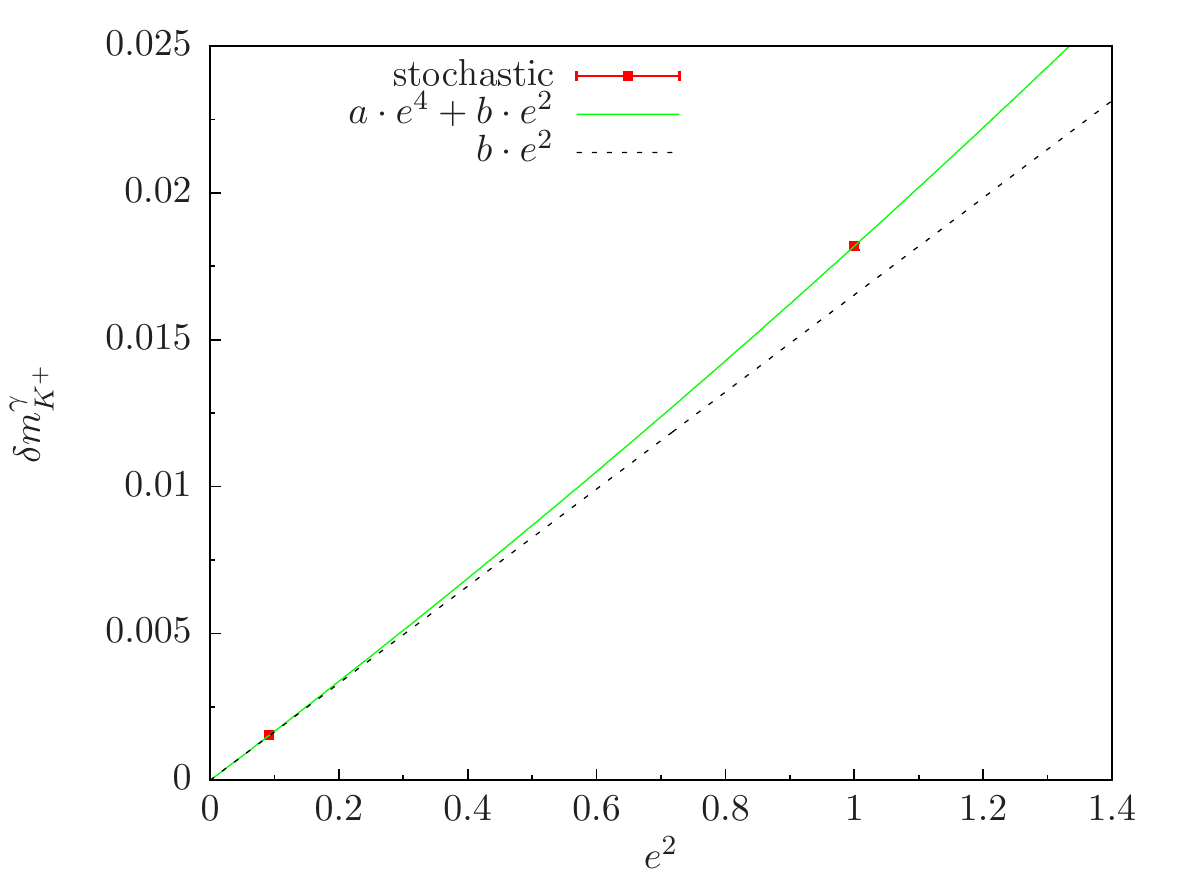}
 \includegraphics[width=0.48\textwidth]{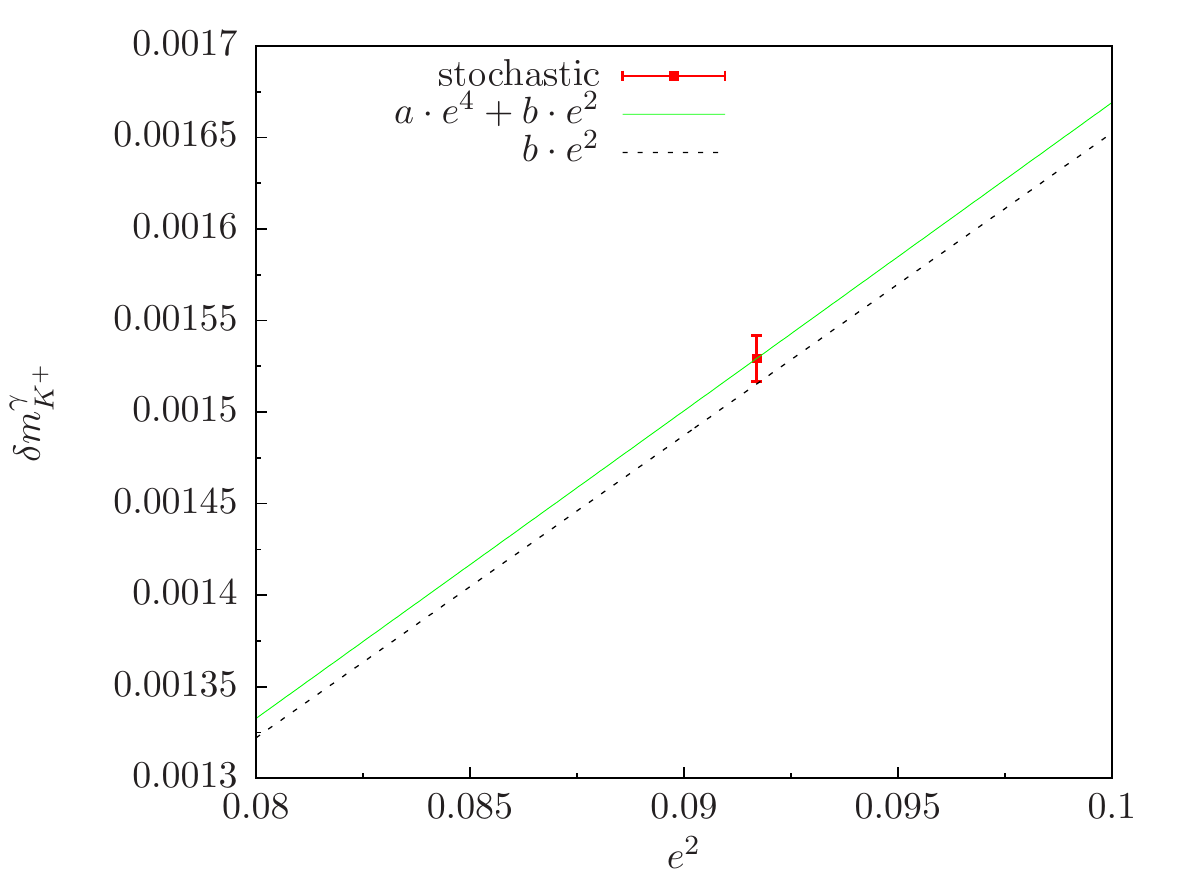}
 \caption{The QED correction to the mass of a charged kaon from the stochastic 
data plotted against the electromagnetic coupling $e^2$.}
\label{fig:K+_vs_e}
 \end{figure}
Figure 
\ref{fig:K+_vs_e} shows the QED correction to the mass of the charged kaon for 
the stochastic data calculated with both values of $\alpha$ plotted against 
$e^2 = 4\pi\alpha$. The solid green curve is a quadratic function of the form 
$a\cdot e^4 + b\cdot e^2$, that was matched to the two data points. The dotted 
grey line is the linear term $b\cdot e^2$ of the quadratic curve, i.e\ the 
leading order QED contribution. The plot on the right-hand side of figure 
\ref{fig:K+_vs_e} shows a zoom around the physical value of the coupling. One 
can clearly see the difference of the leading order contribution $b\cdot e^2$ 
and the full results from the stochastic method, which we find at physical 
$e^2$ 
to be at the same order as the statistical error. Thus, with the statistics 
used in this work, we are able to resolve also the $\alpha^2$ effects.
\par
The $\Oalphasquare$ effect that we obtain from the stochastic data is shown by 
the solid green line on the right-hand side of in figure 
\ref{fig:KaonQEDeffmass}. 
We find the $\Oalphasquare$ contribution to be consistent with the difference 
that we observe between stochastic and perturbative data.
\par
In \cite{Matzelle:2017qsw} the authors discuss finite volume effects for 
the next-to-leading order QED corrections to meson masses. 
However, for our calculation with the physical value of $\alpha$ we find the 
$\Oalphasquare$ effects to be about $1\%$ of the leading QED correction, and 
thus, considering different finite volume effects for the $\Oalphasquare$  
contributions included in the stochastic data is not 
relevant at the current level of precision.

\subsection{HVP}
\begin{figure}[h]
 \centering
 \includegraphics[width=0.48\textwidth]{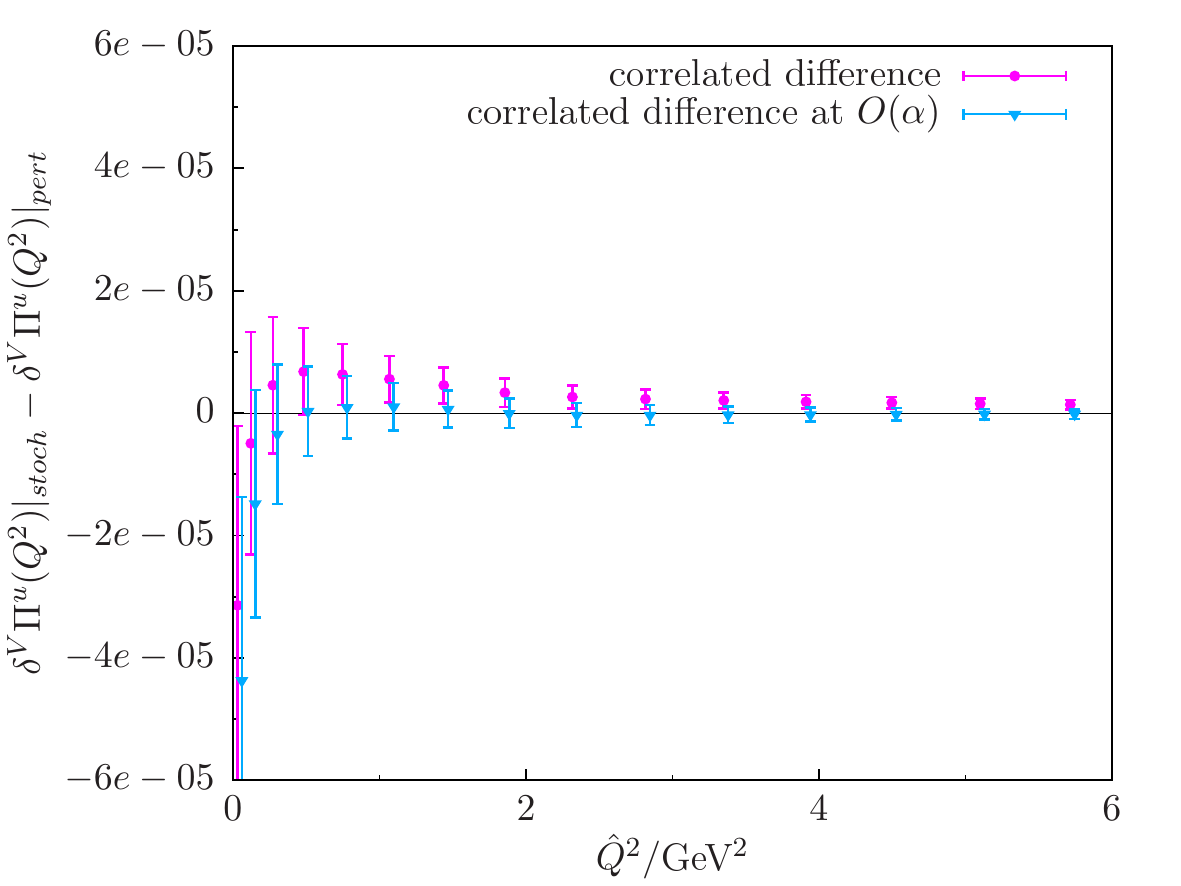}
  \includegraphics[width=0.48\textwidth]{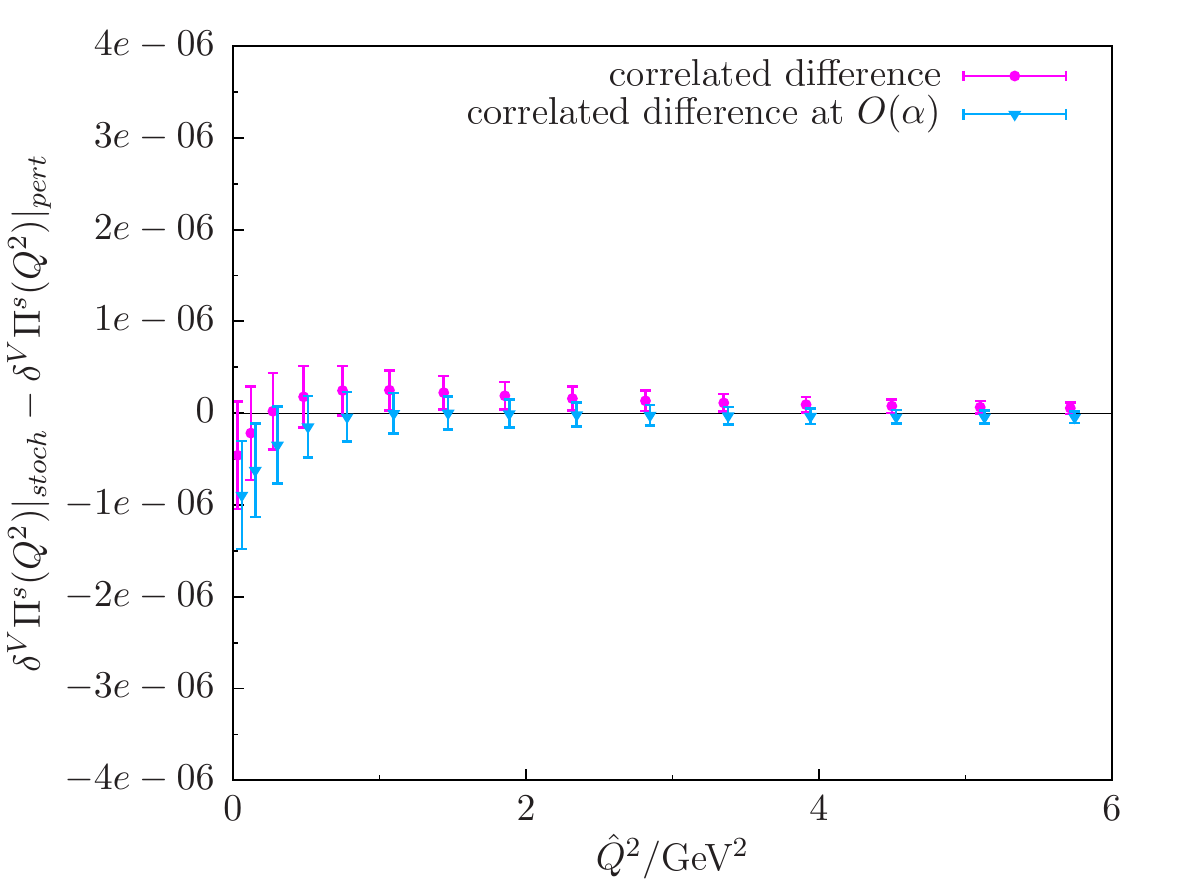}
  \caption{Difference between the QED correction to the HVP from stochastic and 
perturbative method. Purple circles show results using the data from the 
computation at the physical value of the coupling $\alpha$ for the stochastic 
method. Light blue triangles are using only the $\Oalpha$ correction from the 
stochastic data.}
  \label{fig:HVPatOalpha}
\end{figure}

When comparing the results for the HVP from perturbative and stochastic data 
(cf. figures \ref{fig:qedhvp_u} and \ref{fig:qedhvp_s}) we found a deviation 
between both datasets at the level of $1-1.5\sigma$. To check if this deviation 
originates from $\Oalphasquare$ effects, which are only included in the 
stochastic data, we use results from a computation with a larger value of the 
coupling $\alpha = 1/4\pi$. Using the data from 
two different values of $\alpha$ for the stochastic method, we are able to 
extract the $\Oalpha$ contribution at the physical value of the coupling. 
For every value of the four-momentum transfer $Q^2$ we match a quadratic curve 
$a\cdot e^4 + b\cdot e^2$ through the two data points, similarly as described 
for the QED correction to the meson masses above. The leading order QED 
correction is determined by the term linear in $e^2$. The results for the 
leading QED correction from the stochastic data can then be compared with the 
results from the perturbative method. Figure \ref{fig:HVPatOalpha} shows the 
correlated difference between stochastic and perturbative data for up quarks 
(left plot) and strange quarks (right plot). Purple circles show the difference 
using the results from the stochastic data which still include effects to all 
orders in $\alpha$ (i.e.\ the same points that were already shown in figures 
\ref{fig:qedhvp_u} and \ref{fig:qedhvp_s}). The light blue triangles show the 
difference using the $\Oalpha$ contribution from the stochastic data. We find 
the $\Oalpha$ data from the stochastic method to be in agreement with the 
results from the perturbative method over a wide range of $Q^2$. Thus, the 
difference between stochastic and perturbative data found in section 
\ref{subsubsec:hvpresults} is consistent with effects which are of higher 
order in $\alpha$.
\par
When calculating the QED correction to $a_\mu$ using only 
the $\Oalpha$ contribution from the stochastic data, we find the change in 
$\delta^V\!a_\mu$ to be much smaller than the statistical errors itself. Thus, 
we find $\Oalphasquare$ effects to be not relevant for $\delta a_\mu$ at this 
level of precision.  

%% file: appendix_coulomb.tex
\section{Comparison Coulomb and Feynman Gauge}
\label{sec:coulomb}
The quantities calculated in this work (QED corrections to meson masses and 
HVP) are expected to be gauge invariant. Gauge invariance is not broken by in 
the QED$_L$ prescription, i.e.\ subtracting the spatial zero modes, since gauge 
invariance is realized separately on every mode in momentum space.
To numerically check for gauge 
invariance, we compare results using the Feynman gauge with results using 
the Coulomb 
gauge. Coulomb gauge photon fields can be obtained from Feynman gauge photon 
fields using an appropriate projector \cite{Borsanyi:2014jba}
\begin{equation}
 \left(P_{C}\right)_{\mu\nu}  =  
\delta_{\mu\nu}-\left|\vec{\hat{k}}\right|^{-2}\hat{k}_{\mu}\left(0,\vec{\hat{k}
}\right)_{\nu} 
\hspace{1cm}\textrm{with}\hspace{0.3cm}\tilde{A}^\textrm{Coul}_\mu(k) =  
\left(P_{C}\right)_{\mu\nu} \tilde{A}^\textrm{Feyn}_\nu(k)\,.
\end{equation}
The photon propagator in Coulomb gauge is given by
\begin{equation}
\begin{aligned}
& \Delta^\textrm{Coul}_{\mu\nu}(x-y) = 
\left<A^\textrm{Coul}_\mu(x)A^\textrm{Coul}_\mu(y)\right>_\gamma\\
&= 
\frac{1}{N}\sum\limits_{k,\vec{k}\neq 0} 
e^{ik\cdot(x-y)} e^{ik\cdot (\hat{\mu} -\hat{\nu})/2}\,\, 
\frac{1}{\hat{k}^2}\left(\delta_{\mu\nu} - 
\frac{1}{\vec{\hat{k}}^2}\left(\hat{k}_\mu\tilde{\hat{k}}_\nu+\tilde{\hat{k}}
_\mu \hat{k}_\nu-\hat{k}_\mu \hat{k}_\nu\right)\right)
\label{eq:coulomb_prop}
\end{aligned}
\end{equation}
with $\tilde{k}_\mu\equiv(0,\vec{k})_\mu$. The phase factor $\exp(ik\cdot 
(\hat{\mu} 
-\hat{\nu})/2)$ in equation \eqref{eq:coulomb_prop} originates from the 
Fourier transformation with photon fields defined on the mid-links of the 
lattice. Note that this phase factor cancels for diagonal contributions 
$\mu=\nu$.
\par
For the stochastic method we have calculated the QED contributions using the 
same set of statistics with the Feynman and the Coulomb gauge. For the 
perturbative 
method the calculation using the Coulomb gauge is more expensive in our setup, 
since also contributions from $\mu\neq\nu$ have to be determined. Thus, we 
restrict the calculation using the Coulomb gauge with the perturbative method 
to only one source position.
\par
\begin{figure}[h]
\centering
\includegraphics[width=0.48\textwidth]{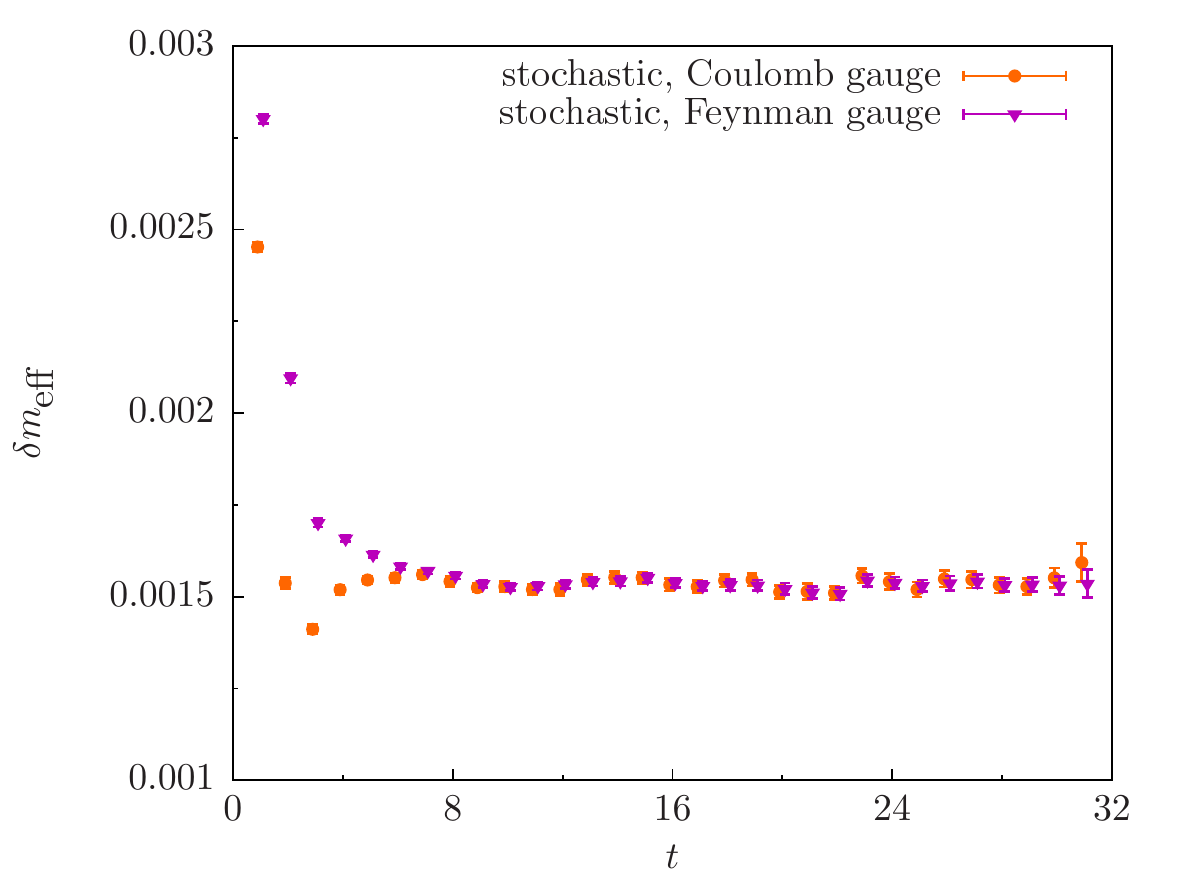}
\includegraphics[width=0.48\textwidth]{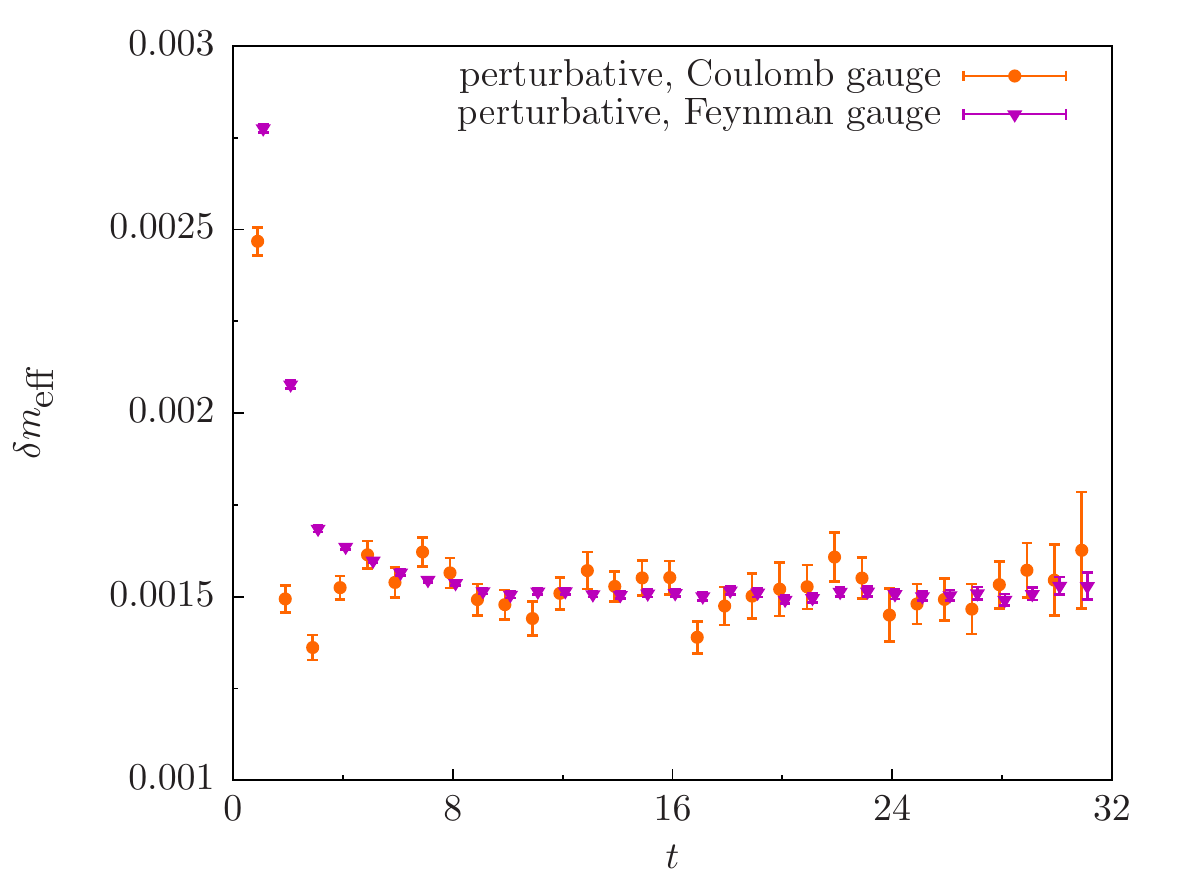}
\caption{Comparison of the QED correction to the effective mass of a charged 
kaon between Feynman (purple triangles) and Coulomb (orange circles) gauge. The 
plot on the left shows data from the stochastic method, the plot on the right 
data from the perturbative method. Note, that the results for the Coulomb 
gauge 
with the perturbative method have been obtained on a subset of the statistics.}
\label{fig:meson_coulomb_feynman}
\end{figure}

In figure \ref{fig:meson_coulomb_feynman} the QED correction to the effective 
mass of a charged kaon is shown using the Coulomb and the Feynman gauge for the 
stochastic method (left) and the perturbative method (right). Purple triangles 
show results in the Feynman gauge, orange circles results in the Coulomb gauge.
\par
We find agreement between the QED correction to the effective mass in 
the Feynman 
and the Coulomb gauge for large $t$, where the QED correction to the meson mass 
is determined. For small $t$ where the data contains contributions from excited 
states, we find deviations between Feynman and Coulomb gauge. The data in this 
region also depends on the creation amplitudes of the states, which are 
not necessarily gauge independent quantities.